\documentclass[twocolumn]{aastex62}

\newcommand{\HI}{\ion{H}{1}}
\newcommand{\HII}{\ion{H}{2}}
\newcommand{\Msol}{$M_\odot$}
\newcommand{\kms}{\mbox{km~s$^{-1}$}}
\def\twco{\mbox{$^{12}$CO}}
\def\ttco{\mbox{$^{13}$CO}}
\defcitealias{Solomon:87}{S87}
\defcitealias{Wong:17}{Paper~I}

\shorttitle{Structure Sizes and Line Widths in the LMC}
\shortauthors{Wong et al.}

\begin{document}

\title{\Large\bf Relations Between Molecular Cloud Structure Sizes and Line Widths in the Large Magellanic Cloud}

\correspondingauthor{Tony Wong}
\email{wongt@illinois.edu}

\author[0000-0002-7759-0585]{Tony Wong}
\affil{Department of Astronomy, University of Illinois, 
Urbana, IL 61801, USA}

\author{Annie Hughes}
\affil{Universit\'{e} de Toulouse, UPS-OMP, 31028 Toulouse, France}
\affil{CNRS, IRAP, Av. du Colonel Roche BP 44346, 31028 Toulouse cedex 4, France}

\author{Kazuki Tokuda}
\affil{Department of Physical Science, Graduate School of Science, Osaka Prefecture University, 1-1 Gakuen-cho, Naka-ku, Sakai, Osaka 599-8531, Japan}
\affil{National Astronomical Observatory of Japan, 2-21-1 Osawa, Mitaka, Tokyo, 181-8588, Japan}

\author{R\'emy Indebetouw}
\affil{Department of Astronomy, University of Virginia, P.O. Box 400325, Charlottesville, VA 22904, USA}
\affil{National Radio Astronomy Observatory, 520 Edgemont Road Charlottesville, VA 22903, USA}

\author{Toshikazu Onishi}
\affil{Department of Physical Science, Graduate School of Science, Osaka Prefecture University, 1-1 Gakuen-cho, Naka-ku, Sakai, Osaka 599-8531, Japan}

\author{Jeffrey B. Bandurski}
\affil{Department of Astronomy, University of Illinois, 
Urbana, IL 61801, USA}

\author[0000-0002-3925-9365]{C.-H. Rosie Chen}
\affil{Max-Planck-Institut f\"ur Radioastronomie, Auf dem H\"ugel 69, D-53121 Bonn, Germany}

\author{Yasuo Fukui}
\affil{Department of Physics, Nagoya University, Chikusa-ku, Nagoya 464-8602, Japan}

\author{Simon C. O. Glover}
\affil{Institut fu\"r Theoretische Astrophysik, Zentrum f\"ur Astronomie der Universit\"at Heidelberg, Albert-Ueberle-Str 2, D-69120 Heidelberg, Germany}

\author{Ralf S. Klessen}
\affil{Institut fu\"r Theoretische Astrophysik, Zentrum f\"ur Astronomie der Universit\"at Heidelberg, Albert-Ueberle-Str 2, D-69120 Heidelberg, Germany}

\author[0000-0001-8898-2800]{Jorge L. Pineda}
\affil{Jet Propulsion Laboratory, California Institute of Technology, 4800 Oak Grove Drive, Pasadena, CA 91109, USA}

\author{Julia Roman-Duval}
\affil{Space Telescope Science Institute, 3700 San Martin Drive, Baltimore, MD 21218, USA}

\author[0000-0003-2248-6032]{Marta Sewi{\l}o}
\affil{CRESST II and Exoplanets and Stellar Astrophysics Laboratory, NASA Goddard Space Flight Center, Greenbelt, MD 20771, USA}
\affil{Department of Astronomy, University of Maryland, College Park, MD 20742, USA}

\author{Evan Wojciechowski}
\affil{Department of Astronomy, University of Illinois, 
Urbana, IL 61801, USA}

\author[0000-0001-6149-1278]{Sarolta Zahorecz}
\affil{Department of Physical Science, Graduate School of Science, Osaka Prefecture University, 1-1 Gakuen-cho, Naka-ku, Sakai, Osaka 599-8531, Japan}
\affil{National Astronomical Observatory of Japan, 2-21-1 Osawa, Mitaka, Tokyo, 181-8588, Japan}

\accepted{\today}

\begin{abstract}
We present a comparative study of the size-line width relation for substructures within six molecular clouds in the Large Magellanic Cloud (LMC) mapped with the Atacama Large Millimeter/submillimeter Array (ALMA).  Our sample extends our previous study, which compared a {\it Planck} detected cold cloud in the outskirts of the LMC with the 30 Doradus molecular cloud and found the typical line width for 1 pc radius structures to be 5 times larger in 30 Doradus.  By observing clouds with intermediate levels of star formation activity, we { find evidence that line width at a given size increases with increasing local and cloud-scale 8\,$\mu$m intensity}.  At the same time, line width at a given size appears to independently correlate with measures of mass surface density.  Our results suggest that both virial-like motions due to gravity and local energy injection by star formation feedback play important roles in determining intracloud dynamics.
\end{abstract}


\section{Introduction}

The physical conditions within giant molecular clouds establish the initial conditions for star formation, thus understanding the factors that determine molecular cloud properties is of major interest.  A correlation between size and line width, of the form $\sigma_v \propto R^{\alpha}$ with $\alpha \approx 0.5$, has long been noted in samples of nearby molecular clouds \citep[hereafter \citetalias{Solomon:87}]{Larson:81,Solomon:87}.  This correlation, hereafter referred to as the $R$--$\sigma_v$ relation, is usually interpreted as the result of turbulent motions in the interstellar medium on all scales \citep{MacLow:04, Falgarone:09}. It closely resembles the turbulent cascade with a power-law slope falling between the \citet{Kolmogorov:41} and \citet{Burgers:39} values for incompressible and highly supersonic turbulence, respectively \cite[see also][]{Falgarone:94,Brunt:02, Kritsuk:13,Federrath:13}. At the same time, a study of \ttco\ emission in the Boston University-FCRAO Galactic Ring Survey by \citet{Heyer:09} showed that the normalization of the relation, $v_0 = \sigma_v/R^{1/2}$, exhibits a linear correlation with mass surface density, $\Sigma=M/\pi R^2$, across more than an order of magnitude in $\Sigma$.  The data are consistent with a state of virial balance between gravity and turbulent motions, except that the observed normalization $v_0$ is about a factor of 2 too large.  Subsequent work by \citet{Field:11} has suggested that the larger than expected $v_0$ may result from external pressure confinement, although to a lesser extent than has been inferred for clouds in the outer Galaxy \citep{Heyer:01} or near the Galactic Center \citep{Oka:01, Shetty:12}.  On the other hand, \citet{Ballesteros:11a} have interpreted the \citet{Heyer:09} result in terms of gravitational collapse near free-fall, which differs from the virial equilibrium prediction by a factor of $\sqrt{2}$ in $v_0$, and is thus roughly consistent with the GRS data.  A third possibility is that errors in the measured or inferred cloud properties create the appearance of excess kinetic energy when in fact clouds are close to being virialized.

Since interstellar turbulence transfers energy across spatial scales, the $R$--$\sigma_v$ relation can be studied on scales much smaller than the full extent of molecular clouds \citep[e.g.,][]{Myers:83}, all the way down to the $\sim$0.1 pc scales at which the thermal contribution to the line width becomes significant \citep{Goodman:98}. Kinetic energy spectra derived from techniques such as principal component analysis (PCA; e.g.\ \citealt{Brunt:02,Brunt:03}) or wavelet transforms \citep[e.g.,][]{Ossenkopf:02} have been interpreted as requiring turbulence to be driven on large ($>$10 pc) scales. The energy can be provided by gas accretion onto the disk of the Galaxy \citep{Klessen:10} or to some degree by large-scale  instabilies and spiral waves \citep{Wada:02}.  Stellar feedback in the form of supernovae, winds or expanding \HII\ regions also produce enough energy to explain the observed energy budget \cite[for reviews, see e.g.][]{MacLow:04, Krumholz:14,  Klessen:16}, and so there is ongoing debate about the astrophysical origin of the observed turbulent motions. Progress is slowed by the fact that {\it a priori} it can be difficult to identify the driving scale of turbulence because energy can cascade to both larger and smaller scales \citep{Vestuto:03}.  Another complication arises from the heterogeneous nature of Galactic data sets: high spatial resolution data are generally limited to the nearest clouds, and for most larger samples of clouds, distance uncertainties and distance-related selection effects are quite significant \citep[e.g.,][]{Traficante:18}.

The Large Magellanic Cloud (LMC) serves as an ideal laboratory to study molecular cloud turbulence on a range of scales and across a range of galaxy environments.  The spatial dynamic range achievable from ground-based observations is rivalled only by observations of the Milky Way and M31, whereas the proximity ($d \approx 50$ kpc; \citealt{Pietrzynski:19}) and low inclination ($i\approx 34\arcdeg$; \citealt{vanderMarel:14}) of the galaxy enables clouds and \HII\ regions to be easily identified.  With the resolution provided by the
Atacama Large Millimeter/submillimeter Array (ALMA), we have been investigating molecular cloud substructure across the LMC\@.  Contrasting two clouds with very different star formation activity at the same spatial resolution, \citet[hereafter \citetalias{Wong:17}]{Wong:17} found that the GMC associated with the actively star-forming 30 Doradus region shows a factor of $\sim$5 higher line width at a given spatial scale than a quiescent GMC in the outskirts of the galaxy.  In this paper, we extend our previous study by analyzing four additional LMC clouds showing intermediate levels of star formation activity.

\begin{figure*}
    \centering
    \includegraphics[width=0.49\textwidth]{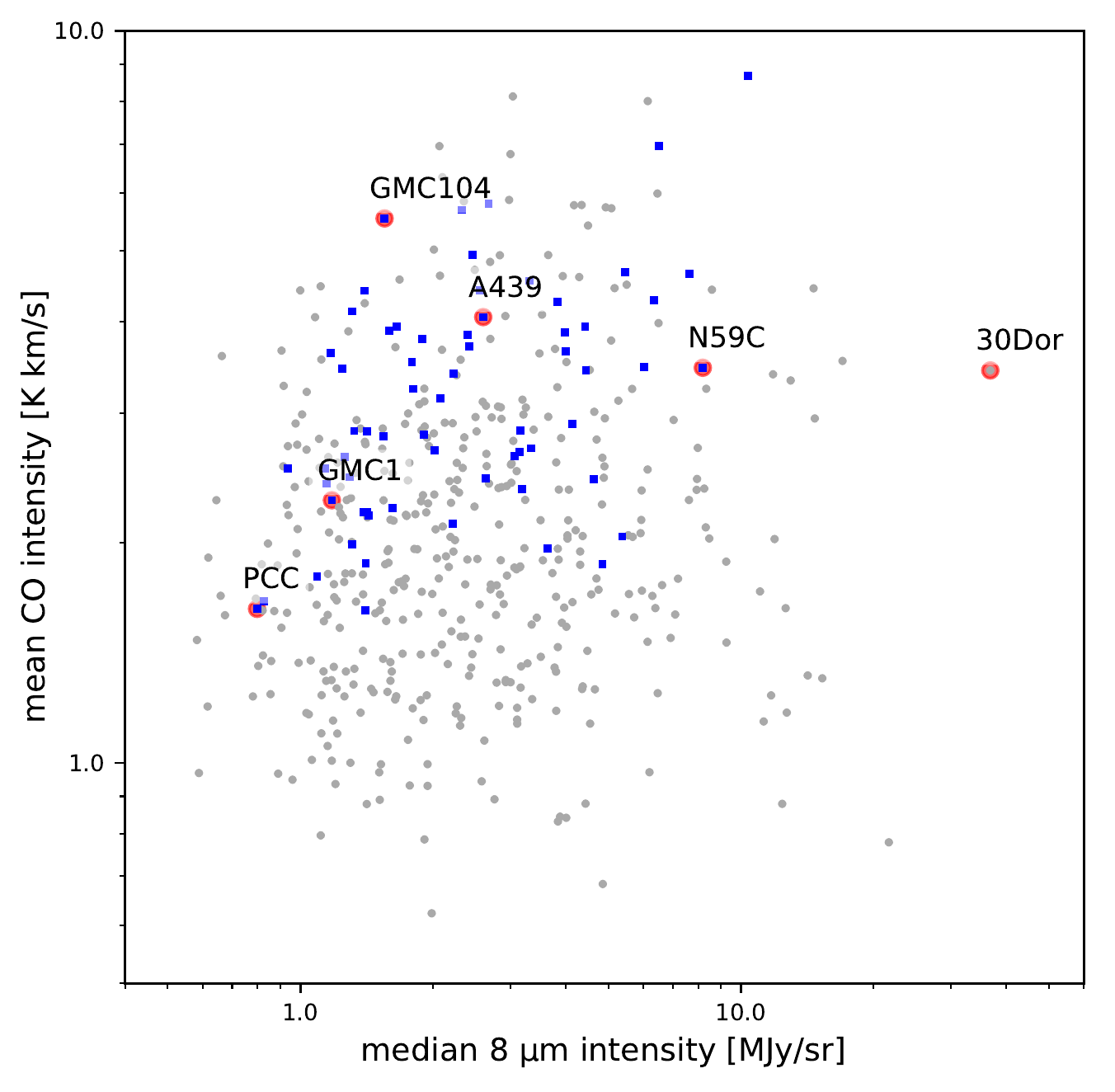}\hfill
    \includegraphics[width=0.49\textwidth]{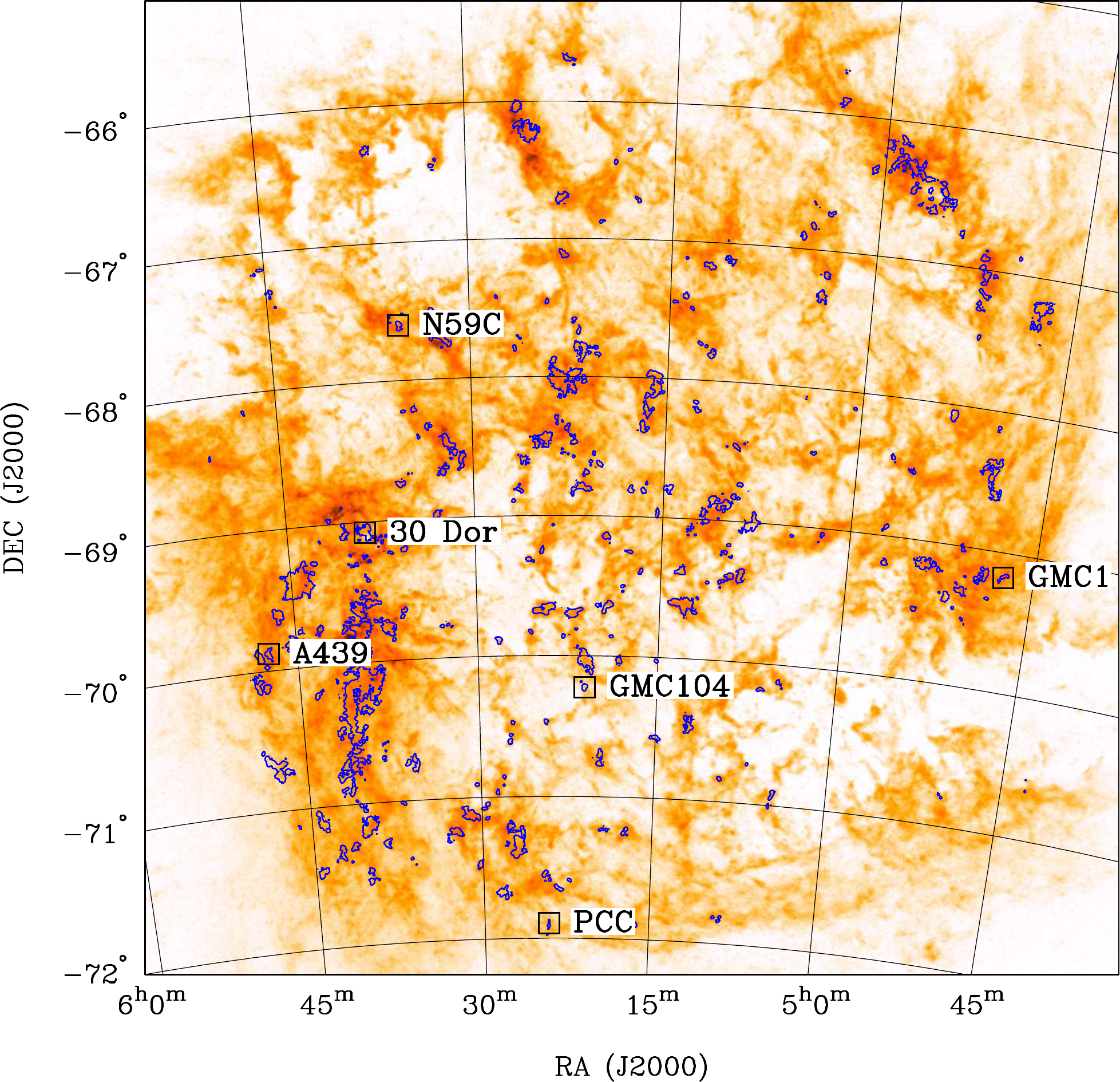}
    \caption{{\it Left}: Distribution of MAGMA CO ``islands'' in mean CO and median 8$\mu$m intensity.  Light gray points correspond to all islands; blue squares are those which match our size selection criteria.  The four clouds newly observed with ALMA, along with the two extreme clouds previously studied in \citetalias{Wong:17}, are indicated as red shaded circles. {\it Right}: Location of the six clouds presented in this work relative to the large-scale structure of CO and \HI\ in the LMC, as mapped by MAGMA (\citealt{Wong:11}, blue contours) and ATCA (\citealt{Kim:98}, color scale).}
    \label{fig:lmchi}
\end{figure*}

\section{Observations}

\subsection{Sample Selection}

Using data from the Magellanic Mopra Assessment (MAGMA) survey, \cite{Wong:11} identified 450 regions of contiguous CO emission (``islands'') in the LMC via the CPROPS emission segmentation software \citep{Rosolowsky:06}.  Additional observations conducted in 2012--3 increased the MAGMA coverage of the LMC by $\sim$20\%, and the MAGMA Data Release 3 (DR3) contains 472 ``islands'' defined using a slightly more stringent set of CPROPS parameters (requiring a 3.5$\sigma$ rather than 3$\sigma$ peak).  To select the cloud sample for ALMA mapping, we examined the joint distribution of CO and 8$\mu$m intensities for the islands, as measured by the MAGMA and {\it Spitzer} SAGE \citep{Meixner:06} programs respectively, considering only the 65 clouds with angular areas between 6.25 and 12.5 arcmin$^2$ (Figure~\ref{fig:lmchi}, {\it left}).  This range of cloud sizes was chosen to maximize the achievable spatial dynamic range while still allowing the mosaic to be completed within a single ALMA observation.  We aimed for a sample exhibiting high CO intensity{\bf, spanning a wide range in 8$\mu$m intensity, and distributed broadly across the galaxy, while also being} complementary to already approved ALMA observations.  Our final sample consisted of four clouds, which we label as GMC1 and GMC104 (designations from the NANTEN cloud catalogue of \citealt{Fukui:08}), A439 (designation from the island catalogue of \citealt{Wong:11}), and N59C (designation based on the associated \HII\ region catalogued by \citealt{Henize:56}).  To summarize, our sample consists of bright CO-emitting clouds selected to have a radius of $\sim$25 pc and to exhibit a wide range of mid-infrared surface brightness.  The locations of our sample clouds within the LMC, including the two additional clouds studied in \citetalias{Wong:17}, are indicated in Figure~\ref{fig:lmchi} ({\it right}).

\begin{deluxetable*}{lccccccccc}[t]
\tablehead{
\colhead{Region} & \colhead{Transition} & \colhead{Ref.\ R.A.} & \colhead{Ref.\ Dec.} & \colhead{Mosaic Size} & \colhead{$\Delta v_{\rm ch}$} & \colhead{$T_{\rm rms,12}$\tablenotemark{a}} & \colhead{$T_{\rm peak,12}$\tablenotemark{b}} & \colhead{$T_{\rm rms,13}$} & \colhead{$T_{\rm peak,13}$}\\
\colhead{} & \colhead{} & \colhead{(J2000)} & \colhead{(J2000)} & \colhead{$(\arcsec \times \arcsec)$} & \colhead{(km s$^{-1}$)}
& \colhead{(K)} & \colhead{(K)} & \colhead{(K)} & \colhead{(K)}}
\tablecaption{Summary of Map Parameters at a Common Resolution of 3\farcs5\label{tab:regs}}
\startdata
30 Dor & $J$=2--1 & 5$^{\rm h}$38$^{\rm m}$47\fs0 & $-69$\degr 04\arcmin 36\arcsec & 50 $\times$ \phn50 & 0.5 & 0.05 & 43.6 & 0.05 & 14.1\\
N59C & $J$=1--0 & 5$^{\rm h}$35$^{\rm m}$18\fs8 & $-67$\degr 36\arcmin 12\arcsec & 160 $\times$ 260 & 0.2 & 0.19 & 26.9 & 0.14 & \phn6.2\\
A439 & $J$=1--0 & 5$^{\rm h}$47$^{\rm m}$26\fs1 & $-69$\degr 52\arcmin 46\arcsec & 150 $\times$ 165 & 0.2 & 0.17 & 13.2 & 0.11 & \phn4.8\\
GMC104 & $J$=1--0 & 5$^{\rm h}$21$^{\rm m}$04\fs0 & $-70$\degr 13\arcmin 29\arcsec & 150 $\times$ 160 & 0.2 & 0.21 & 17.4 & 0.10 & \phn6.4\\
GMC1 & $J$=1--0 & 4$^{\rm h}$47$^{\rm m}$30\fs8 & $-69$\degr 10\arcmin 32\arcsec & 320 $\times$ 130 & 0.2 & 0.19 & 14.6 & 0.12 & \phn5.2\\
PCC & $J$=2--1 & 5$^{\rm h}$24$^{\rm m}$09\fs2 & $-71$\degr 53\arcmin 37\arcsec & 80 $\times$ 220 & 0.2 & 0.13 & 10.4 & 0.14 & \phn3.9\\
\enddata
\tablenotetext{a}{rms noise in a channel map of width $\Delta v_{\rm ch}$.}
\tablenotetext{b}{Peak brightness temperature in cube.}
\end{deluxetable*}

\subsection{ALMA Cycle 4 Data}\label{sec:almadata}

Observations toward four molecular clouds in the LMC were obtained in ALMA Cycle 4 under project code 2016.1.00193.S (PI: Wong).  Each cloud was observed in two frequency settings, the first covering the \twco(1--0) line and the second covering \ttco(1--0), C$^{18}$O(1--0), CS(2--1), and C$^{34}$S(2--1).  The velocity resolution after Hanning smoothing was 61 kHz ($\sim$0.16 \kms) across a bandwidth of 59 or 117 MHz.  Although the CS(2--1) line was detected in all four clouds, in this paper we consider only the \twco\ and \ttco\ data.  Observations of \twco\ were conducted in both the 12m and 7m arrays, providing sensitivity to structures up to 45\arcsec\ in size, while observations of \ttco\ were conducted in the 12m array only, which is sensitive to structures up to 30\arcsec\ in size at the observing frequency of 110 GHz.  

While we did not revisit or modify the system calibration, we re-imaged the calibrated visibilities using the CASA package \citep{McMullin:07}.  For each spectral line, the visibility data were imaged together using the {\tt tclean} task in CASA 5.0.0.  The imaging grid was set to have square pixels of width 0.5\arcsec\ and velocity channels spaced by 0.2 \kms.  The size of the imaging grid was 800 pixels square for all clouds besides GMC1, for which extending the grid to 1000 pixels in R.A. was necessary.  The flexible visibility weighting scheme of \citet{Briggs:95} was used.  For most clouds we obtained satisfactory imaging results using the {\tt auto-multithresh} procedure in {\tt tclean} with the {\tt clark} deconvolver.  In brief, the procedure continuously updates the deconvolution mask based on the current residual image.  The initial masking of the image is controlled by the parameters {\sl pbmask}, {\sl sidelobethreshold} and {\sl noisethreshold}, which we set to 0.2, 2, and 3 respectively.  Regions smaller than the beam are pruned from the mask ({\sl minbeamfrac}=1) and the resulting mask is convolved with a Gaussian 4 times the synthesized beam size ({\sl smoothfactor}=4) and clipped at 10\% of the smoothed peak ({\sl cutthreshold}=0.1).  We also expand the mask to an enclosing low-level contour of 1.5$\sigma$ ({\sl lownoisethreshold}=1.5) to allow recovery of fainter extended emission.  We found slightly different {\tt tclean} parameters worked best for generating the deconvolution mask for the 7m data ({\sl noisethreshold}=5, {\sl minbeamfrac}=0.5, {\sl smoothfactor}=1; all other parameters left unchanged).  Deconvolution proceeded down to a 1$\sigma$ stopping threshold.

For combining the 12m and 7m data for \twco, a hybrid imaging procedure was adopted, in order to ensure that the larger field-of-view 7m data were properly deconvolved.  The 12m and 7m data were separately deconvolved using the {\tt auto-multithresh} procedure in {\tt tclean}, as described above, and then the combined visibilities were imaged and deconvolved using the union of the 12m and 7m masks, with no further mask adjustment.  This final deconvolution was performed using the {\tt multiscale} deconvolver \citep{Cornwell:08} with scales of 0, 4, and 12 pixels, a stopping threshold of 2$\sigma$, and a {\it smallscalebias} parameter of 0.6 (the default value).

For one cloud (GMC104), significant sidelobes persisted after applying our standard imaging procedure, and interactive masking needed to be applied to finalize the imaging.  We found that this procedure resulted in a $\sim$50\% increase in the total integrated CO flux, but a negligible impact on the properties of the structures that are measured following decomposition (\S\ref{sec:decomp}).  This is not surprising, given that the structure properties are primarily sensitive to the brightest emission peaks, where the deconvolved flux is well-constrained.

\begin{figure*}
    \centering
    \includegraphics[width=0.48\textwidth]{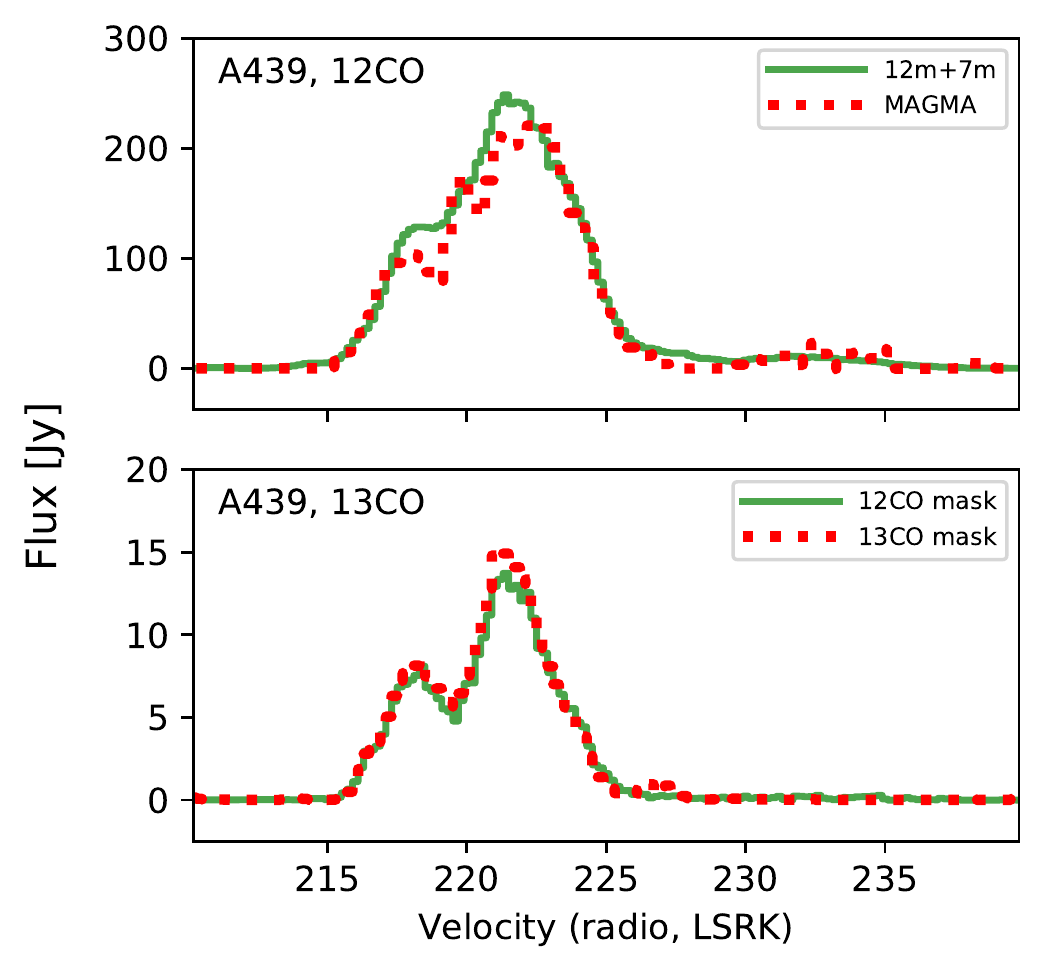}
    \includegraphics[width=0.48\textwidth]{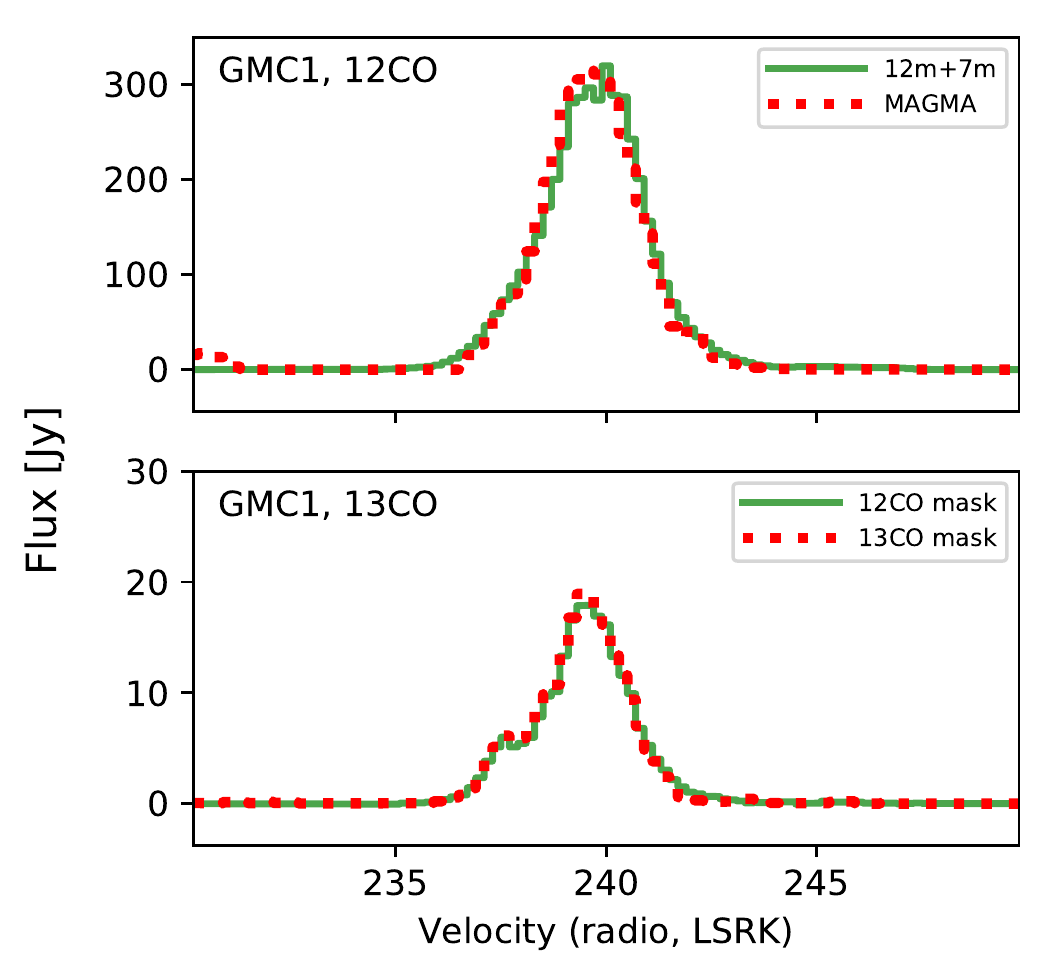}\\
    \includegraphics[width=0.48\textwidth]{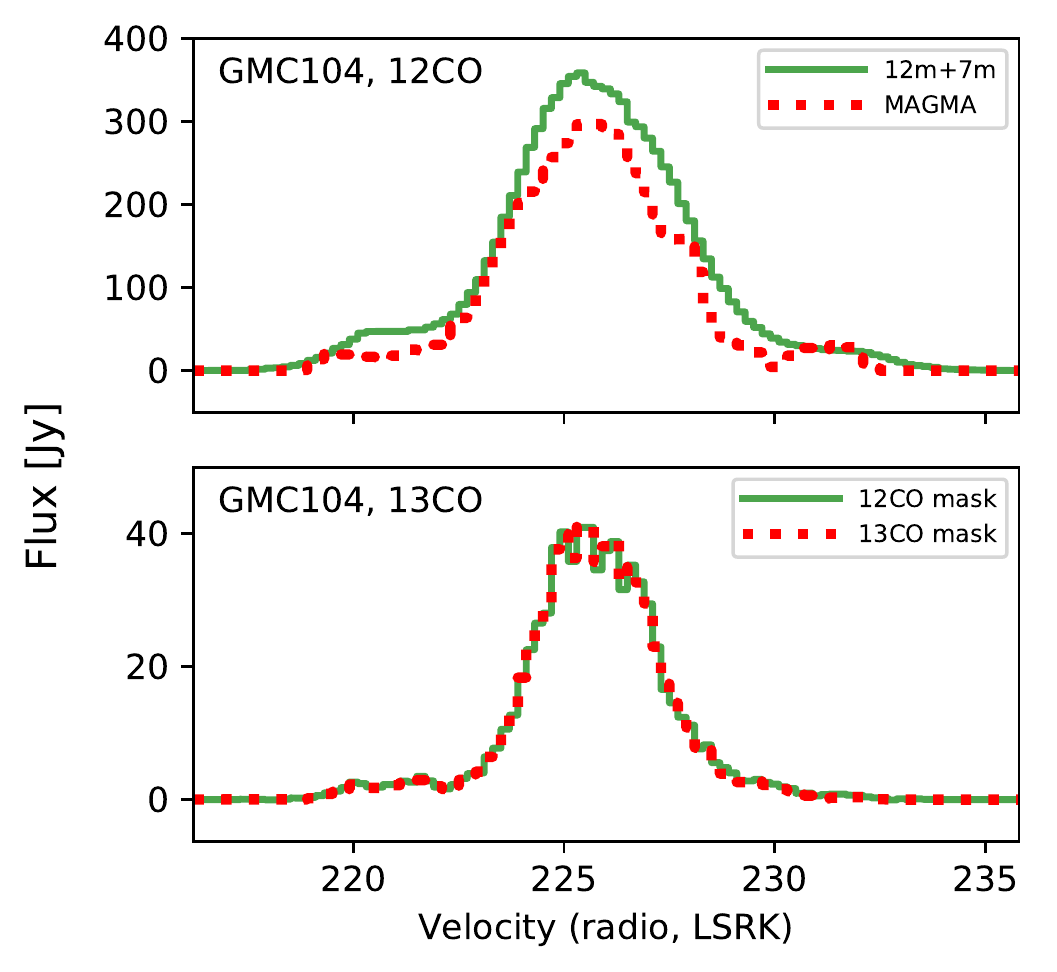}
    \includegraphics[width=0.48\textwidth]{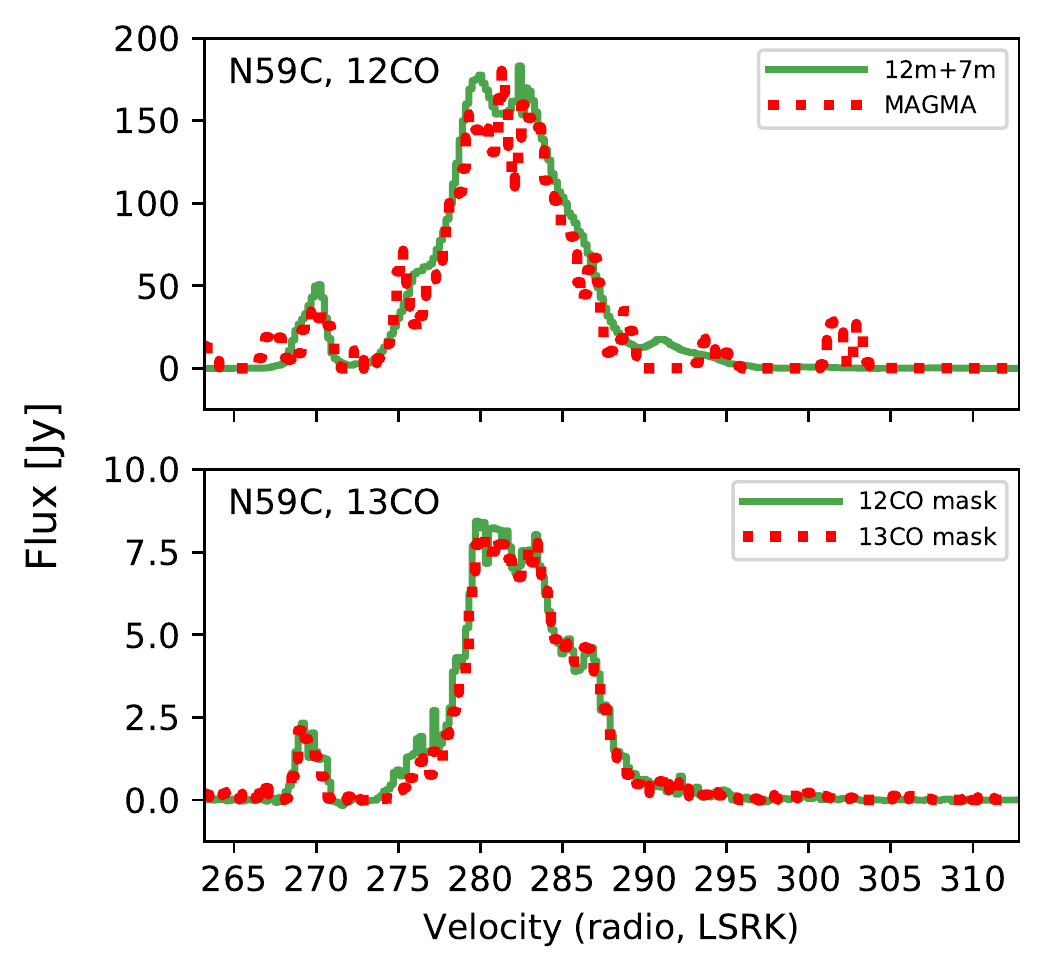}\\
    \caption{Comparison of flux spectra.  The top panel of each pair shows the integrated \twco\ spectrum obtained within the dilated mask for the ALMA cube compared with the MAGMA cube after a similar masking procedure is applied.  The bottom panel compares the \ttco\ spectrum integrated over the \twco\ mask to that obtained with a mask based on the \ttco\ data.}
    \label{fig:fluxcomp}
\end{figure*}

For purposes of comparison, we smoothed the cubes to a common resolution of 3\farcs5 {(0.8 pc)}.  The rms noise and peak detected signal within the maps at this resolution are summarized in Table~\ref{tab:regs}.  To reduce noise when computing flux spectra or moment maps, we apply a signal mask to the cube prior to integration in the spatial or velocity dimensions.  The mask was obtained by starting at the 3.5$\sigma$ contour and expanding to the surrounding 2$\sigma$ contour, then extending the mask by 1 channel towards the blue- and redshifted ends of each spectrum.  We refer to this signal mask as the ``dilated'' mask.  We compare the ALMA-derived \twco\ flux spectra with those obtained from the single-dish MAGMA survey in Figure~\ref{fig:fluxcomp}.  The MAGMA spectra were obtained by applying the ALMA mosaic gain to the MAGMA map and employing the same masking procedure that was applied to the ALMA cubes.  Since the differences are comparable to the $\sim$20\% calibration uncertainties in the much lower sensitivity MAGMA maps, we have not attempted to merge the interferometer and single-dish data.  Note that we lack single-dish or ACA 7m data for the \ttco\ line, so the spectra compared in the lower panels of Figure~\ref{fig:fluxcomp} are both from the ALMA cube, but using different signal masks (a dilated mask derived from the \twco\ cube and then transferred to the \ttco, and one derived from the \ttco\ cube directly). The agreement indicates that the achieved signal-to-noise ratio is high enough that the measured fluxes are not sensitive to the details of the masking approach.

\subsection{Archival Data for 30 Dor and PCC}

We include archival ALMA data presented in Paper I for comparison with the data from our new ALMA Cycle 4 observations.  The archival data were obtained over somewhat smaller fields of view in the $J = 2 \rightarrow 1$ transition of \twco\ and \ttco.  They include total power (TP) but not 7m data.

Data for the 30\,Dor-10 molecular cloud (hereafter ``30 Dor'') were collected in ALMA Cycle 0 under project code 2011.0.00471.S \citep{Indebetouw:13}.  The ALMA data, covering a field 50\arcsec\ $\times$ 50\arcsec\ in size, were combined with total power data from APEX using the {\tt feather} task in CASA to recover large-scale flux.  The native resolution of the datacubes was 2\farcs38 $\times$ 1\farcs54 for \twco\ and 2\farcs47 $\times$ 1\farcs59 for \ttco\ with 0.5 \kms\ channels.  The cubes were then smoothed to a 3\farcs5 circular beam for comparison with the other clouds.

For the quiescent cloud PGCC G282.98$-$32.40 \citep{Planck:XXVIII}, which we refer to as the ``Planck Cold Cloud (PCC)'', ALMA Cycle 2 observations were obtained in 2014 and 2015 under project code 2013.1.00832.S \citepalias{Wong:17}.  The observations cover a region of 220\arcsec\ $\times$ 80\arcsec\ and include total power data from ALMA merged using the {\tt feather} task in CASA.  The native resolution of the datacubes was 1\farcs72 $\times$ 1\farcs19 for \twco\ and 1\farcs81 $\times$ 1\farcs24 for \ttco\ with 0.2 \kms\ channels.  Again, the cubes were then smoothed to a 3\farcs5 circular beam for comparison with the other clouds.

By including both Band 3 and Band 6 observations within our sample, we are assuming that the $J = 1 \rightarrow 0$ and $J = 2 \rightarrow 1$ lines trace similar structures within GMCs, and thus a direct comparison of results obtained from the different lines is possible.  This is reasonable given the relatively modest excitation requirements for both lines ($E_{10} = 5.5$ K, $E_{21} = 11$ K) and the relatively small (factor of $\lesssim$2) departures of the $I_{\rm CO(2-1)}/I_{\rm CO(1-0)}$ ratio from unity in both Galactic \citep[e.g.][]{Sakamoto:95,Nishimura:15} and extragalactic \citep[e.g.][]{Leroy:09} studies.  In particular, \citet{Sorai:01} measure a luminosity-weighted CO(2--1)/CO(1--0) brightness temperature ratio of $\sim$0.9 across the LMC.\@ We note, however, that variations in the 
$^{13}$CO(2--1)/$^{13}$CO(1--0) intensity ratio are expected to be larger than variations in the CO(2--1)/CO(1--0) ratio, given the lower optical depth of the \ttco\ lines \citep[e.g.][]{Nishimura:15}.  To more rigorously test our assumption that $J$ level does not strongly affect structure properties would require a comparative structural analysis of $J = 1 \rightarrow 0$ and $J = 2 \rightarrow 1$ data for the same cloud, which we defer to a future paper.

\subsection{IR and Dust Comparison Images}

We use published mid-infrared and far-infrared images of the LMC to measure global characteristics of the GMCs in our sample.  The {\it Spitzer} 8$\mu$m and 24$\mu$m images from the SAGE legacy program \citep{Meixner:06} are employed as indicators of star formation activity, and longer wavelength {\it Herschel} imaging from the HERITAGE key program \citep{Meixner:13} are employed to trace dust temperature and mass.  Specifically, for 8$\mu$m emission we use the 2\arcsec-pixel mosaic of point source subtracted residual images ({\tt SAGE\_LMC\_IRAC8.0\_2\_resid.fits}), while for 24$\mu$m emission we use the 2\farcs49-pixel mosaic ({\tt SAGE\_LMC\_MIPS24\_E12.fits}).  We use the dust temperature ($T_{\rm dust}$) and dust column density ($N_{\rm dust}$) maps generated by \citet{Utomo:19} at 13 pc (53\arcsec) resolution from the HERITAGE observations.  Because of their coarser resolutions, the $\lambda \ge 24\,\mu$m images are used only to describe the overall GMC properties presented in \S\ref{sec:bound}.  To do this, the boundary of each GMC is defined by a dilated mask derived from the MAGMA data starting from 3.5$\sigma$ peaks and extended to the 2$\sigma$ edge, regridded to match the SAGE pixel grid.  Within each GMC boundary we obtain the median 8$\mu$m and 24$\mu$m intensity and a mean value for $N_{\rm dust}$ and $T_{\rm dust}$.

As the resolution of the 8$\mu$m images is similar to that of the ALMA data, we also measure the mean 8$\mu$m intensity in CO-emitting structures at a resolution matched to the CO data.  For this purpose, the SAGE residual image is convolved to 3\farcs5 assuming a native resolution of 2\arcsec, and then regridded to match each ALMA image.

\section{Results}

\begin{figure*}
    \centering
    \includegraphics[height=2.3in]{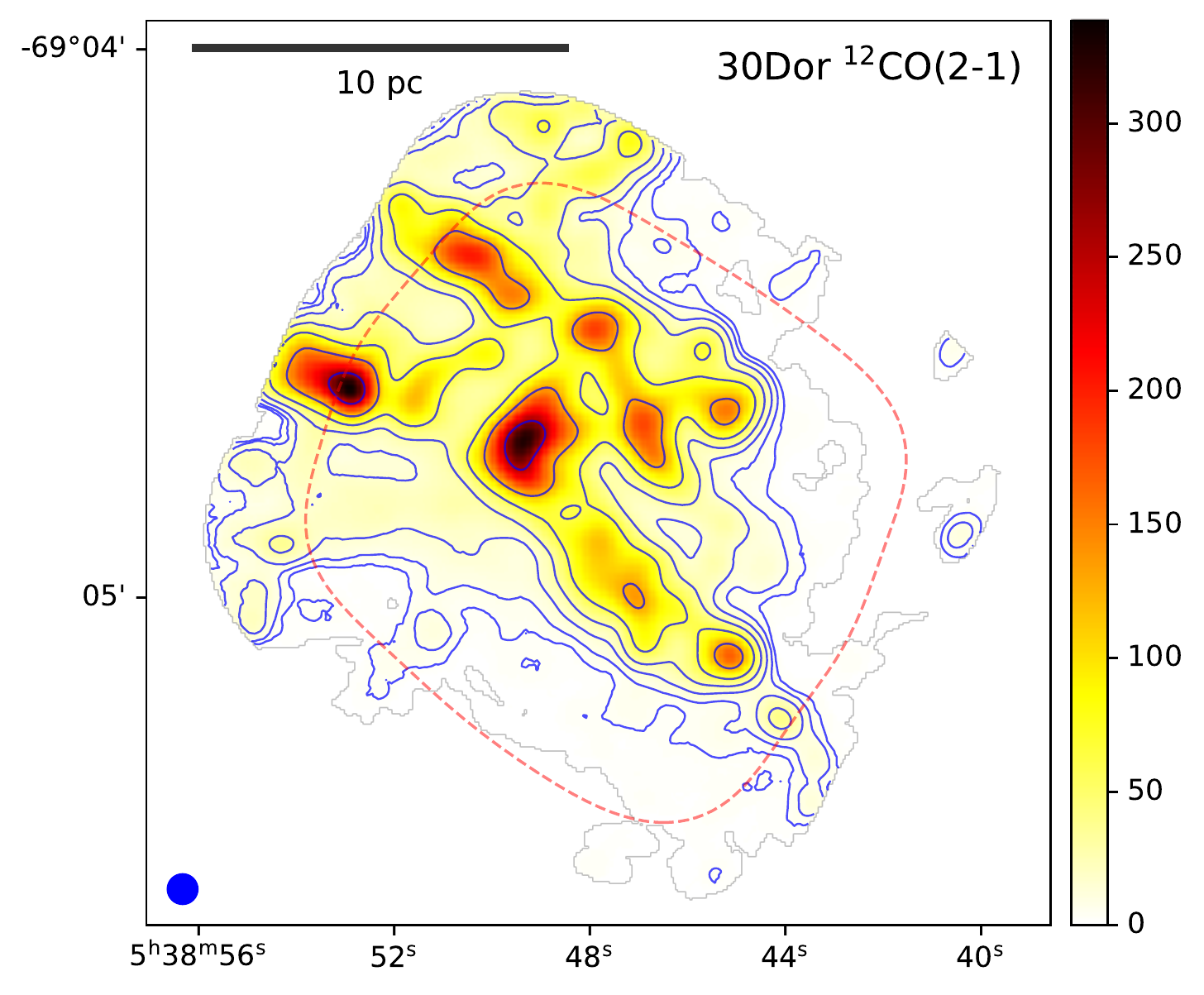}
    \includegraphics[height=2.3in]{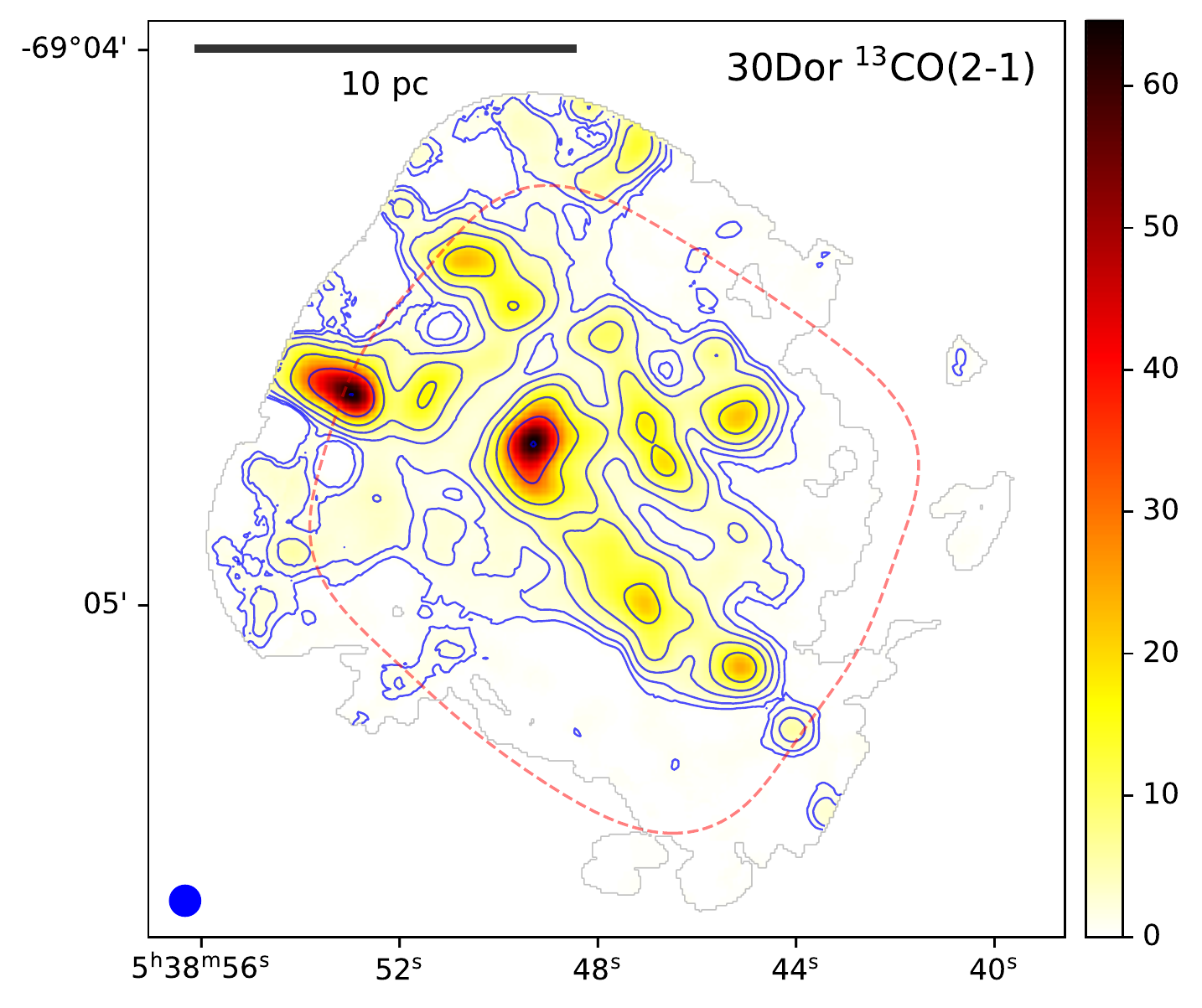}\\
    \includegraphics[height=3in]{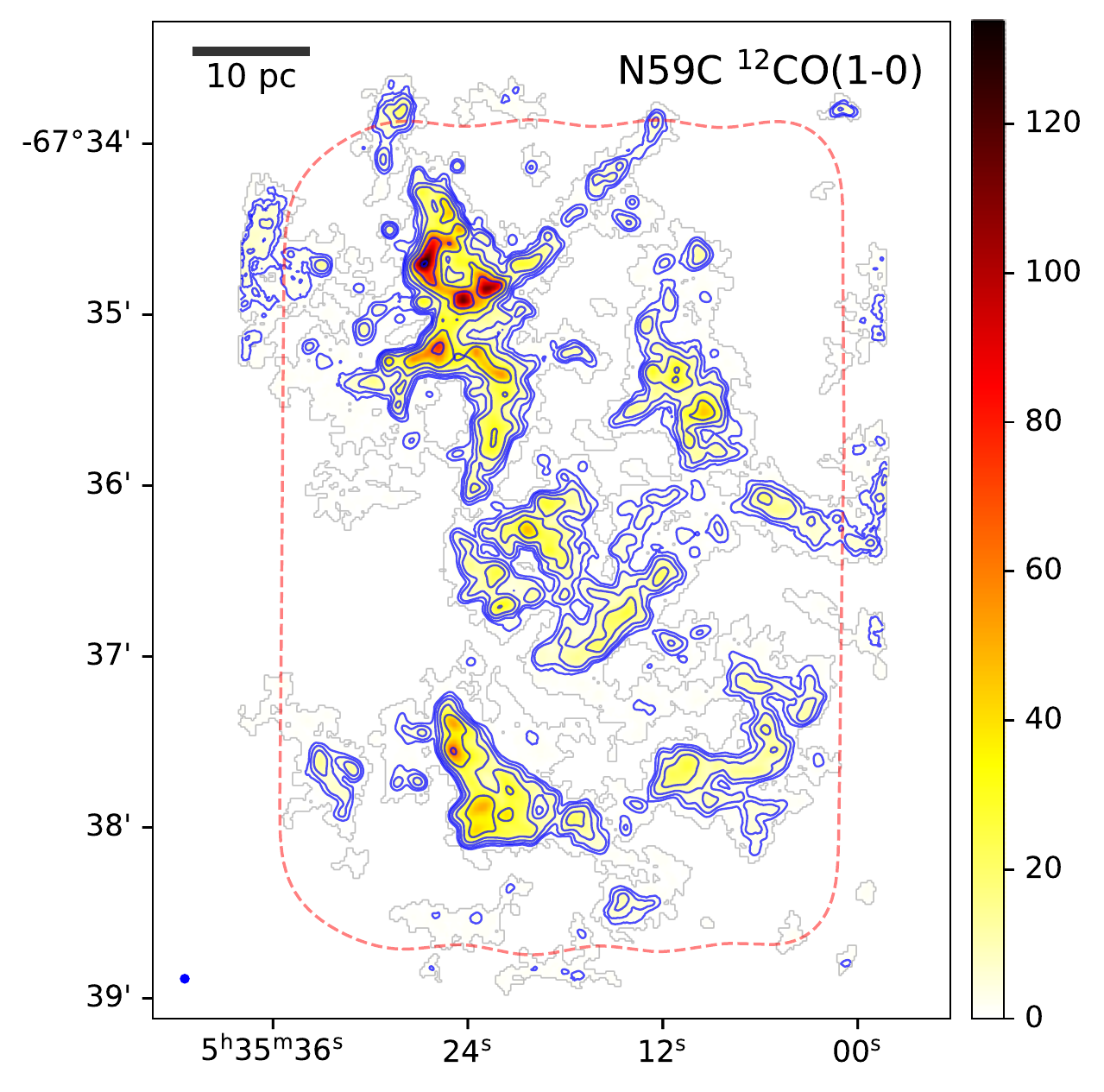}
    \includegraphics[height=3in]{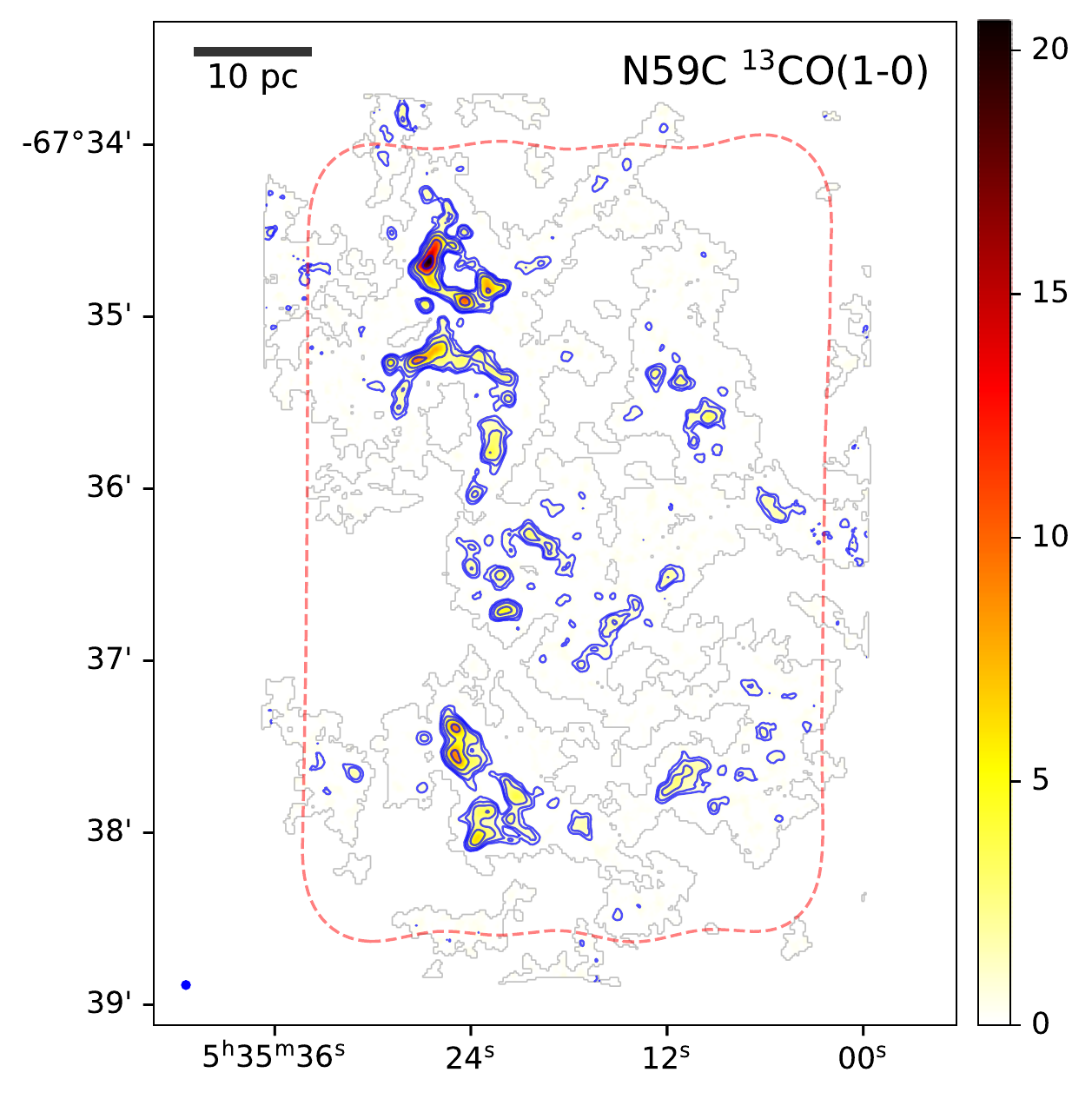}\\
    \includegraphics[height=2.5in]{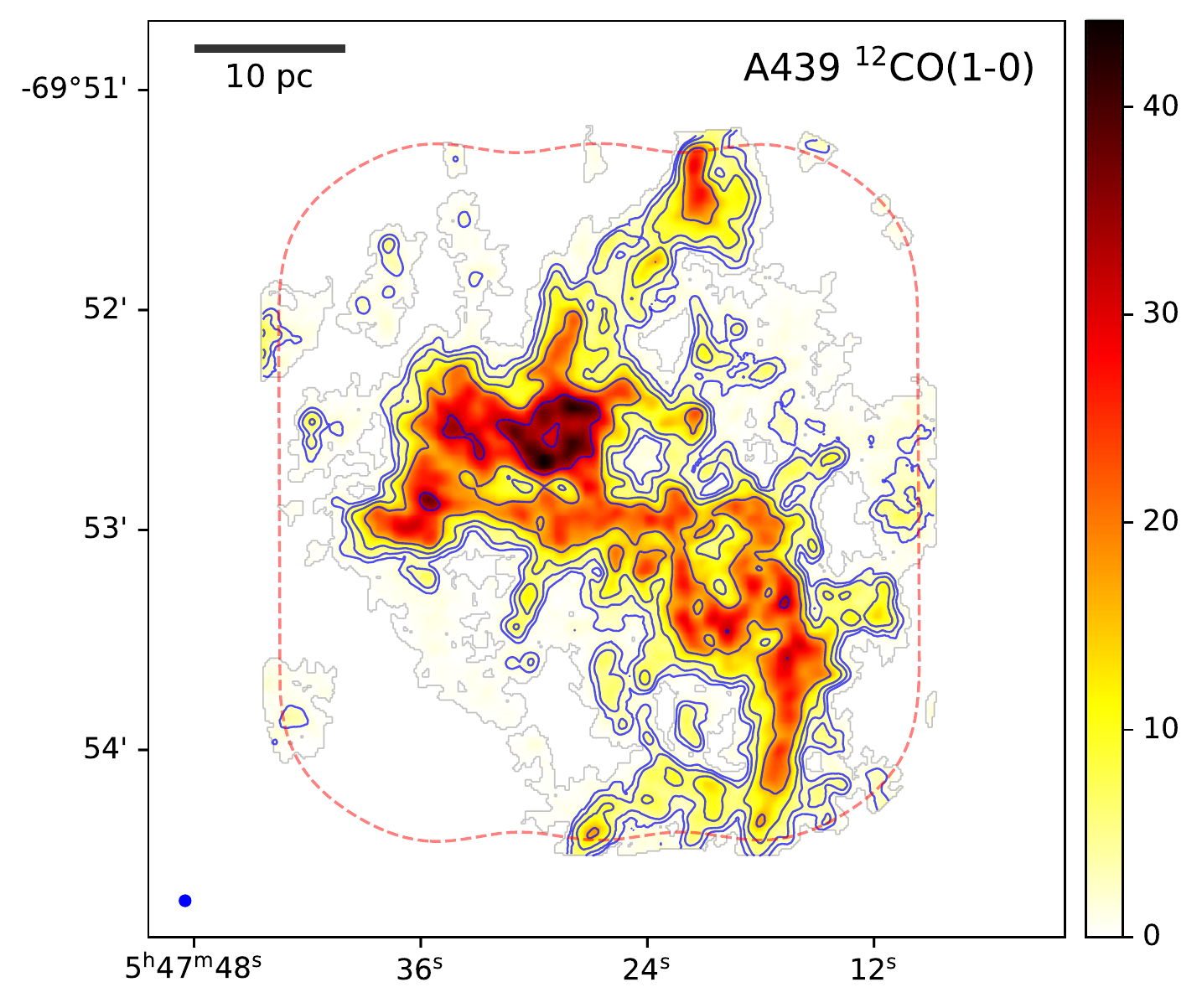}
    \includegraphics[height=2.5in]{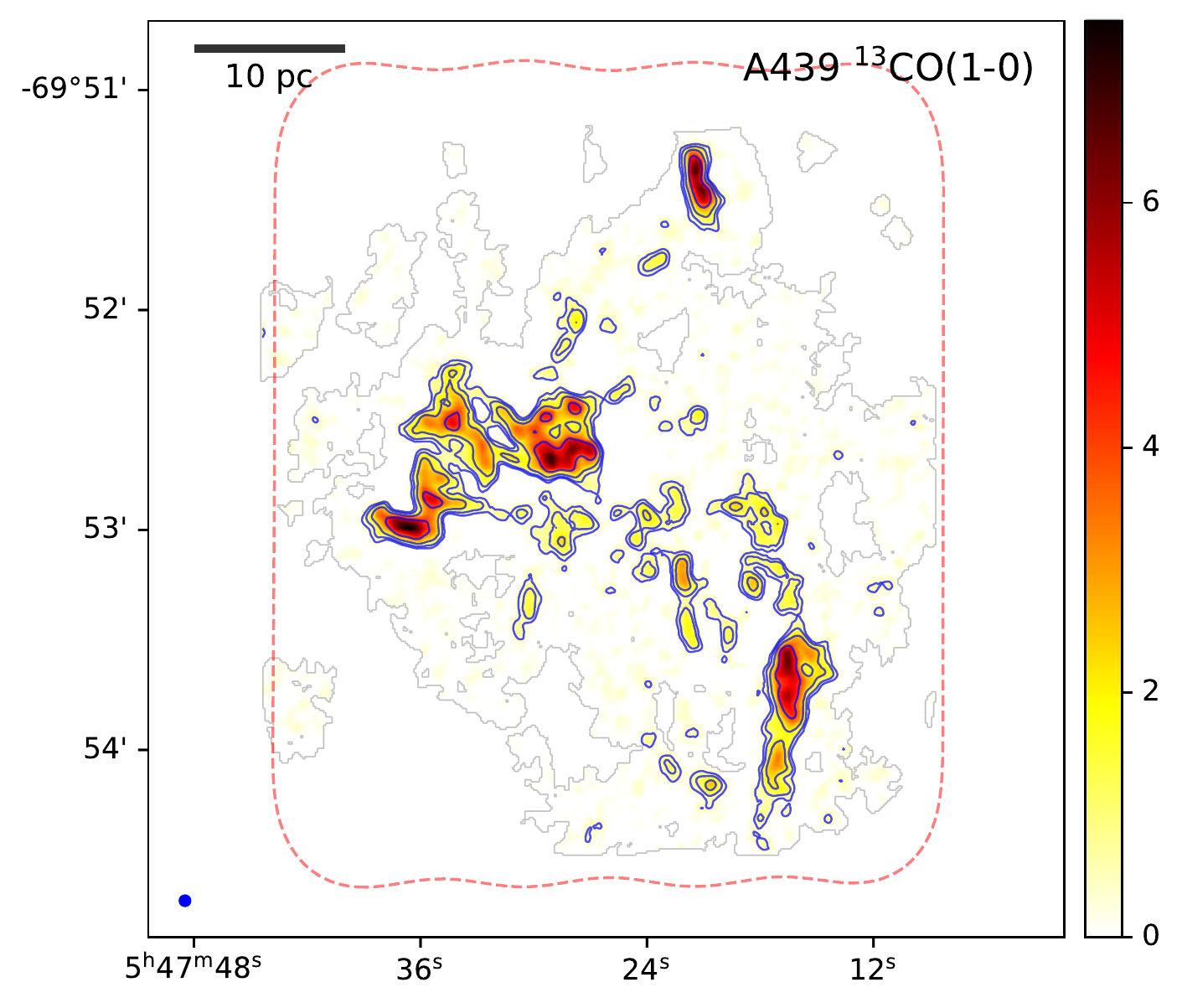}\\
    \caption{Integrated intensity maps for \twco\ ({\it left panels}) and \ttco\ ({\it right panels}) in the 30 Dor, N59C, and A439 clouds.  Color scale units are K \kms.  The common 3\farcs5 (0.8 pc) beam is shown in the lower left corner.  Contour levels are $2^n$ K \kms\ for \twco\ and $2^{n-2}$ K \kms\ for \ttco, where $n$=2, 3, \ldots\ for 30 Dor and $n$=1, 2, \ldots\ for N59C and A439.  The red dashed contour indicates where the mosaic sensitivity falls to 50\%.}
    \label{fig:mom0maps1}
\end{figure*}

\begin{figure*}
    \centering
    \includegraphics[height=2.4in]{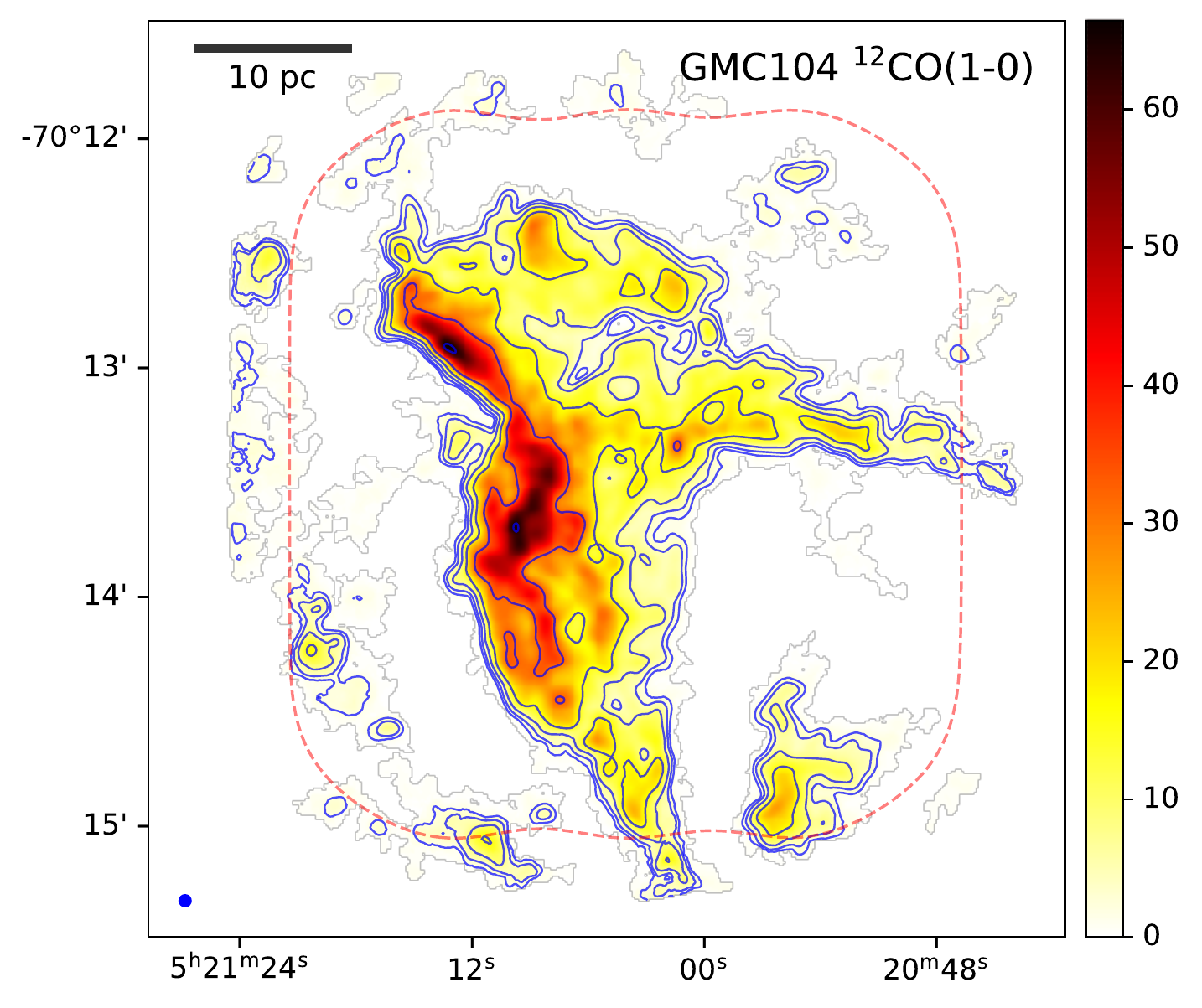}
    \includegraphics[height=2.4in]{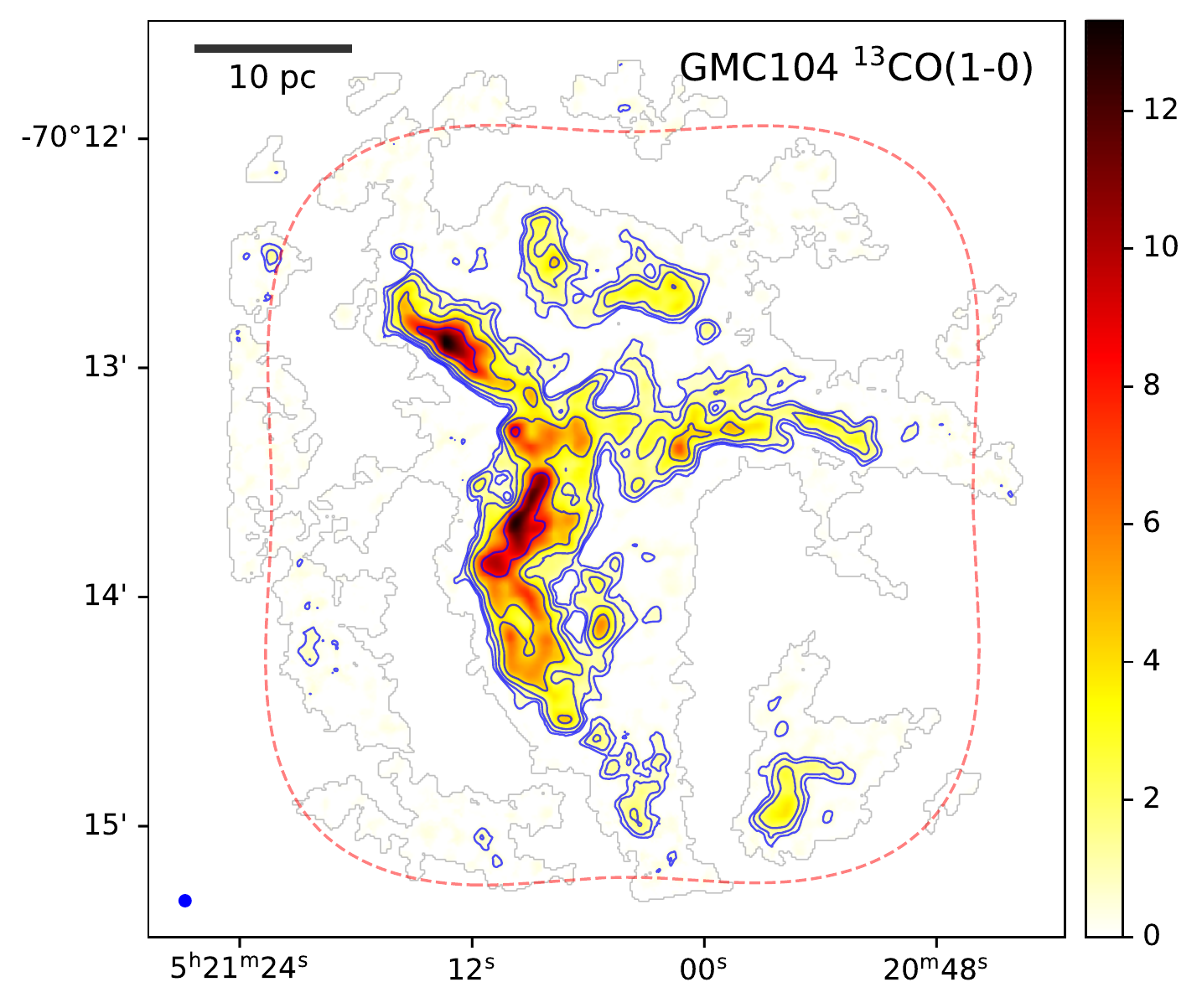}\\
    \includegraphics[height=2.6in]{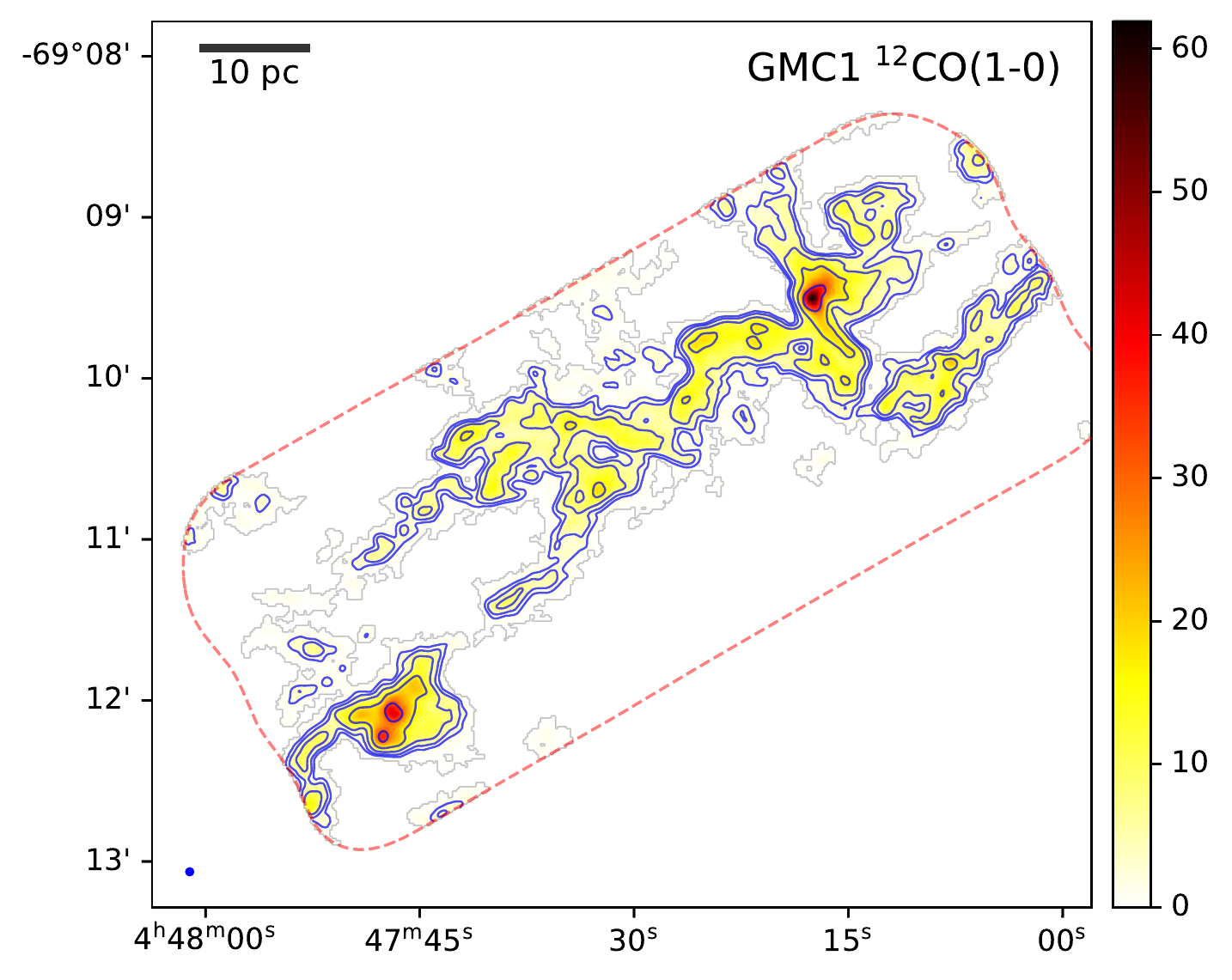}
    \includegraphics[height=2.6in]{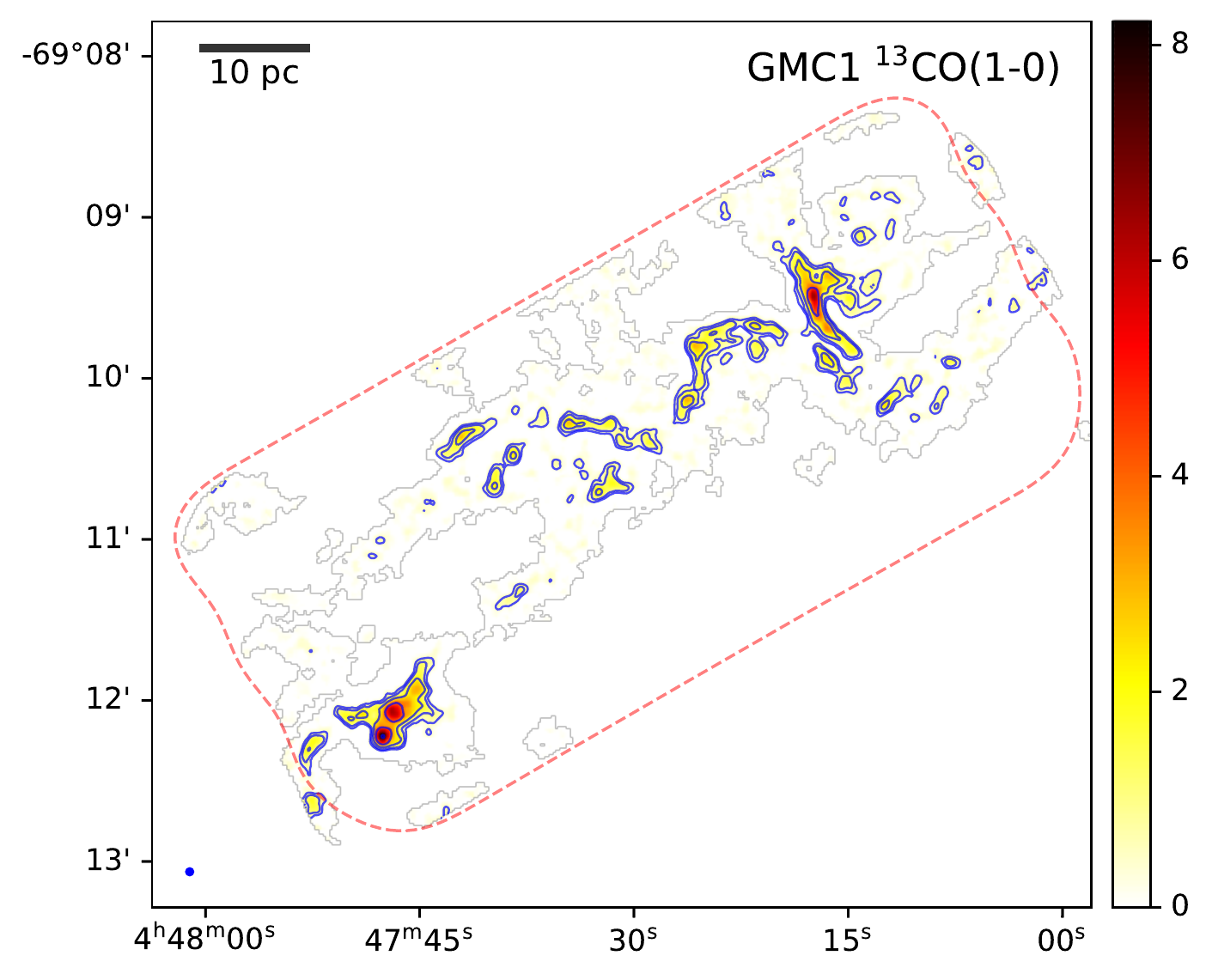}\\
    \includegraphics[height=3.5in]{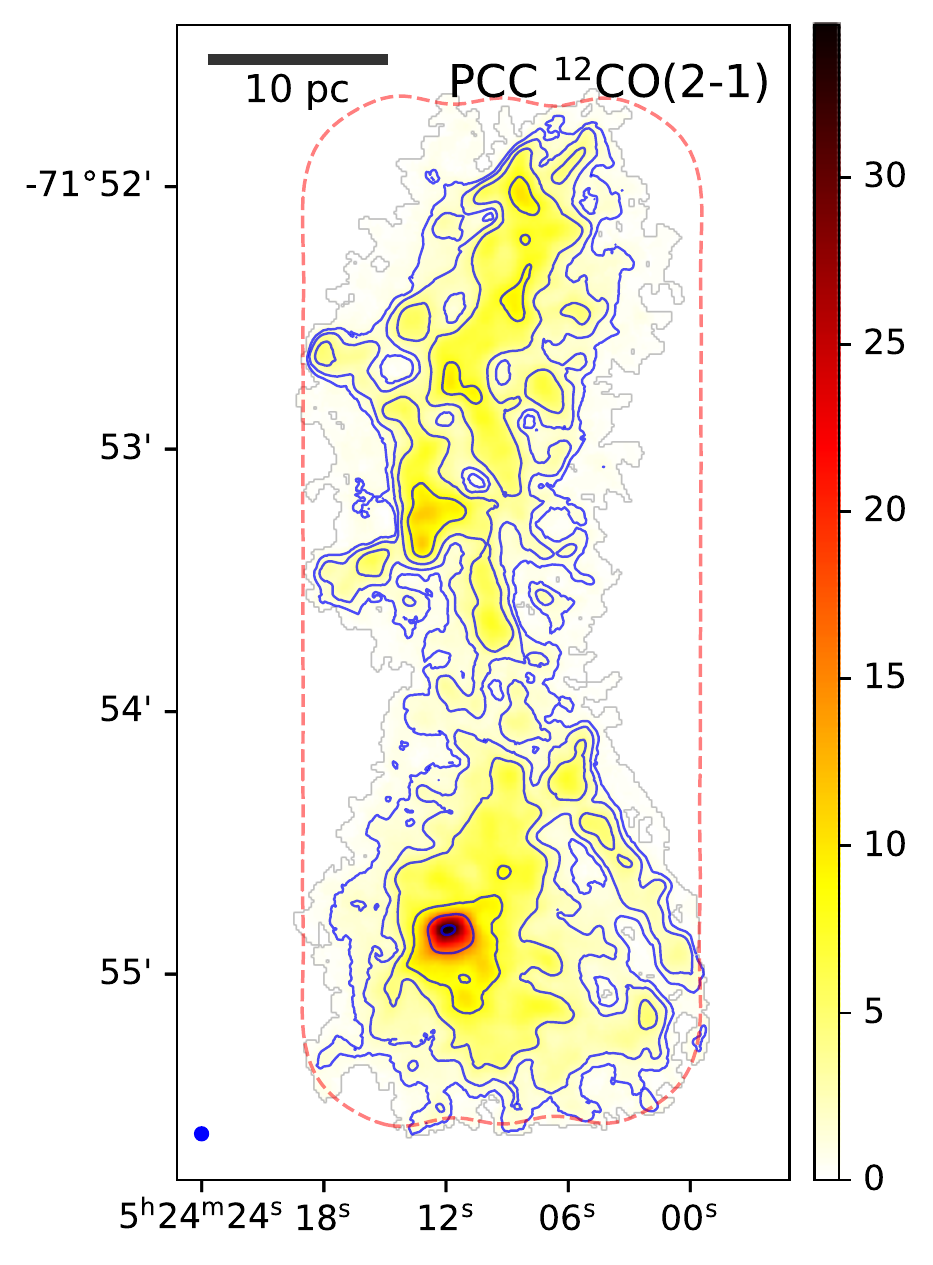}
    \includegraphics[height=3.5in]{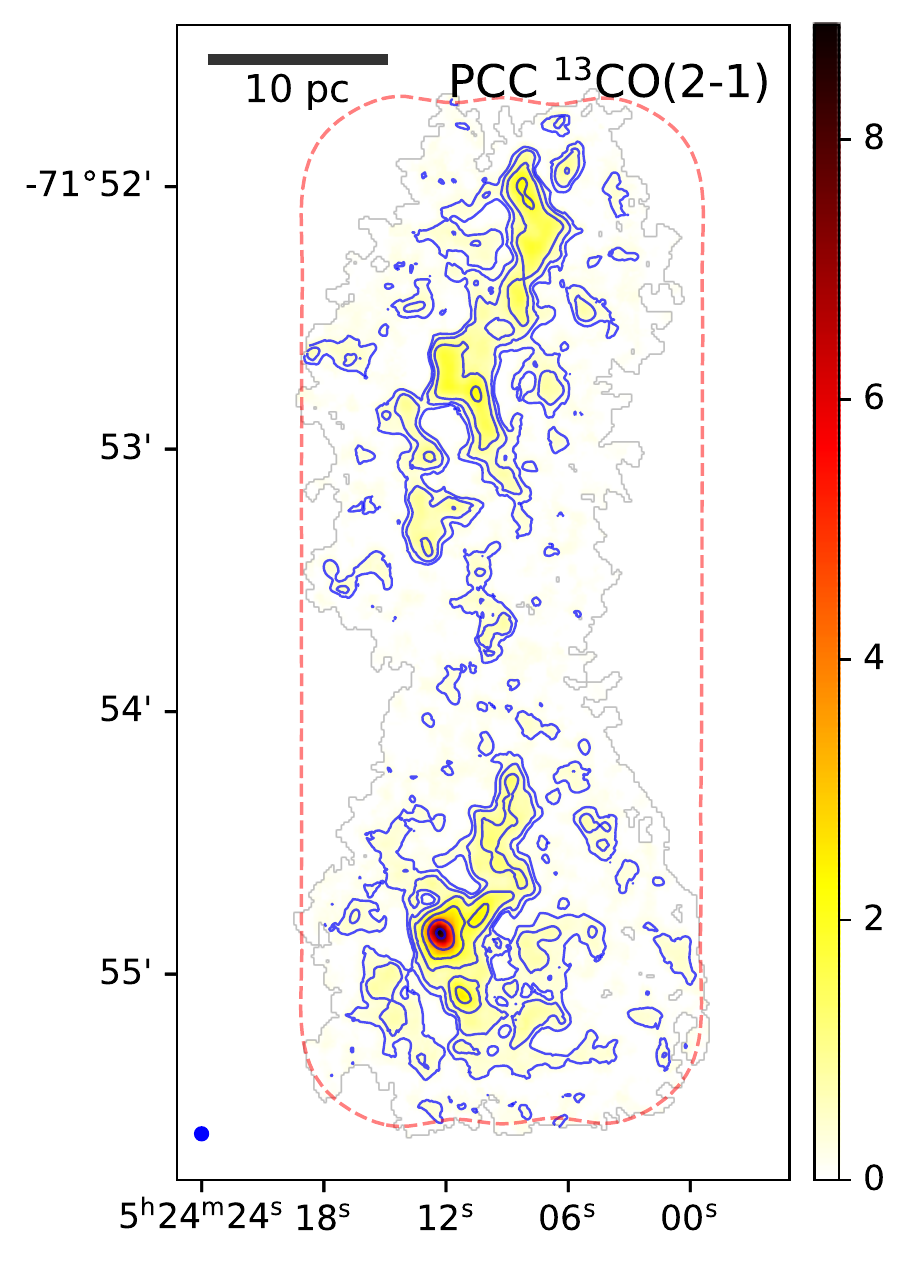}\\
    \caption{Integrated intensity maps for \twco\ ({\it left panels}) and \ttco\ ({\it right panels}) in the GMC104, GMC1, and PCC clouds.  Color scale units are K \kms.  The common 3\farcs5 (0.8 pc) beam is shown in the lower left corner.  Contour levels are $2^n$ K \kms\ for \twco\ and $2^{n-2}$ K \kms\ for \ttco, where $n$=0, 1, \ldots\ for PCC and $n$=1, 2, \ldots\ for GMC104 and GMC1.  The red dashed contour indicates where the mosaic sensitivity falls to 50\%.}
    \label{fig:mom0maps2}
\end{figure*}

\subsection{Line Intensity Maps}\label{sec:mommaps}

We present the integrated intensity maps of \twco\ and \ttco\ emission for the full sample of six clouds in Figures~\ref{fig:mom0maps1} and \ref{fig:mom0maps2}.  These were obtained by applying the dilated masks described in \S\ref{sec:almadata} to the cubes and summing in velocity.  Since the mosaic field of view (indicated by the red dashed line) differs across the sample, the FWHM size of the common-resolution beam (3\farcs5, corresponding to 0.8 pc) is indicated in the lower left of each plot.
In addition to matching the spatial resolution of the cubes, we have added Gaussian noise at the beam scale to reach a uniform 1$\sigma$ noise of 0.25 K for \twco\ and 0.15 K for \ttco.  These values were scaled down by $\sqrt{2.5}$ for 30 Dor, for which the channel map spacing is 0.5\,\kms\ rather than 0.2\,\kms.  Since line widths in the 30 Dor region are substantially larger ($\gtrsim$1 \kms) than in other parts of the LMC (\citealt{Indebetouw:13}; \citetalias{Wong:17}), the mismatch in channel width and noise should not have a significant impact on our results.

The ALMA maps reveal a wealth of structure, which we plan to investigate more deeply in future papers, but what is immediately apparent is that the emission structures are nested rather than discrete, and that much of the brightest emission occurs in filaments.  In these and subsequent plots, we will present the results for the six clouds in the following order: 30 Dor, N59C, A439, GMC104, GMC1, and PCC.  This ordering is by decreasing median 8$\mu$m intensity (Figure~\ref{fig:lmchi}, {\it left}), and reflects a sequence of decreasing star formation activity.

\subsection{LTE Analysis}\label{sec:lte}

Following \citetalias{Wong:17}, we conduct a simple local thermodynamic equilibrium (LTE) analysis to infer the \ttco\ column density from the observed \twco\ and \ttco\ emission.  The analysis assumes that both lines share a common excitation temperature \citep[e.g.,][]{Nishimura:15}.
We assume the $^{12}$CO(2--1) line is optically thick at line center and not subject to beam dilution, so that for a given line of sight the excitation temperature is uniform and given by \citep[e.g.,][]{Bourke:97}:
\begin{equation}
J(T_{\rm ex}) = f_{\rm bm}^{-1}\,T_{\rm 12,pk} + J(T_{\rm cmb})\;,
\end{equation}
where the beam filling fraction $f_{\rm bm}$ is assumed to be 1, $T_{\rm 12,pk}$ is the peak temperature of the \twco\ line profile, and
\begin{equation}
J(T) \equiv \frac{h\nu/k}{\exp(h\nu/kT)-1}\;.
\end{equation}
The beam-averaged $^{13}$CO optical depth is then calculated from the brightness temperature $T_{13}$ at each position and velocity in the cube by solving
\begin{equation}\label{eqn:t13}
T_{13} = f_{\rm bm}[J(T_{\rm ex}) - J(T_{\rm cmb})][1-\exp(-\tau_{13})]\;,
\end{equation}
again assuming $f_{\rm bm} = 1$. Since $\tau_{13}$ varies linearly with $T_{13}$ in the optically thin limit (for a given value of $T_{\rm ex}$), we allow negative values of $\tau_{13}$ due to noise.  Given our assumption of a single $T_{\rm ex}$ at each sky position, and because of the limited range of $T_{\rm ex}$ for most clouds (as discussed below), these noise values tend to average out when integrated in the cube.

\begin{figure*}
    \centering
    \includegraphics[width=0.75\textwidth]{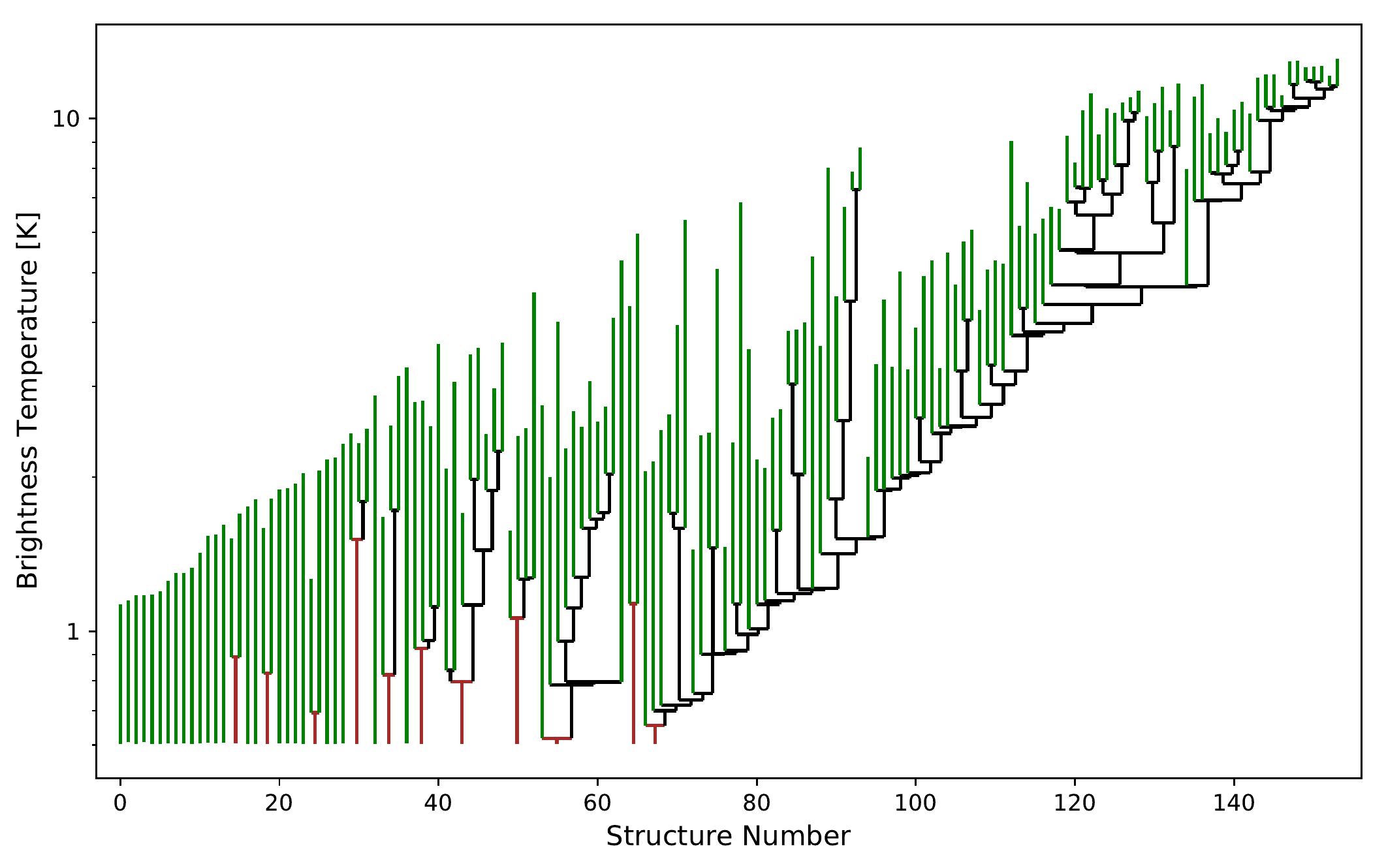}
    \caption{Dendrogram structure tree for \twco\ in the A439 cloud.  Each structure is represented by a vertical line that spans the intensity range of the pixels that are uniquely assigned to that structure.  Trunks (maroon lines) are the largest contiguous structures, leaves (green lines) are those with no resolvable substructure, and branches (black lines) span the hierarchy in between.}
    \label{fig:dendrogram}
\end{figure*}

For any given $T_{\rm ex}$, there is a maximum allowed value of $T_{13}$ beyond which $\tau_{13}$ becomes undefined.  For $T_{\rm ex}=6$ K we require $T_{13}<2.8$ K for $J=1\rightarrow0$ and $T_{13}<2$ K for $J=2\rightarrow1$.  To reduce the number of undefined values we impose a minimum value on $T_{\rm ex}$ of $T_{\rm floor}$ = 6 K under the assumption that lower inferred values reflect beam dilution of \twco\ (i.e., $f_{\rm bm}<1$).  We chose $T_{\rm floor}$ based on an examination of the distribution of inferred $T_{\rm ex}$ for pixels detected at high signal-to-noise ratio (SNR$>$5) in \ttco.
Pixels with $T_{\rm ex}<6$ K are extremely rare and account for 0.2\% or less of the high SNR pixels in any given cloud.  As a result, imposition of the floor has little effect on the high SNR column densities, while avoiding problems with artificially high column densities due to low nominal $T_{\rm ex}$ values in low SNR regions.

Following determination of $T_{\rm ex}$ and $\tau_{13}$, the total $^{13}$CO column density in cm$^{-2}$, summed over all rotational levels, is calculating using \citep{Garden:91}:
\begin{eqnarray}
N(^{13}{\rm CO}) = \frac{3k}{8\pi^3B\mu^2}\exp\left[\frac{hBJ(J+1)}{kT_{\rm ex}}\right]\nonumber\\
\frac{T_{\rm ex}+hB/3k}{1-\exp(-h\nu/kT_{\rm ex})}\frac{\int \tau_{13}\,dv}{J+1}\;,
\end{eqnarray}
where $J$ is the rotational quantum number of the lower state, $B$ is the rotational constant for \ttco\ (55.1 GHz), and $\mu$ is the dipole moment of \ttco\ (0.112 debye).

The calculated uncertainties in $N(^{13}{\rm CO})$ mainly reflect the uncertainties due to map noise. 
The assumption of a single $T_{\rm ex}$ that can be derived from $T_{\rm 12,pk}$, as well as the definition of the 3D masks used for integration, introduce additional systematic uncertainties.  
The most important effect is beam dilution, which reduces $T_{\rm 12,pk}$ when smoothing to a common resolution of 3\farcs5; our resulting underestimate of $T_{\rm ex}$ would lead us to overestimate $\tau_{13}$ (Equation~\ref{eqn:t13}), although this is mitigated somewhat by the fact that we would have also overestimated $f_{\rm bm}$.  The net effect on $N$($^{13}$CO) of lowering $\tau_{13}$ while raising $T_{\rm ex}$ is difficult to assess, and would be better constrained by measuring the \ttco\ excitation with additional $J$ transitions.

We derive an LTE-based estimate of clump masses and column densities by scaling $N$($^{13}$CO) to $N$(H$_2$) using a fixed abundance ratio of 
\begin{equation}
\frac{N(\rm H_2)}{N(\rm ^{13}CO)} = 3 \times 10^6\;. 
\end{equation}
Although the \ttco\ abundance in the LMC is not well-constrained by observations, and may be subject to spatial variations, our adopted value is consistent with values inferred or adopted in previous work \citep{Heikkila:99,Mizuno:10,Fujii:14}.  Henceforth we denote LTE-based mass surface densities as $\Sigma_{\rm LTE}$.

\subsection{Structural Decomposition}\label{sec:decomp}

\begin{figure*}
    \centering
    \includegraphics[width=0.32\textwidth]{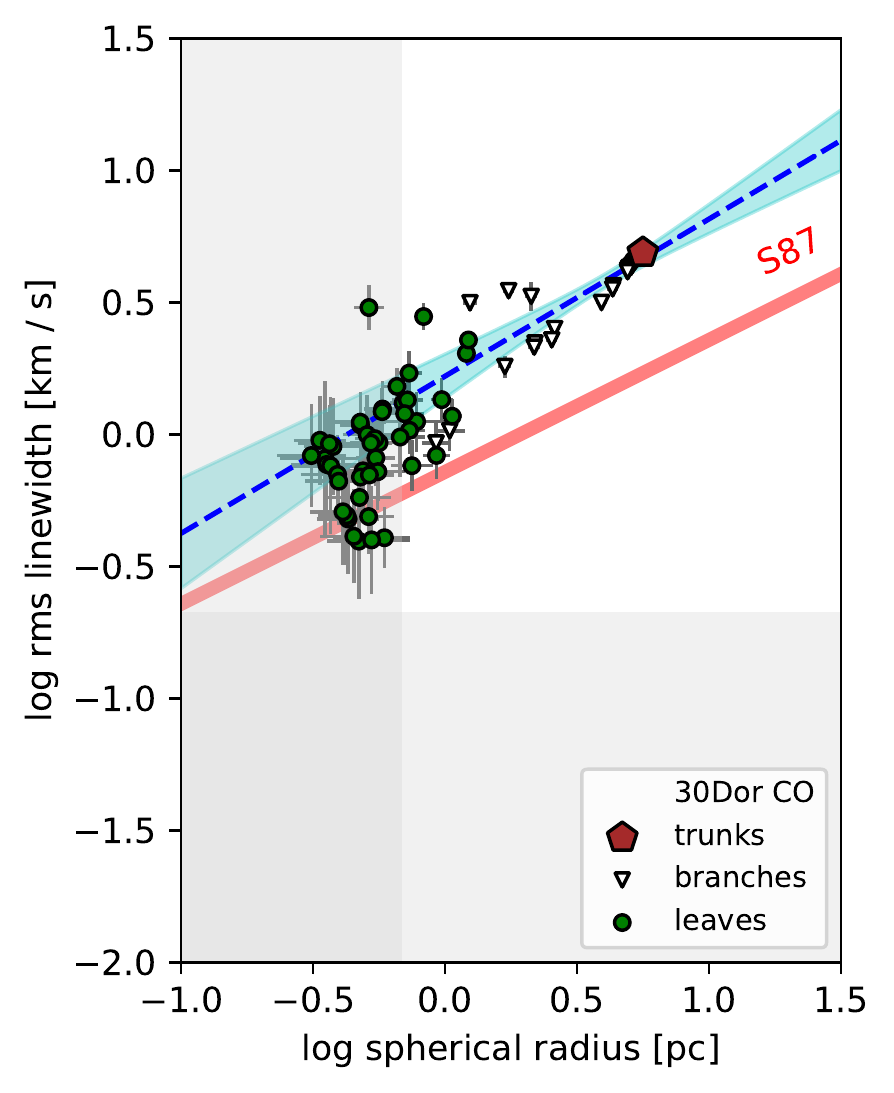}
    \includegraphics[width=0.32\textwidth]{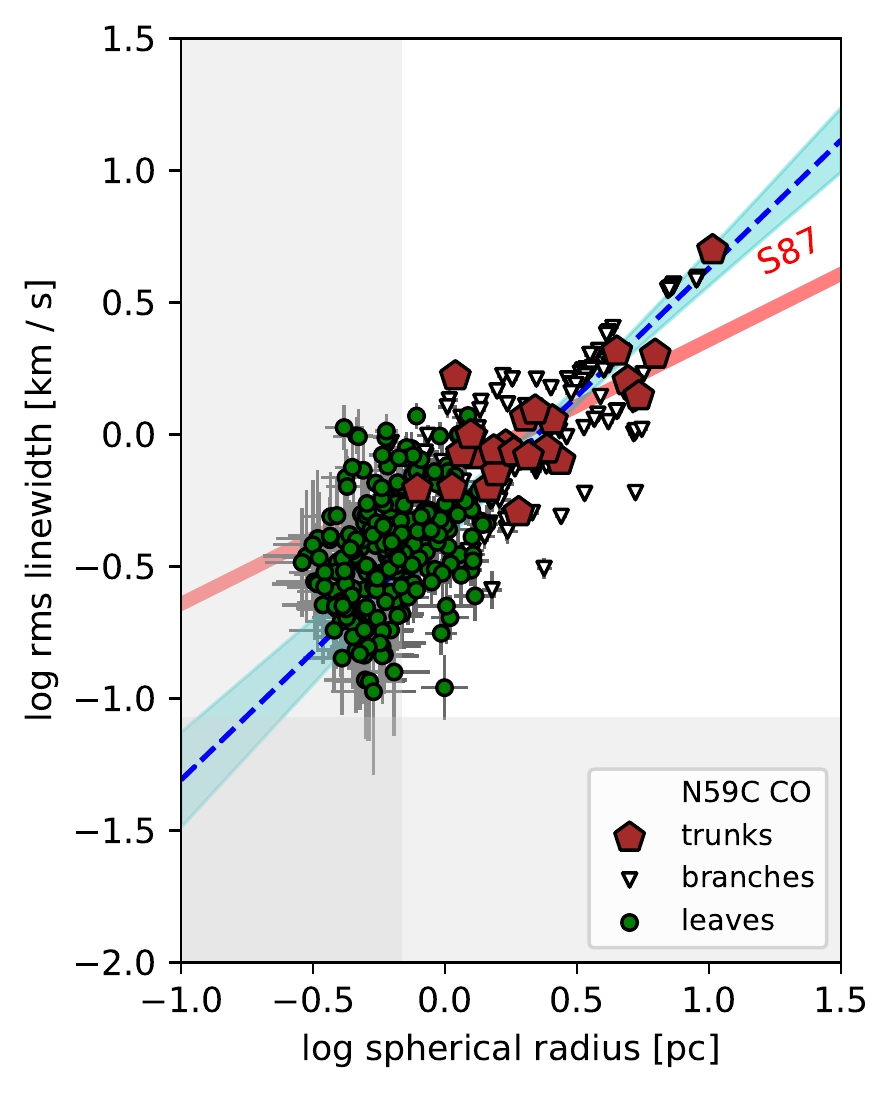}
    \includegraphics[width=0.32\textwidth]{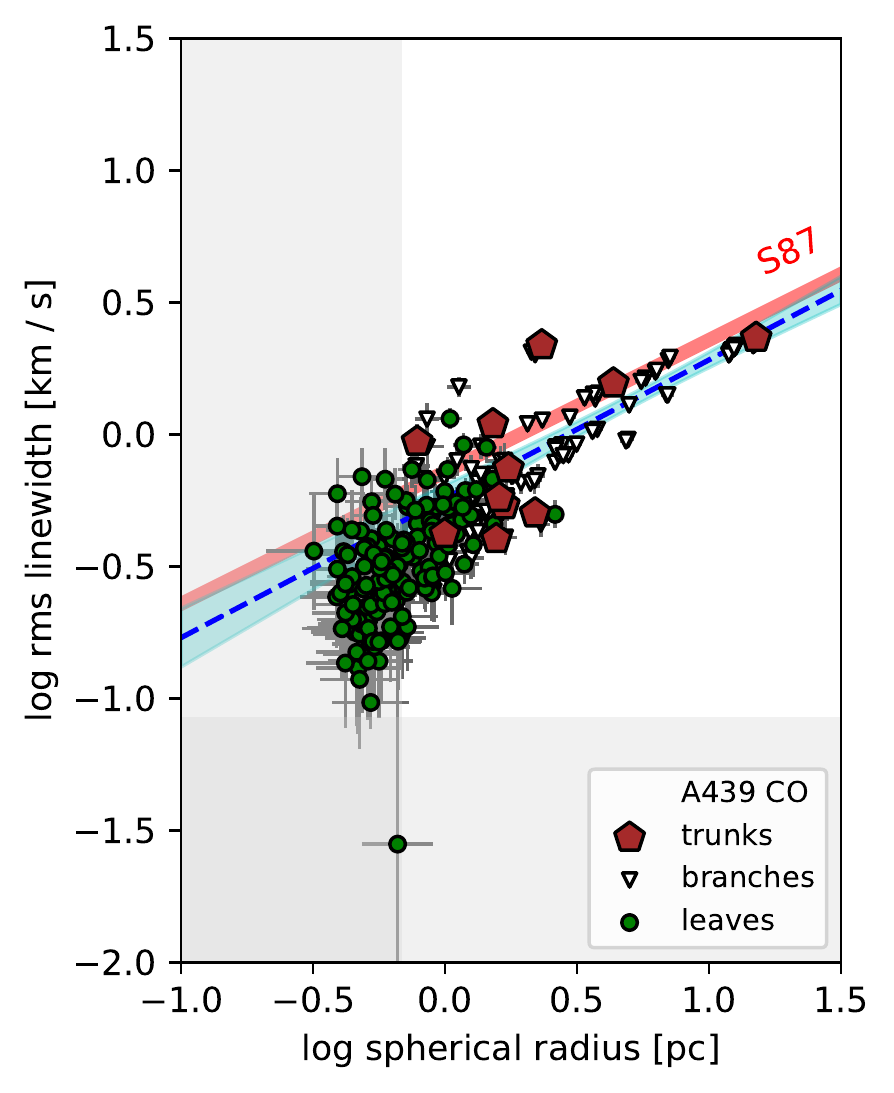}\\
    \includegraphics[width=0.32\textwidth]{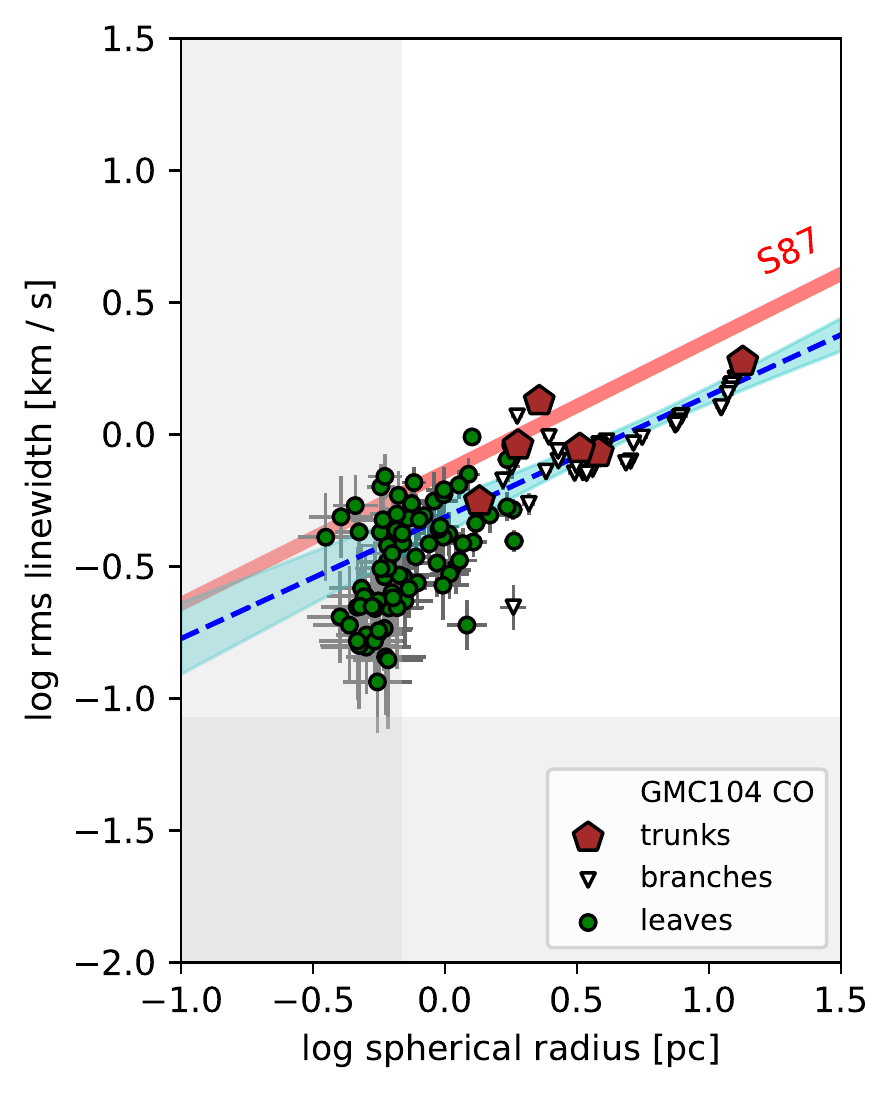}
    \includegraphics[width=0.32\textwidth]{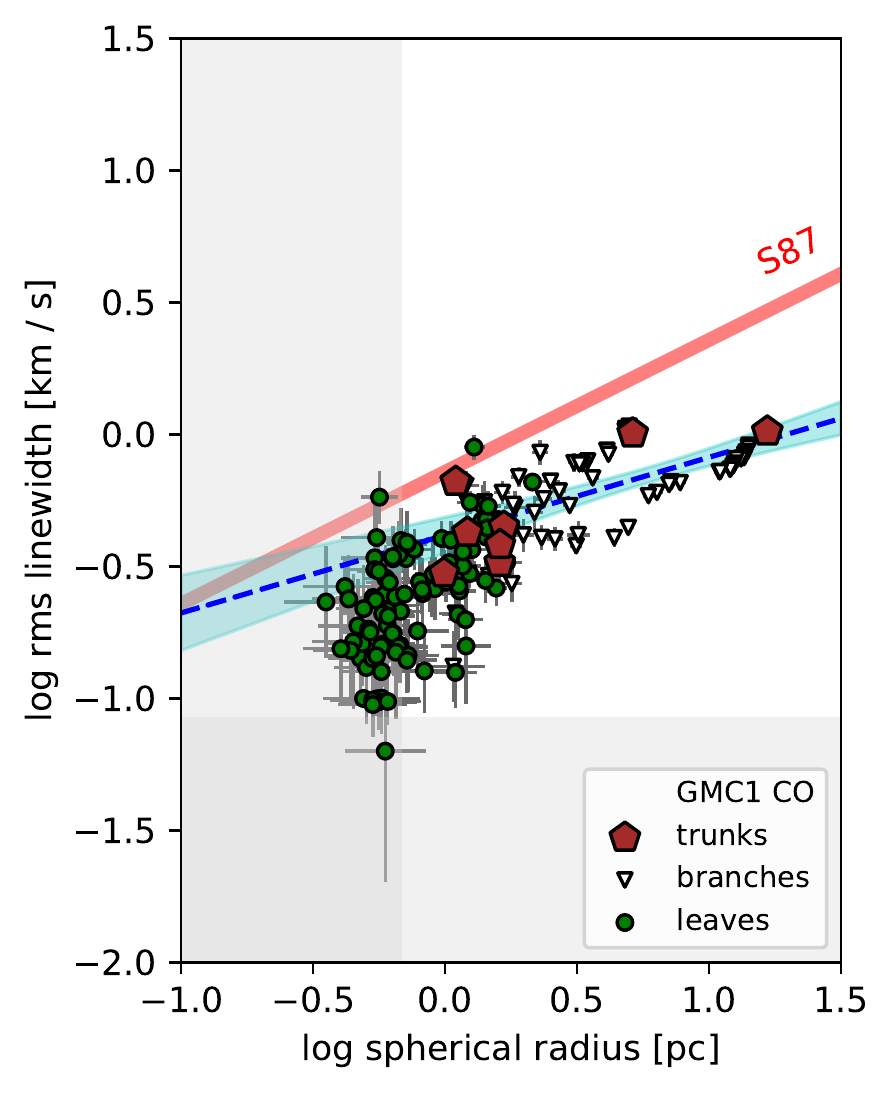}
    \includegraphics[width=0.32\textwidth]{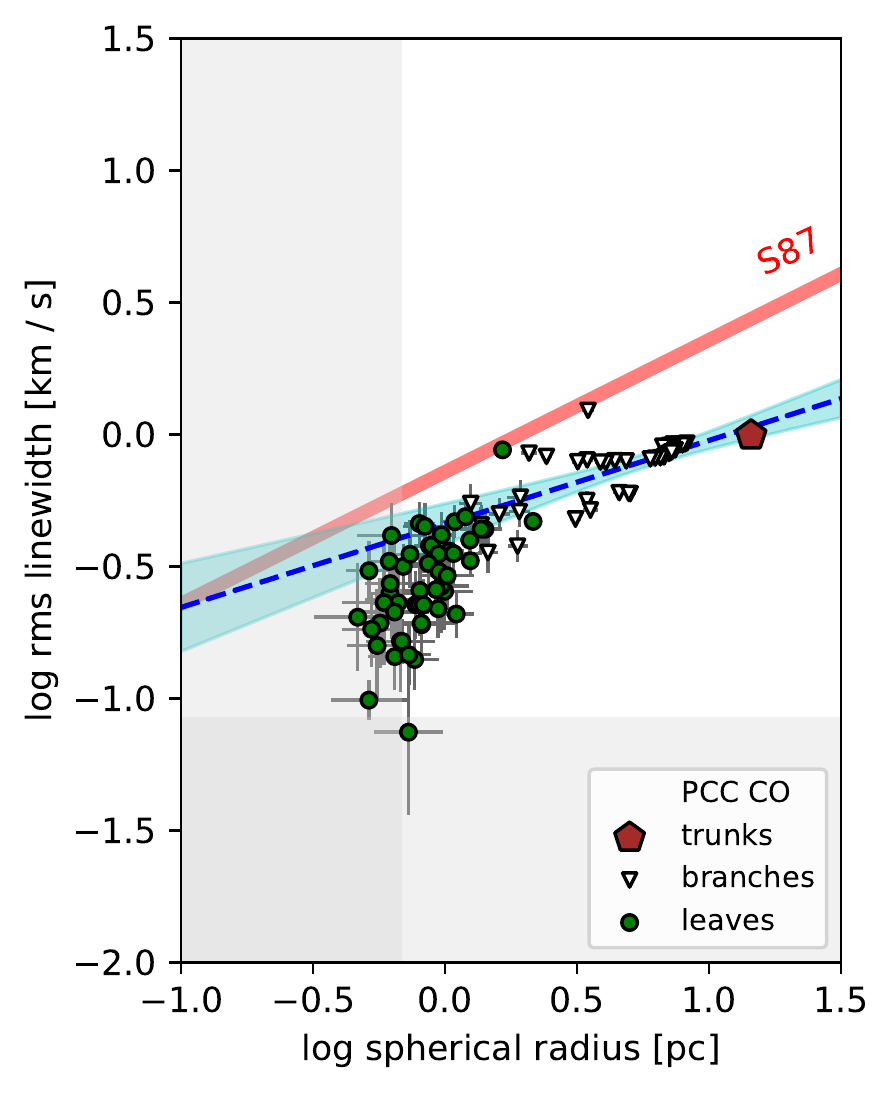}\\
    \caption{Size-line width relations for \twco\ structures in the six molecular clouds.  Data for each cloud are shown in a separate panel.  Dendrogram structure types (trunks, branches, and leaves) are distinguished by different plot symbols.  Power law fits, with 3$\sigma$ confidence intervals, are shown as blue dashed lines with associated shading.  The Galactic relation of \citetalias{Solomon:87} is shown as a pink line.  Gray shaded regions at low $\sigma_v$ and $R$ are poorly resolved and excluded from fitting.}
    \label{fig:rdv12}
\end{figure*}

\begin{figure*}
    \centering
    \includegraphics[width=0.32\textwidth]{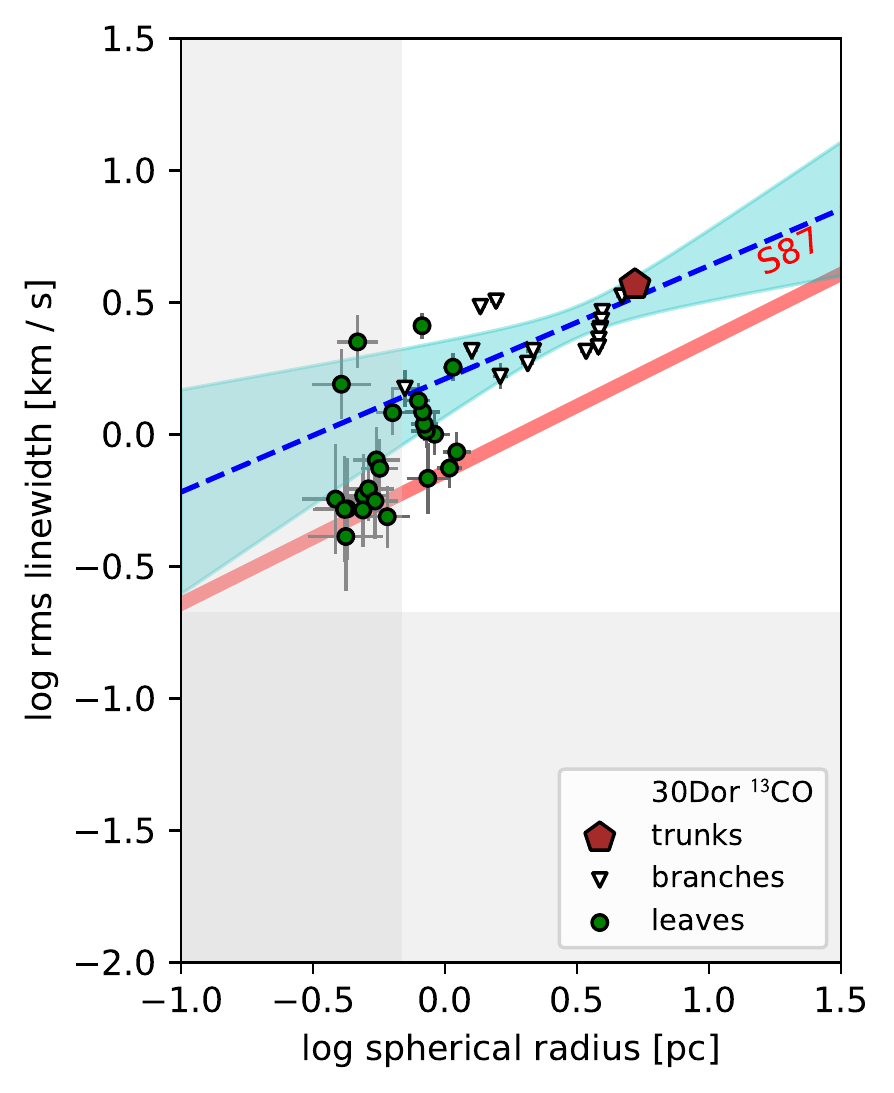}
    \includegraphics[width=0.32\textwidth]{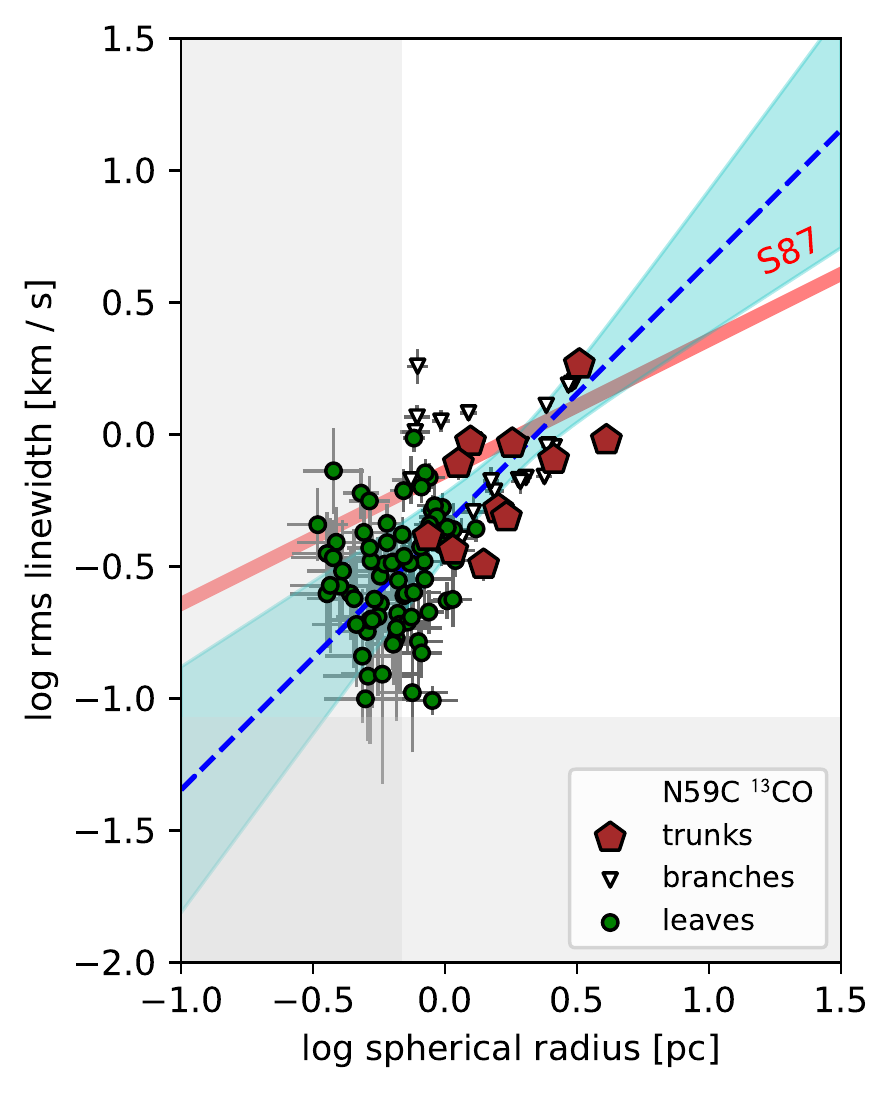}
    \includegraphics[width=0.32\textwidth]{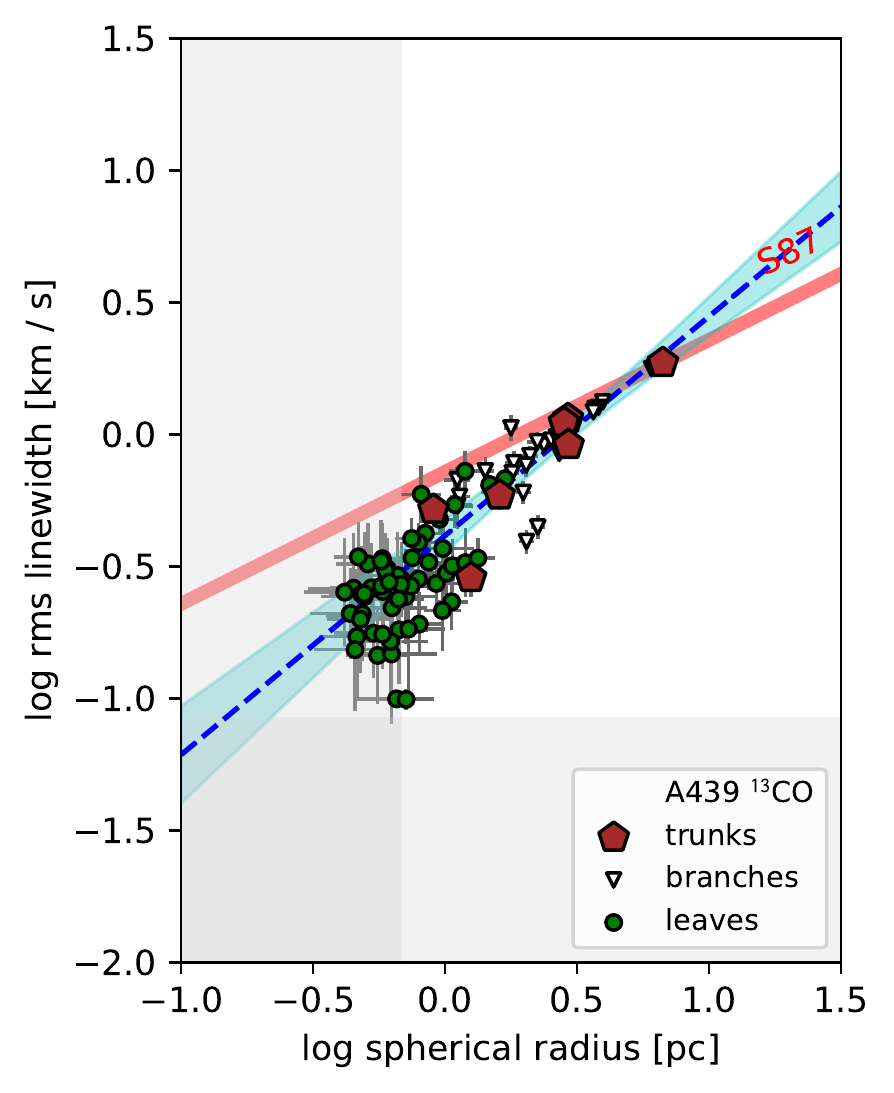}\\
    \includegraphics[width=0.32\textwidth]{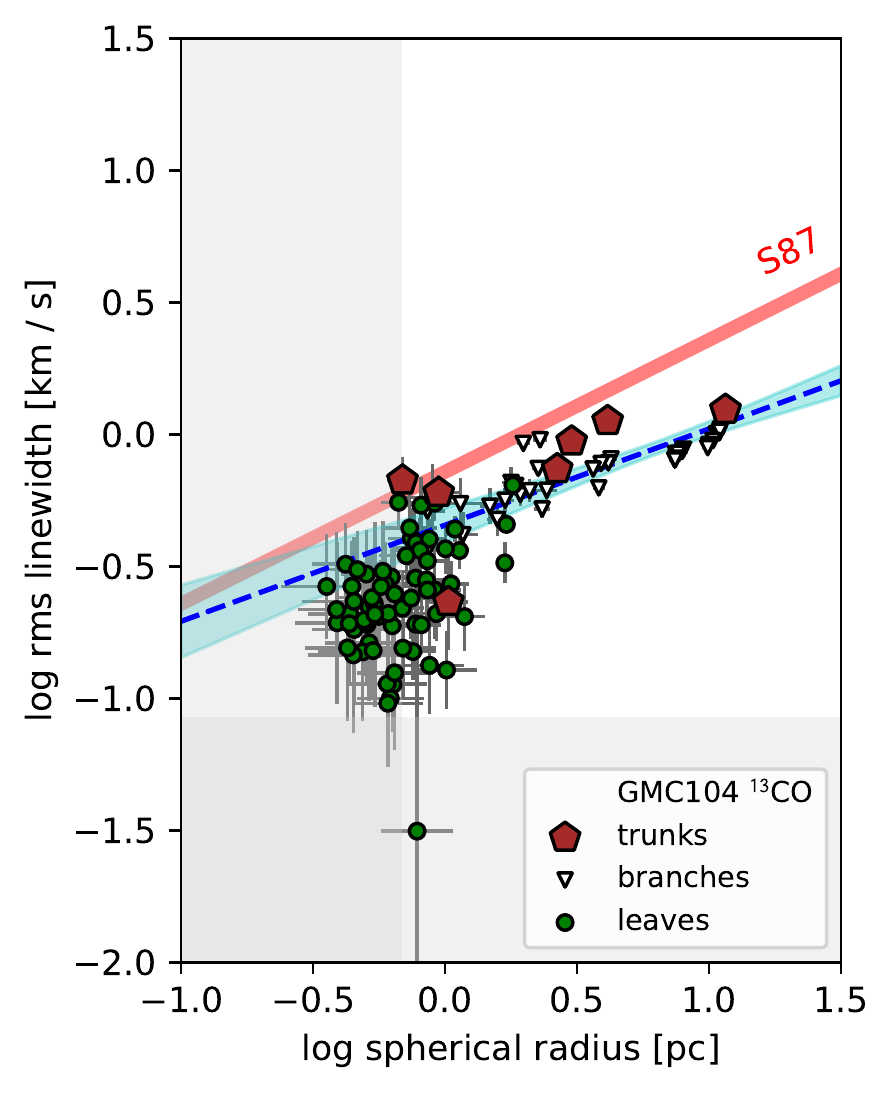}
    \includegraphics[width=0.32\textwidth]{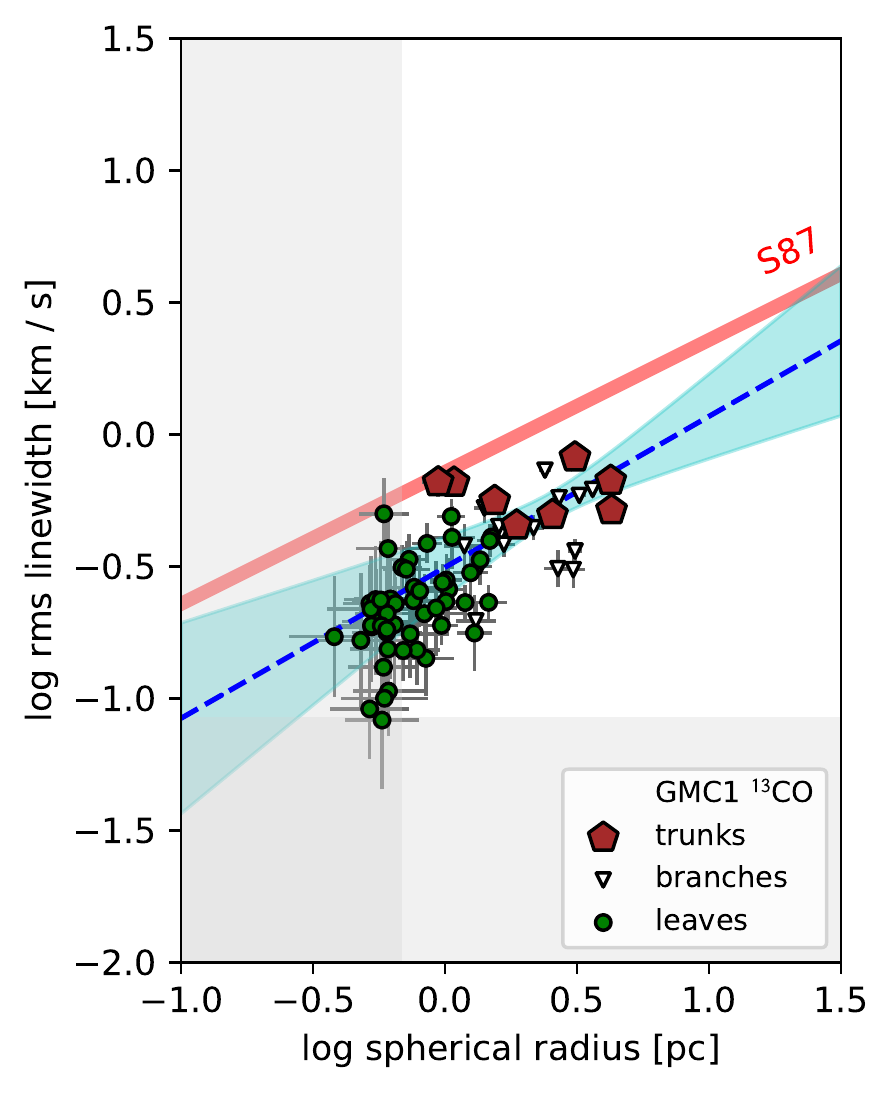}
    \includegraphics[width=0.32\textwidth]{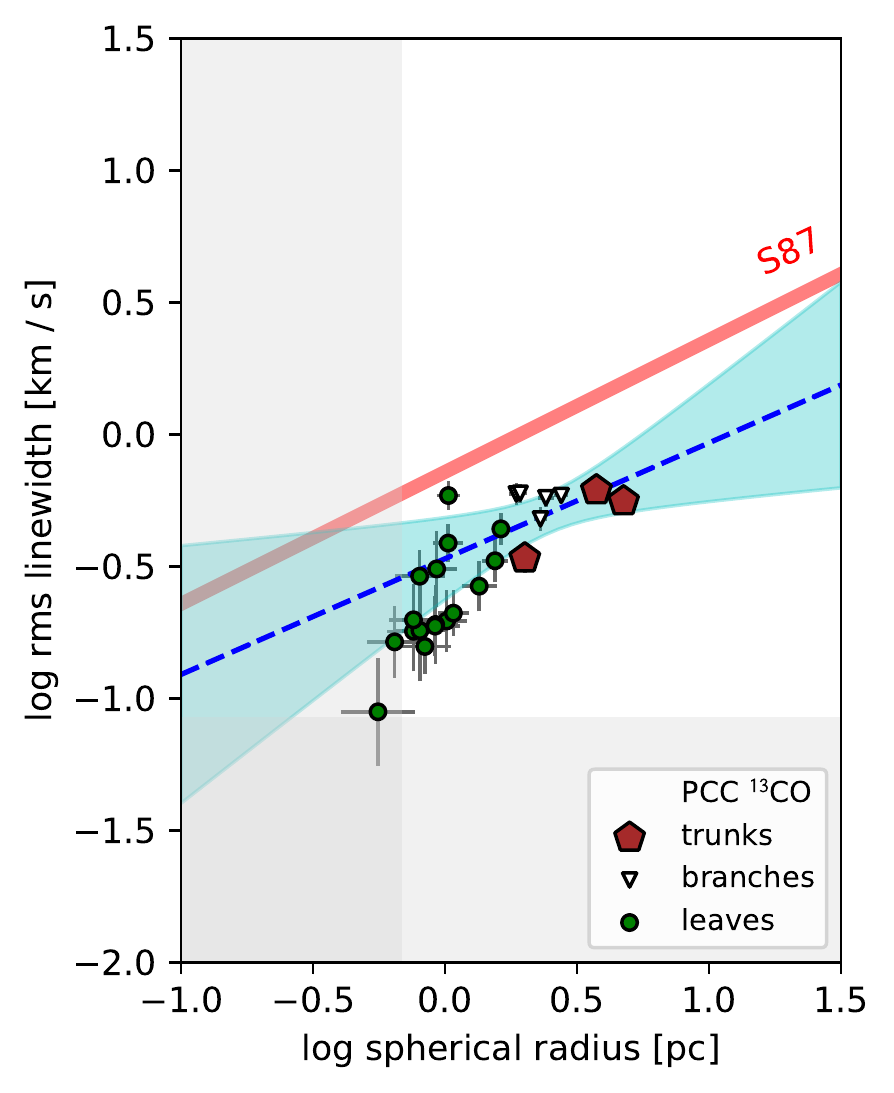}\\
    \caption{Size-line width relations for \ttco\ structures in the six molecular clouds. Plot symbols and overlays are the same as in Figure~\ref{fig:rdv12}.}
    \label{fig:rdv13}
\end{figure*}

To identify significant emission structures in the datacubes we used the Python package {\tt astrodendro}\footnote{\url{http://www.dendrograms.org}}, which decomposes emission into a hierarchy of structures \citep{Rosolowsky:08,Shetty:12,Colombo:15}.  Our procedure follows closely that used in our previous analysis of the PCC and 30~Dor clouds \citepalias{Wong:17}.  The algorithm identifies local maxima in the cube above the 3$\sigma_{\rm rms}$ level that are also at least 2.5$\sigma_{\rm rms}$ above the merge level with adjacent structures.  Each local maximum is required to span at least two synthesized beams in area.  Isosurfaces surrounding the local maxima are categorized as trunks, branches, or leaves according to whether they are the largest contiguous structures (trunks), are intermediate in scale (branches), or have no resolved substructure (leaves).  For a given cloud, the trunks do not overlap other trunks and leaves do not overlap other leaves, but every trunk can be decomposed into leaves (and usually branches).  The resulting dendrogram for CO in the A439 cloud is shown in Figure~\ref{fig:dendrogram}, with leaves, branches, and trunks colored green, black, and maroon, respectively.

The basic properties of the identified structures are also determined by {\tt astrodendro}, including their spatial and velocity centroids ($\bar{x}, \bar{y}, \bar{v}$), the integrated flux $S$, rms line width $\sigma_v$ (defined as the intensity-weighted second moment of the structure along the velocity axis), the position angle of the major axis (as determined by principal component analysis) $\phi$, and the rms sizes along the major and minor axes, $\sigma_{\rm maj}$ and $\sigma_{\rm min}$. 
All properties are determined using the ``bijection'' approach discussed by \citet{Rosolowsky:08}, which associates all emission bounded by an isosurface with the identified structure.
From these basic properties we have calculated additional properties, including the effective rms spatial size, $\sigma_r = \sqrt{\sigma_{\rm maj}\sigma_{\rm min}}$, the spherical radius $R = 1.91 \sigma_r$, following \citetalias{Solomon:87}, the luminosity $L=Sd^2$, adopting $d=50$ kpc \citep{Pietrzynski:19}, the virial mass $M_{\rm vir}=5\sigma_v^2R/G$, derived from solving the equilibrium condition
\begin{equation}\label{eq:vireq}
2{\cal T} + {\cal W} = 2\left(\frac{3}{2}M\sigma_v^2\right) - \frac{3}{5}\frac{GM^2}{R} = 0\,,
\end{equation}
and the luminosity-based mass (from $^{12}$CO)
\[\frac{M_{\rm CO}}{M_\odot} = 4.3X_2\, \frac{L_{\rm CO}}{\rm K\;km\;s^{-1}\;pc^2}\,,\]
where $X_2=1$ for a standard (Galactic) CO to H$_2$ conversion factor \citep{Bolatto:13a}.  In this paper we have adopted $X_2=2.4$ for the CO(1--0) line based on the virial analysis of the MAGMA GMC catalog by \citet{Hughes:10}, and $X_2=3$ for the CO(2--1) line assuming a CO(2--1)/CO(1--0) line ratio of 0.8.  The line ratio is known to vary with cloud conditions (\citealt{Sorai:01}; \citetalias{Wong:17}), with values $\sim$0.5 for clouds in the outskirts of the LMC and rising to $\sim$1.2 near 30 Dor, so our adoption of a constant value is only approximate.

Our expression for virial mass is not fully self-consistent given that \citetalias{Solomon:87} assumed a truncated $\rho \propto r^{-1}$ density profile in deriving $R = 1.91 \sigma_r$, whereas we have assumed a constant density sphere in deriving the potential energy ${\cal W}$.  For $\sigma_v$ in \kms\ and $\sigma_r$ in pc, our expression for $M_{\rm vir}$ simplifies to 
$2.22 \times 10^3\, \sigma_v^2 \sigma_r$ \Msol, which is about 10\% higher than the equivalent expression in \citetalias{Solomon:87}.  For simplicity, and consistent with the crudeness of our treatment of the virial theorem, we choose not to correct for this offset.

As with any emission segmentation approach, one can justifiably question the reality of the recovered structures in the dendrogram \citep[see discussion in][]{Shetty:10,Beaumont:13}.  Since we are principally interested in recovering the line widths of structures spanning a range of different sizes, our results should be consistent with other approaches that measure emission properties within bounded regions of the data cube---i.e., that avoid extrapolation of properties beyond the structure boundaries.  Any approach that seeks to recover line widths at different size scales must ultimately compare nested rather than disjoint structures, and so is unlikely to yield dramatically different results.  Note that we do not attempt to generate histograms (e.g., clump mass spectra) of the structure properties.  Even if limited to a set of disjoint structures (e.g., the subset of dendrogram leaves), such histograms would be difficult to interpret given the continuous nature of the CO emission (Figure~\ref{fig:mom0maps1} and \ref{fig:mom0maps2}) and the fact that leaf structures will tend to be similar by construction.  Uniformly sampled property measurements \citep[e.g.,][]{Hughes:13,Leroy:16} are probably better suited for obtaining parameter distributions.

\subsection{Size vs.\ Line Width Relations}\label{sec:rdv}

\begin{deluxetable*}{cD@{ $\pm$}DD@{ $\pm$}Drr|D@{ $\pm$}DD@{ $\pm$}Drr}
\tablehead{
\colhead{} & \multicolumn{10}{c}{\twco} & \multicolumn{10}{c}{\ttco}\\
\cline{2-11} \cline{12-21}
\colhead{Cloud} & \multicolumn{4}{c}{$a_1$} & \multicolumn{4}{c}{$a_0$} & \colhead{$\chi^2_\nu$} & \colhead{$\varepsilon$\tablenotemark{a}} & \multicolumn{4}{c}{$a_1$} & \multicolumn{4}{c}{$a_0$} & \colhead{$\chi^2_\nu$} & \colhead{$\varepsilon$\tablenotemark{a}}}
\tablecaption{Power Law Fit Parameters: $\log \sigma_v = a_1 \log R + a_0$\label{tab:fitpar}}
\decimals
\startdata
30Dor & 0.60 & 0.04 & 0.22 & 0.03 & 4.30 & 0.10 & 0.43 & 0.08 & 0.21 & 0.04 & 10.86 & 0.14 \\
N59C & 0.97 & 0.04 & -0.34 & 0.02 & 17.63 & 0.19 & 1.00 & 0.12 & -0.35 & 0.04 & 11.41 & 0.24 \\
A439 & 0.53 & 0.02 & -0.24 & 0.02 & 9.09 & 0.15 & 0.83 & 0.04 & -0.38 & 0.02 & 3.07 & 0.13 \\
GMC104 & 0.46 & 0.02 & -0.31 & 0.02 & 6.77 & 0.12 & 0.36 & 0.02 & -0.34 & 0.02 & 4.76 & 0.15\\
GMC1 & 0.29 & 0.03 & -0.38 & 0.02 & 9.59 & 0.16 & 0.57 & 0.08 & -0.50 & 0.04 & 5.98 & 0.15\\
PCC & 0.32 & 0.03 & -0.34 & 0.03 & 6.56 & 0.13 & 0.44 & 0.10 & -0.47 & 0.05 & 4.19 & 0.14\\
All  & 0.65 & 0.03 & -0.33 & 0.02 & 60.28 & 0.25 & 0.54 & 0.04 & -0.31 & 0.03 & 38.99 & 0.25 \\
\enddata
\tablenotetext{a}{rms scatter in $\log \sigma_v$ relative to the best-fit line.  Units are dex.}
\end{deluxetable*}

Figures~\ref{fig:rdv12} and \ref{fig:rdv13} show the individual $R$--$\sigma_v$ relations for the six clouds in \twco\ and \ttco\ respectively.  Each relation is fit by a power-law model of the form
\begin{equation}
    \log \sigma_v = a_1\log R + a_0\;,
\end{equation}
with the slope ($a_1$) and intercept ($a_0$) of the fitted line reported in Table~\ref{tab:fitpar}.  Following \citetalias{Wong:17}, fitting is performed using the {\tt kmpfit} module of the Python package {\tt Kapteyn} \citep{KapteynPackage}, which treats errors in both axes using the effective variance method.  Points in the gray shaded regions are excluded from fitting due to resolution limitations.  However, because a sharp truncation of the data can skew the fit results, we have also repeated the fitting without excluding the shaded regions.  The resulting slopes and intercepts are usually within the quoted 1$\sigma$ uncertainties, but are sometimes discrepant by nearly 3$\sigma$.  We have therefore chosen to plot a (conservative) 3$\sigma$ confidence interval as blue shading in Figures~\ref{fig:rdv12} and \ref{fig:rdv13}, while still listing the 1$\sigma$ uncertainties in Table~\ref{tab:fitpar}.  We note that the fit parameters are quite sensitive to the selection of data points to be fitted and their relative uncertainties, and thus fits to particular subsets of the data, or to differently weighted data, could differ from the reported fits by much more than 1$\sigma$.  Moreover, a power law is generally a poor fit to the data, as reflected by the reduced $\chi^2$ values all being substantially greater than 1 (Table~\ref{tab:fitpar}).

In Figures~\ref{fig:rdv12} and \ref{fig:rdv13} the fiducial relation of \citetalias{Solomon:87} for Galactic clouds is shown in pink.  The LMC clouds appear roughly consistent with this relation, although smaller structures tend to exhibit a wide range of line widths.  Comparing the \twco\ and \ttco\ relations for a given cloud, we find no systematic differences between the fitted slopes or intercepts.  {Particularly for the \ttco} line, however, there {appears to be a} progression from large to small line widths at a given size when {comparing clouds with} decreasing star formation activity.  This is consistent with the discrepancy in the $R$--$\sigma_v$ relations for 30 Dor and PCC previously noted in \citetalias{Wong:17}, but the four additional clouds clearly demonstrate a continuous variation across the sample.  {Variations in the fitted slope, in particular a fairly steep slope in the case of N59C, make this trend less obvious when examining only the fitted values of $a_0$; the trend is revealed more clearly when referencing the $R$--$\sigma_v$ relation to a common slope (\S\ref{sec:cldavg}).}

\begin{figure*}
    \centering
    \includegraphics[height=0.36\textwidth]{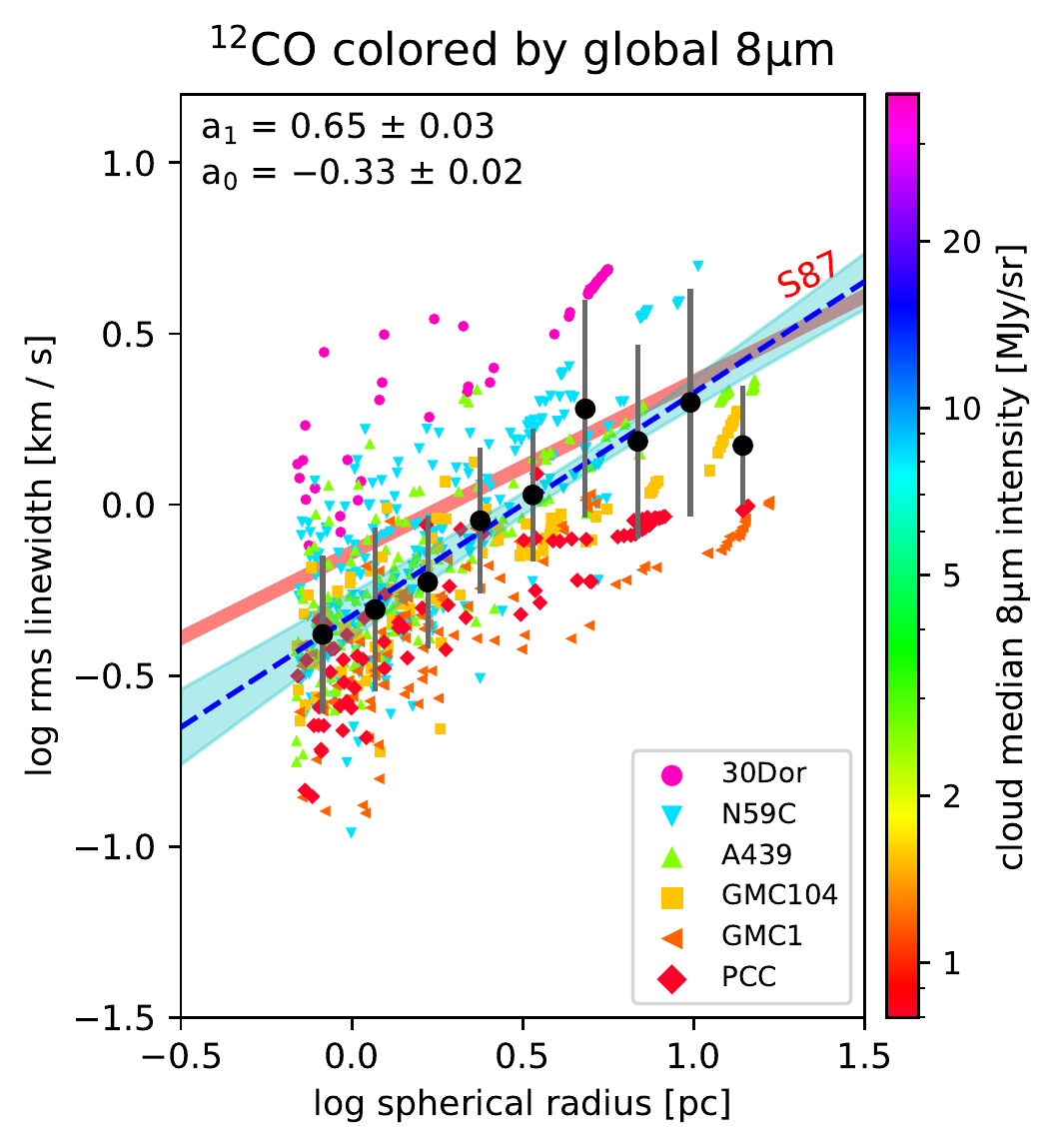}
    \hspace{0.5in}
    \includegraphics[height=0.36\textwidth]{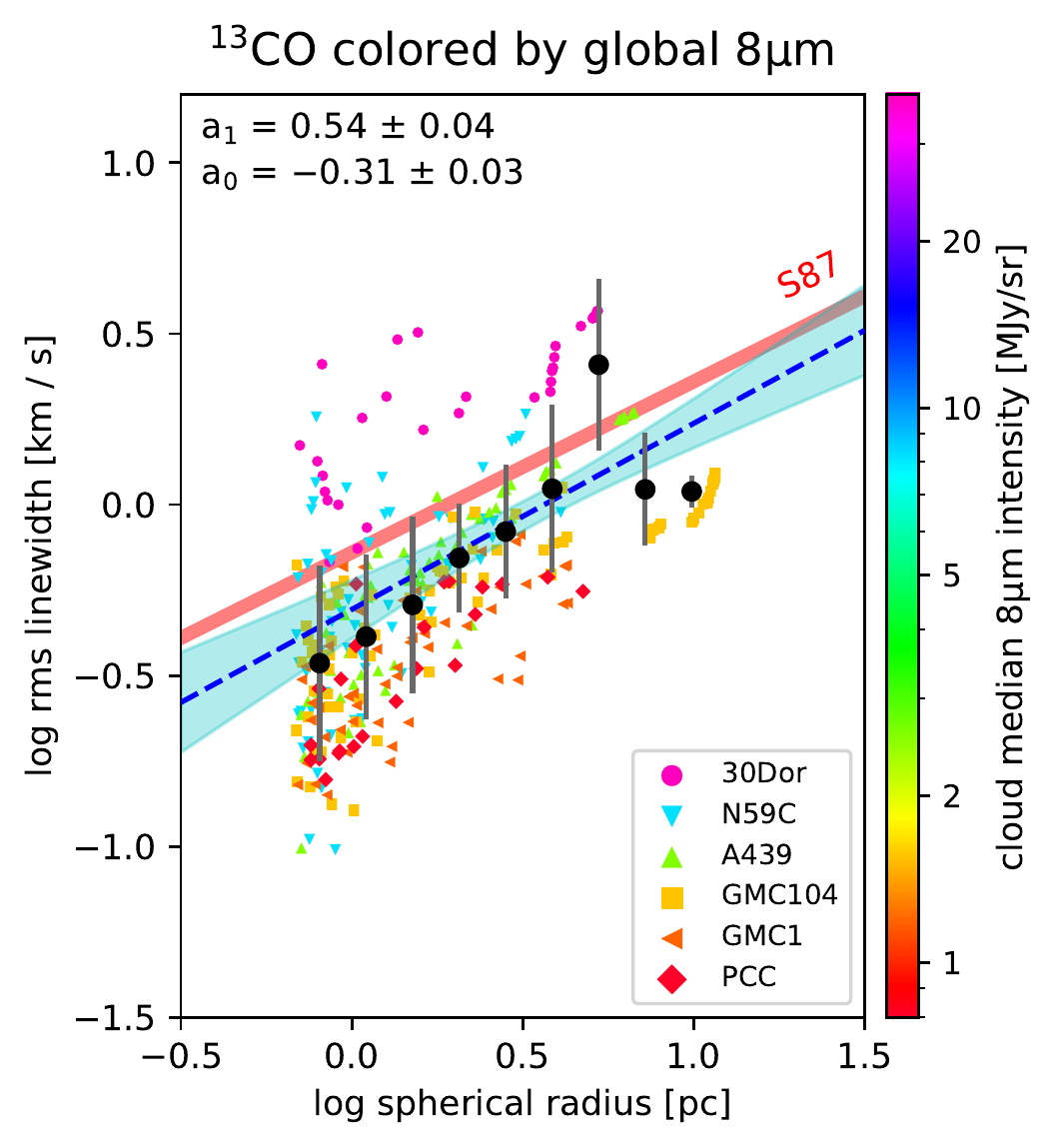}\\
    \includegraphics[height=0.36\textwidth]{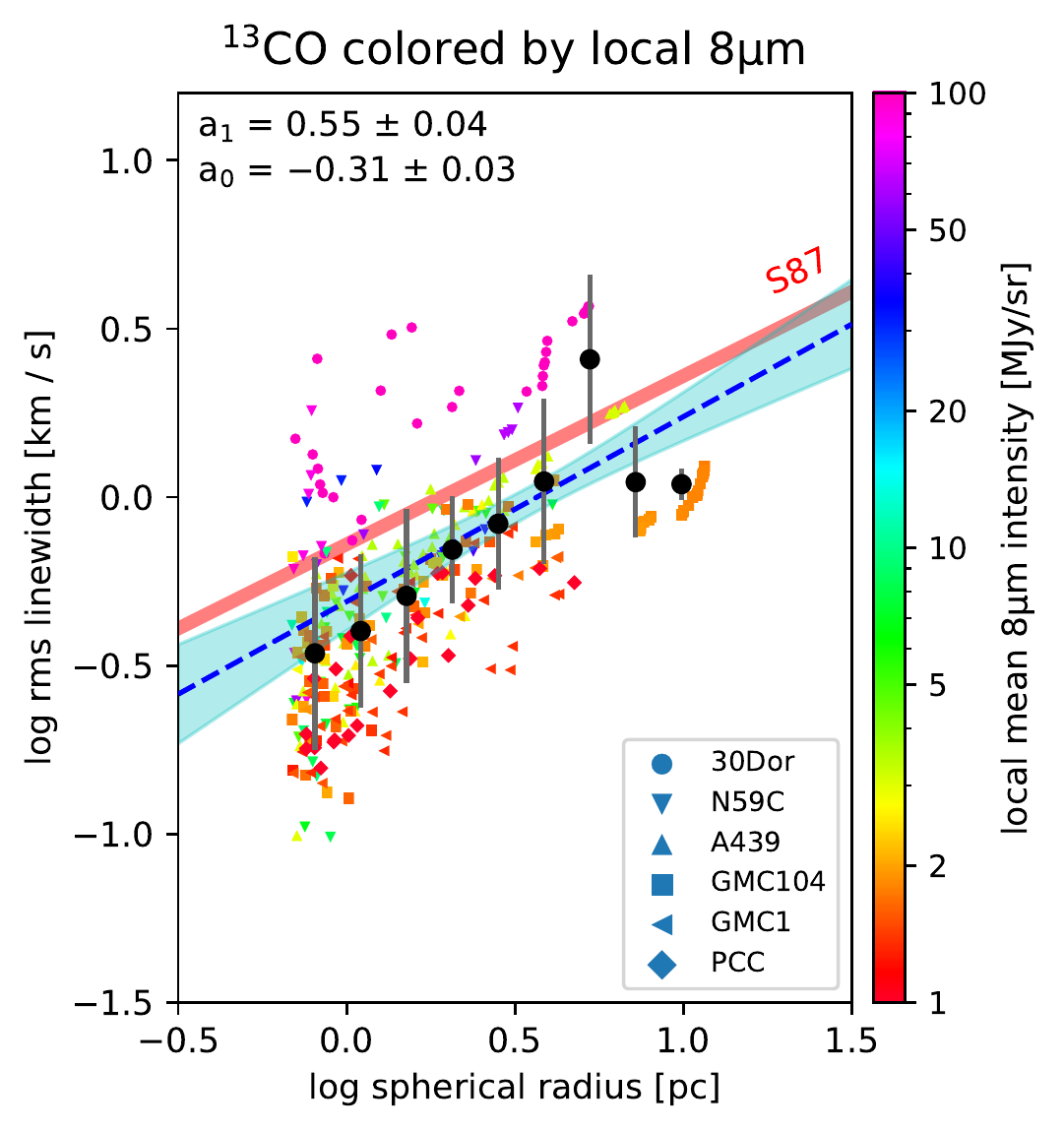}
    \hspace{0.5in}
    \includegraphics[height=0.36\textwidth]{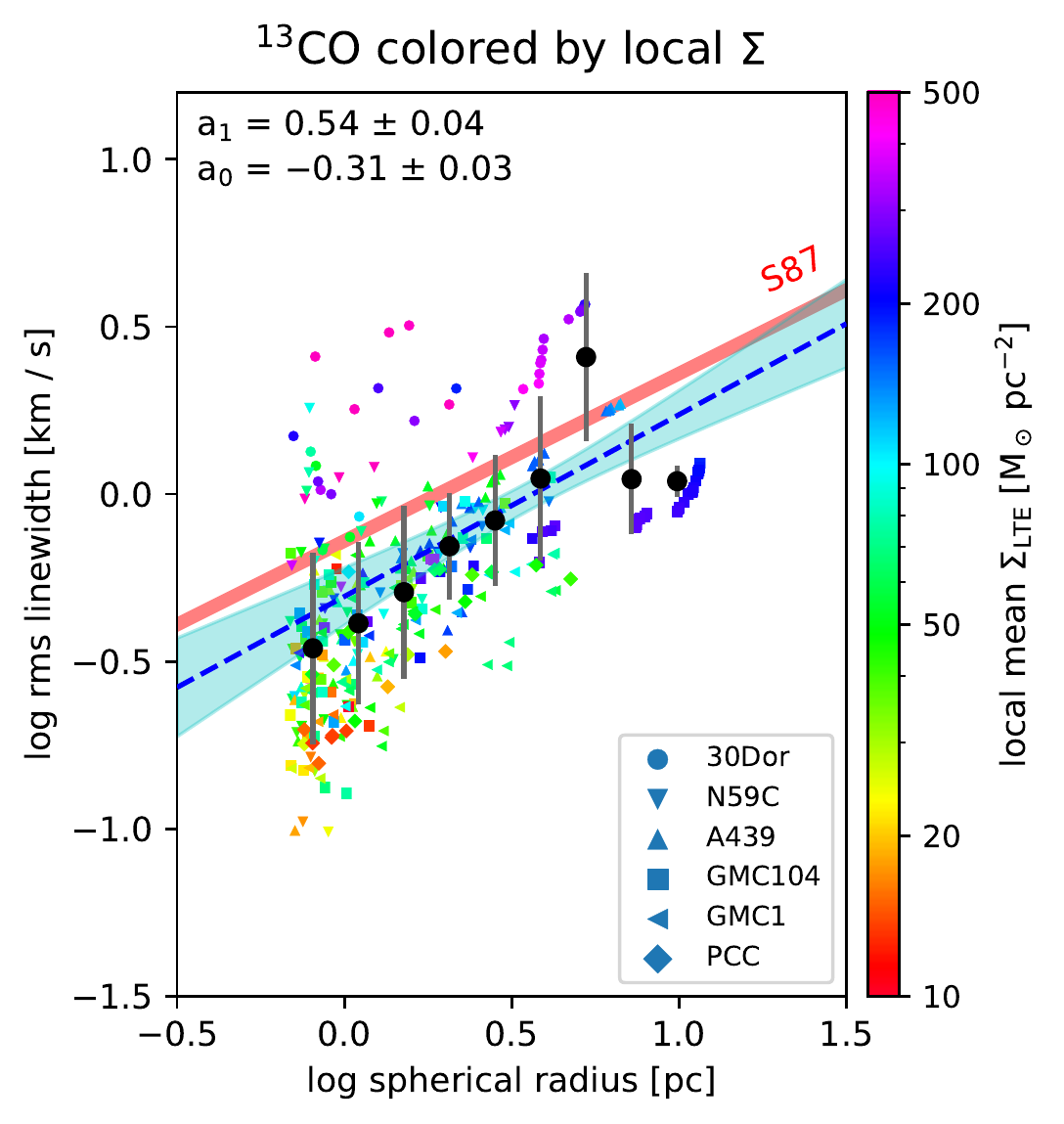}\\
    \caption{Aggregated $R$--$\sigma_v$ relations for the six molecular clouds, color-coded by 8\,$\mu$m intensity.  The fitted slope and intercept for the aggregate sample is shown at the upper left of each panel; for comparison, the \citetalias{Solomon:87} relation (pink line) has a slope of 0.5 and an intercept of $-0.14$.  {The mean and standard deviation in evenly spaced bins are shown as black points, but are not used in fitting.}  The upper left panel shows properties for \twco\ structures while the remaining panels show properties for \ttco\ structures, using different color codes as indicated by the label on the color bar.  The line width at fixed size increases with 8\,$\mu$m intensity, both locally and on a cloud-scale basis, and locally with LTE-based surface density.}
    \label{fig:comp:rdv}
\end{figure*}

Figure~\ref{fig:comp:rdv} provides a summary view of the $R$--$\sigma_v$ relation across the six clouds, with points for each cloud colored according to the median 8$\mu$m intensity within the cloud.  The aggregate relation for \twco\ can be described by
\begin{equation}
    \log \sigma_v\, {\rm [km\, s^{-1}]} = 0.65\log R\, {\rm [pc]} - 0.33\;,
\end{equation}
with an rms scatter of $\varepsilon = 0.25$ dex in $\sigma_v$ (see Table~\ref{tab:fitpar}).  {In these and subsequent plots, the fitting is still performed on individual points weighted by their uncertainties, with binned values plotted to illustrate overall trends.} The aggregate scatter is substantially larger than the scatter for any single cloud, indicating that differences between clouds are substantial.  Again, these differences are in the sense of higher line widths for 8$\mu$m-bright clouds (blue and magenta colors) and smaller line widths for 8$\mu$m-faint clouds (orange and red colors).  {This is seen for the \twco-identified structures ({\it upper left panel}), the \ttco-identified structures} ({\it upper right panel}), and persists also when the marker colors are based on the local rather than cloud-scale 8$\mu$m intensity ({\it lower left panel}).  These {first} three panels of Figure~\ref{fig:comp:rdv} provide direct evidence for a coupling between the IR brightness and the CO velocity dispersion.

\begin{figure*}
    \centering
    \includegraphics[height=0.31\textwidth,trim=0 0 55 0,clip=true] {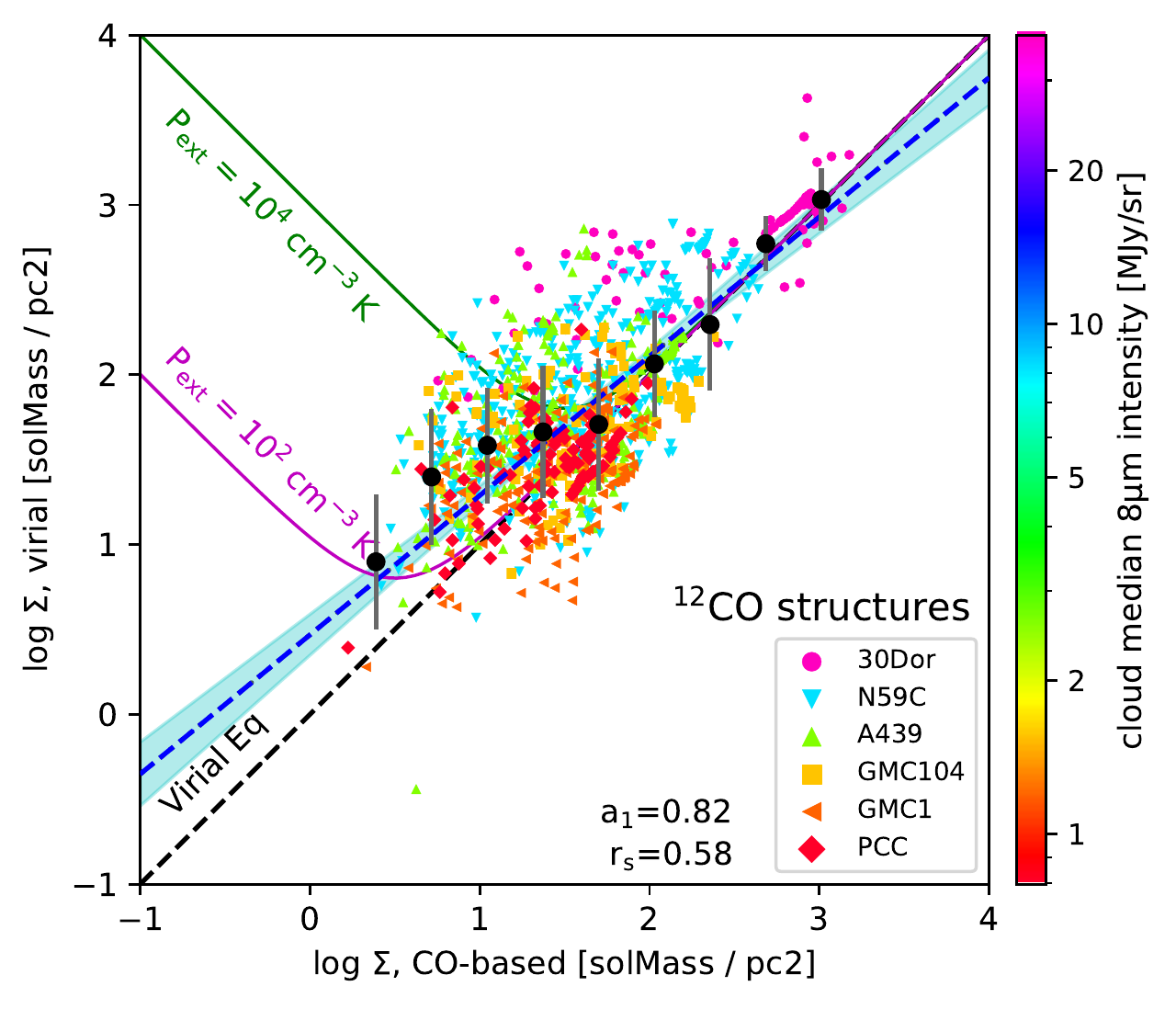}\hfill
    \includegraphics[height=0.31\textwidth,trim=0 0 55 0,clip=true] {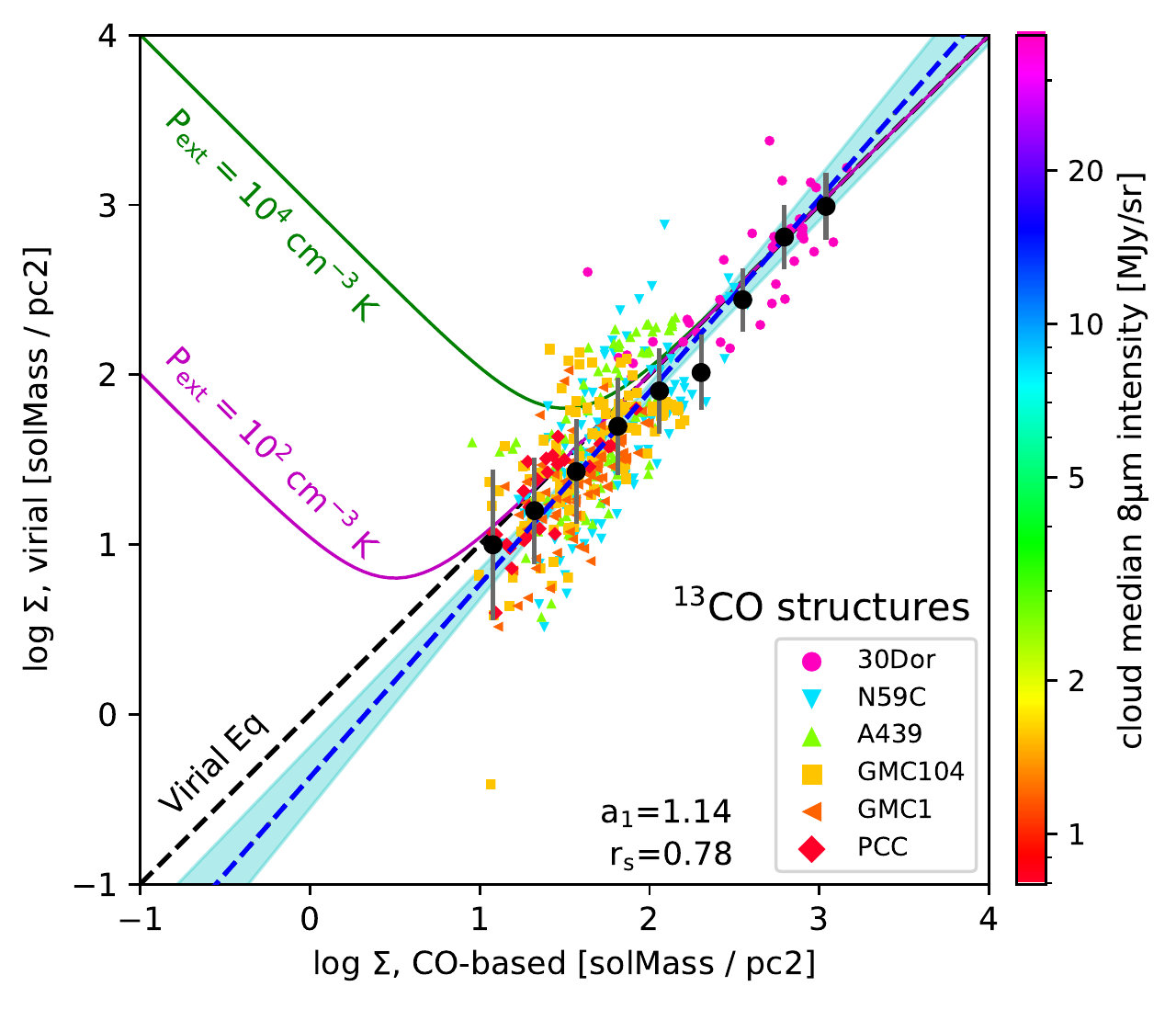}\hfill
    \includegraphics[height=0.31\textwidth]{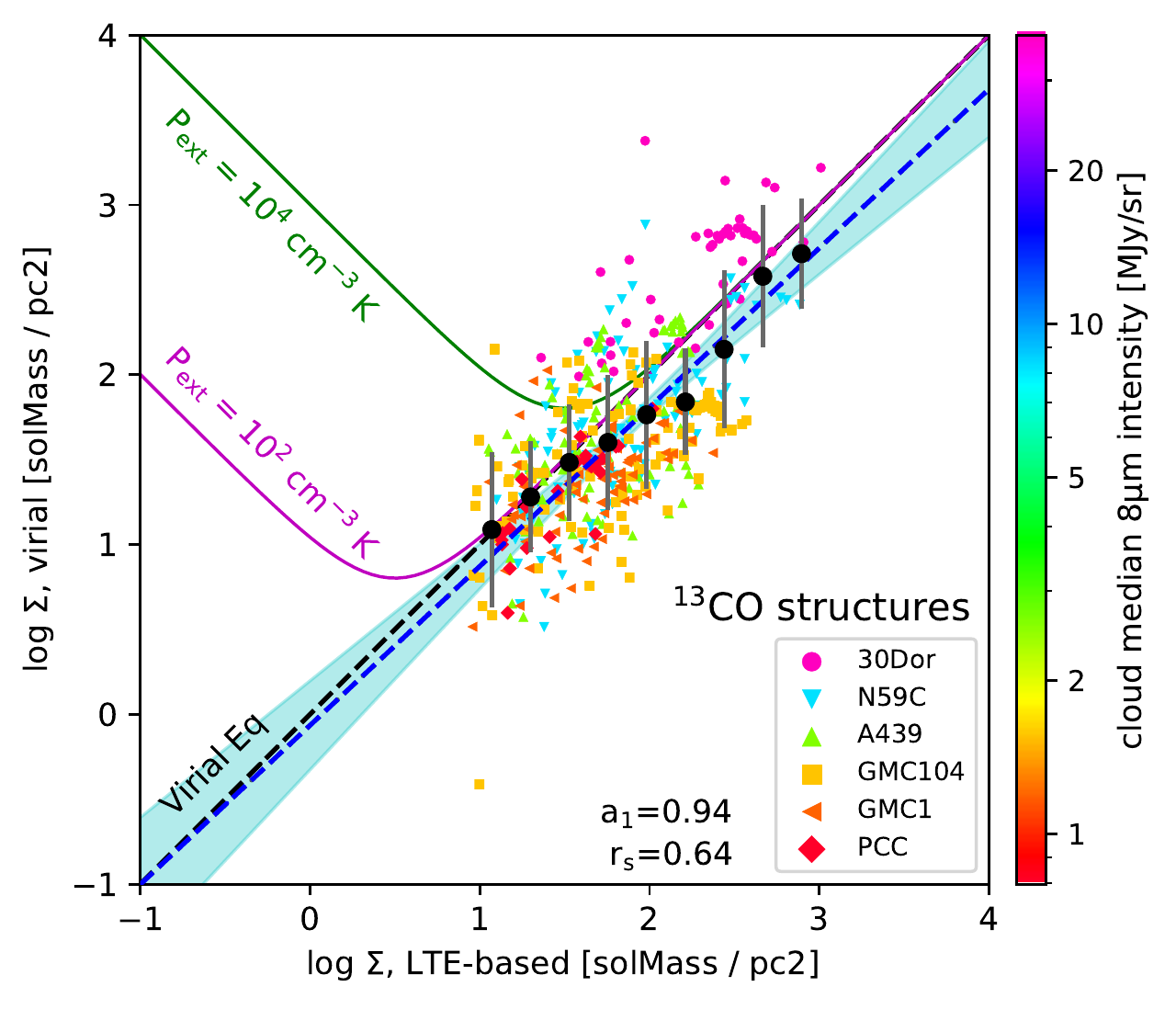}\\
    \caption{Comparisons of virial-based and CO-based surface densities for \twco\ structures ({\it left}) and \ttco\ structures ({\it middle}) in the six clouds.  For the \ttco\ structures, the CO-based surface densities are obtained by averaging the \twco\ intensities over the cube pixels that define each structure.  {\it Right:} Comparisons of virial and LTE-based surface densities for \ttco\ structures in the six clouds.  Fitted slopes and Spearman correlation coefficients are given at the lower right of each panel.  The model curves represent simple virial equilibrium (dashed line) and pressure-bounded equilibria at two different values of $P_{\rm ext}$ (Eq.~\ref{eqn:pext}).}
    \label{fig:comp:bnd}
\end{figure*}

\begin{figure*}
    \centering
    \includegraphics[height=0.41\textwidth,trim=0 0 55 0,clip=true]{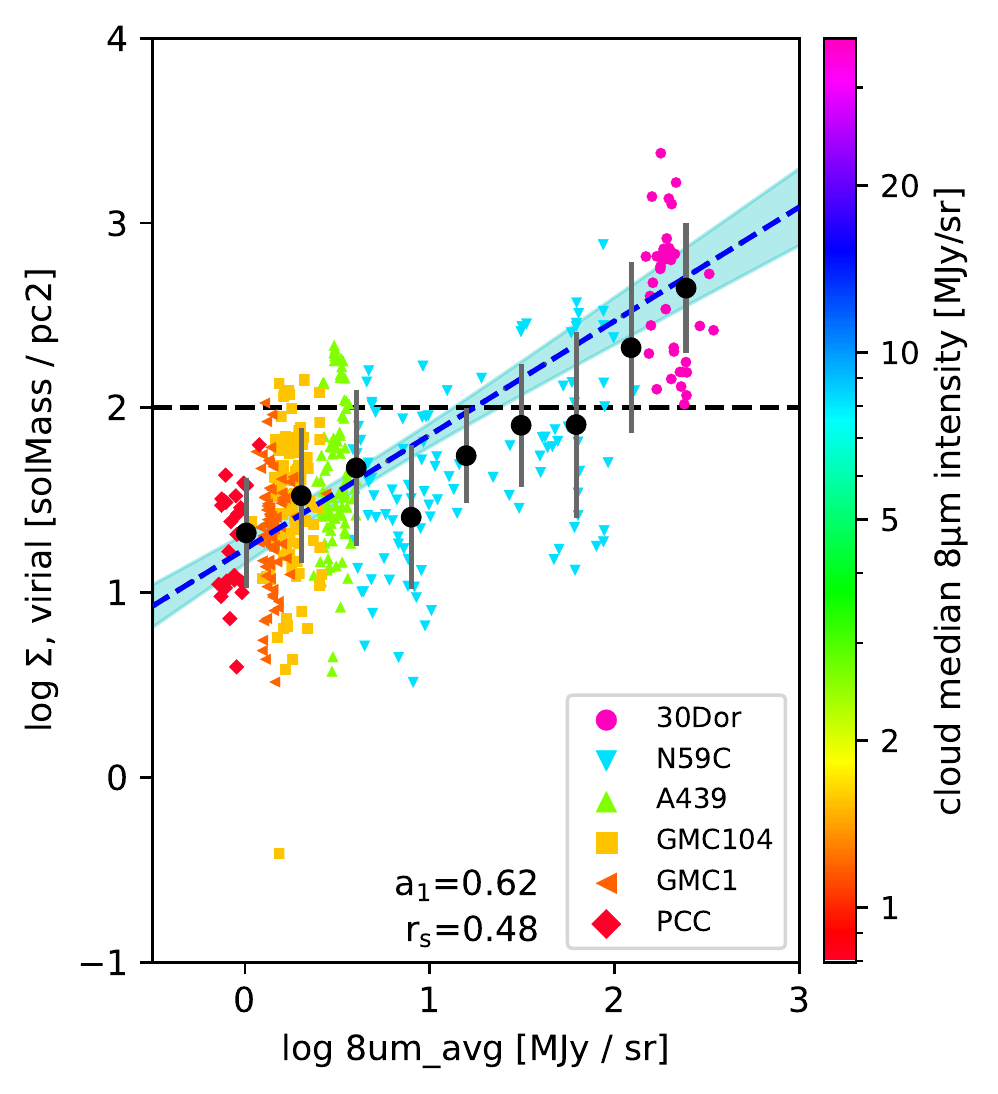}\hfill
    \includegraphics[height=0.41\textwidth,trim=0 0 55 0,clip=true]{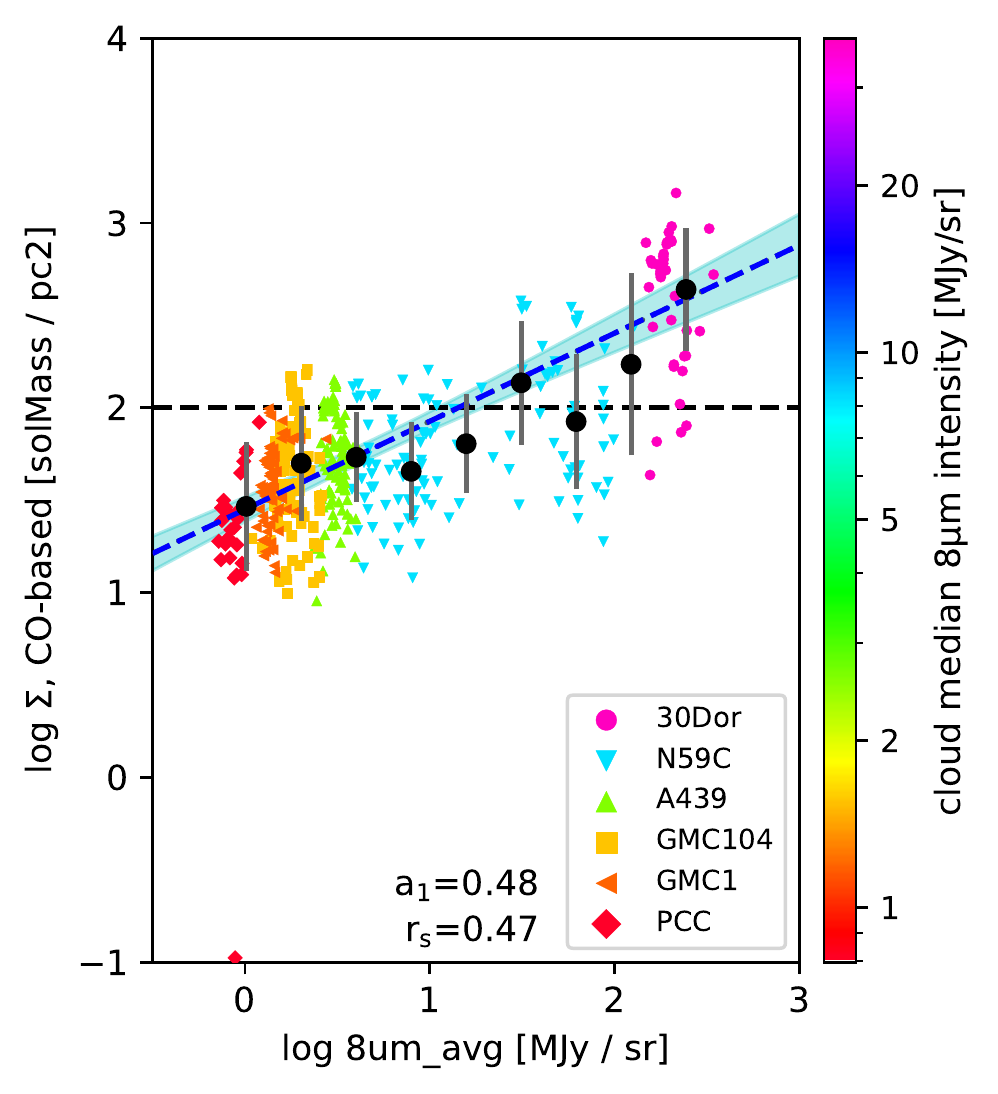}\hfill
    \includegraphics[height=0.41\textwidth]{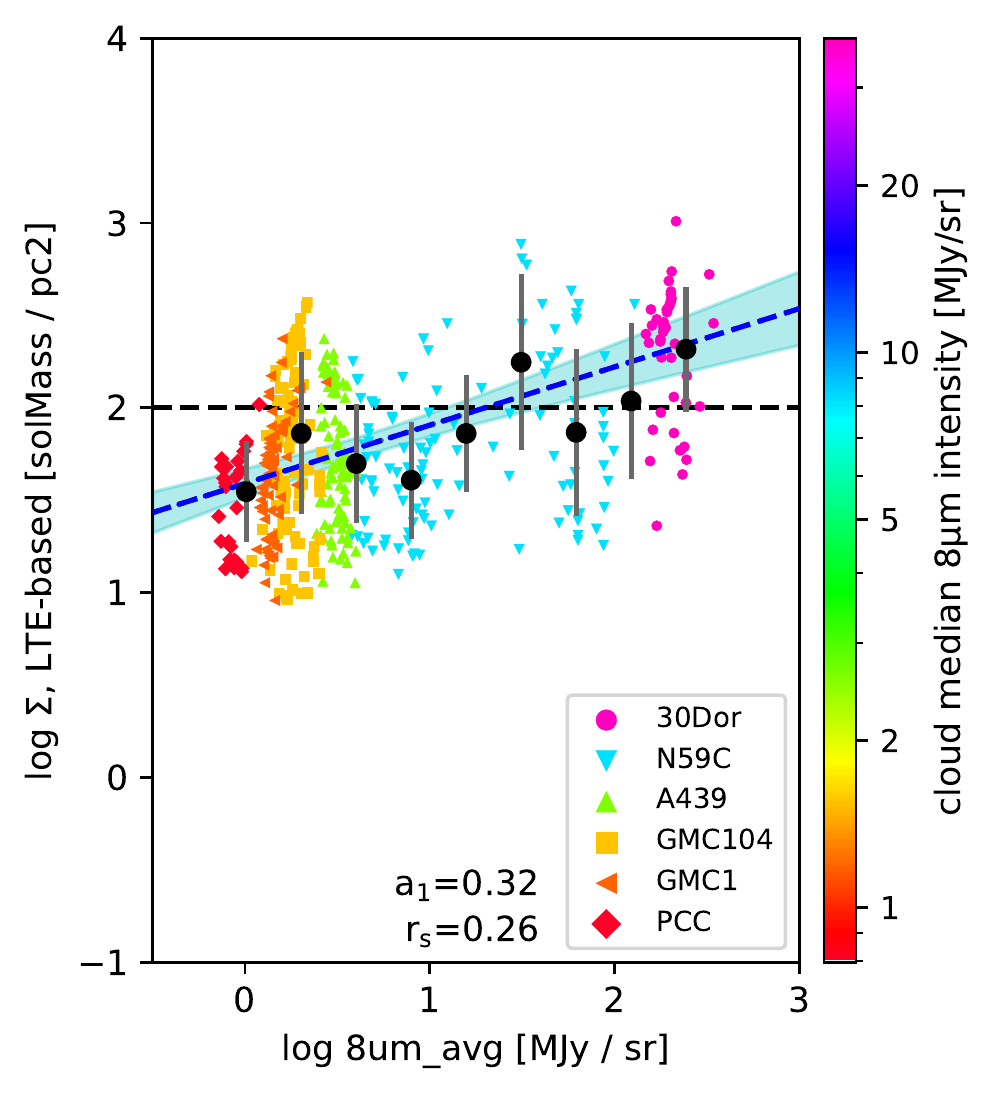}\\
    \caption{Correlation of mean local 8\,$\mu$m brightness with surface density $\Sigma$ for \ttco\ structures, with $\Sigma$ estimated from ({\it left}) virial equilibrium, ({\it middle}) $X_{\rm CO}$, and ({\it right}) LTE column density.  Values of $\Sigma$ determined from the virial and CO-based methods, which are more sensitive to the line width, show a stronger correlation with local 8\,$\mu$m brightness, as evidenced by larger slopes ($a_1$) and Spearman rank correlation coefficients ($r_s$).  A fiducial surface density of 100 $M_\odot$ pc$^{-2}$ is shown as the dashed horizontal line for comparison.}
    \label{fig:comp:8um}
\end{figure*}

\subsection{IR Brightness or Cloud Surface Density?}\label{sec:bound}

We have seen that the six clouds in our sample all show a consistent trend of increased line width at a given size with increased IR surface brightness.  A natural interpretation of this trend is that IR brightness tracks star formation activity and that higher star formation activity results in stronger stirring of the ISM due to feedback.  Here we consider an alternative interpretation, which posits that molecular clouds and their substructures lie close to simple virial equilibrium as defined by Equation~(\ref{eq:vireq}).  Increased line width then reflects higher mass surface densities, which also tend to correlate with higher star formation activity because the molecular gas is more susceptible to star formation or simply more abundant.  Figure~\ref{fig:comp:rdv} ({\it bottom right}) shows that when structures are color coded by the local surface density derived from LTE analysis, there is shift in surface density from the bottom to top of the plot which closely resembles the shift in IR brightness seen in the other panels.

The normalization of the $R$--$\sigma_v$ relation, $v_0 = \sigma_v/R^{1/2}$, is related to a virial mass surface density,
\[\Sigma_{\rm vir} = \frac{M_{\rm vir}}{\pi R^2} \propto \frac{\sigma_v^2}{R} = v_0^2\;,\]
so the plot of virial vs.\ luminous surface density (the so-called boundedness plot, Figure~\ref{fig:comp:bnd}) provides another view of the trends seen in the size-line width relation.  As expected, the 8$\mu$m-bright clouds are seen at higher $\Sigma_{\rm vir}$ in Figure~\ref{fig:comp:bnd}, i.e.\ toward the top of the figure.  It is clear, however, that $\Sigma_{\rm vir}$ (and thus $v_0$) is also correlated with observational measures of surface density, such as $\Sigma_{\rm CO}$ and $\Sigma_{\rm LTE}$; indeed, most of the \ttco\ structures lie close to the line of simple virial equilibrium ($\Sigma_{\rm vir}=\Sigma_{\rm obs}$, the diagonal dashed lines in each panel of Figure~\ref{fig:comp:bnd}).  We have also drawn lines of constant external pressure $P_{\rm ext}$, derived from setting the right-hand side of Equation~(\ref{eq:vireq}) equal to $4\pi R^3 P_{\rm ext}$ \citep{Field:11}:
\begin{equation}
\Sigma_{\rm vir} - \Sigma = \frac{20}{3\pi G}\frac{P_{\rm ext}}{\Sigma}\;.
\label{eqn:pext}
\end{equation}
We note that the observed \twco\ and \ttco\ structures show considerable scatter around the virial equilibrium line, with no clear ``threshold'' density above which structures tend to be virialized.  Furthermore, they do not appear consistent with confinement by a single value of $P_{\rm ext}$, although a role for a variable external pressure cannot be excluded.

\begin{figure}
    \includegraphics[width=0.47\textwidth]{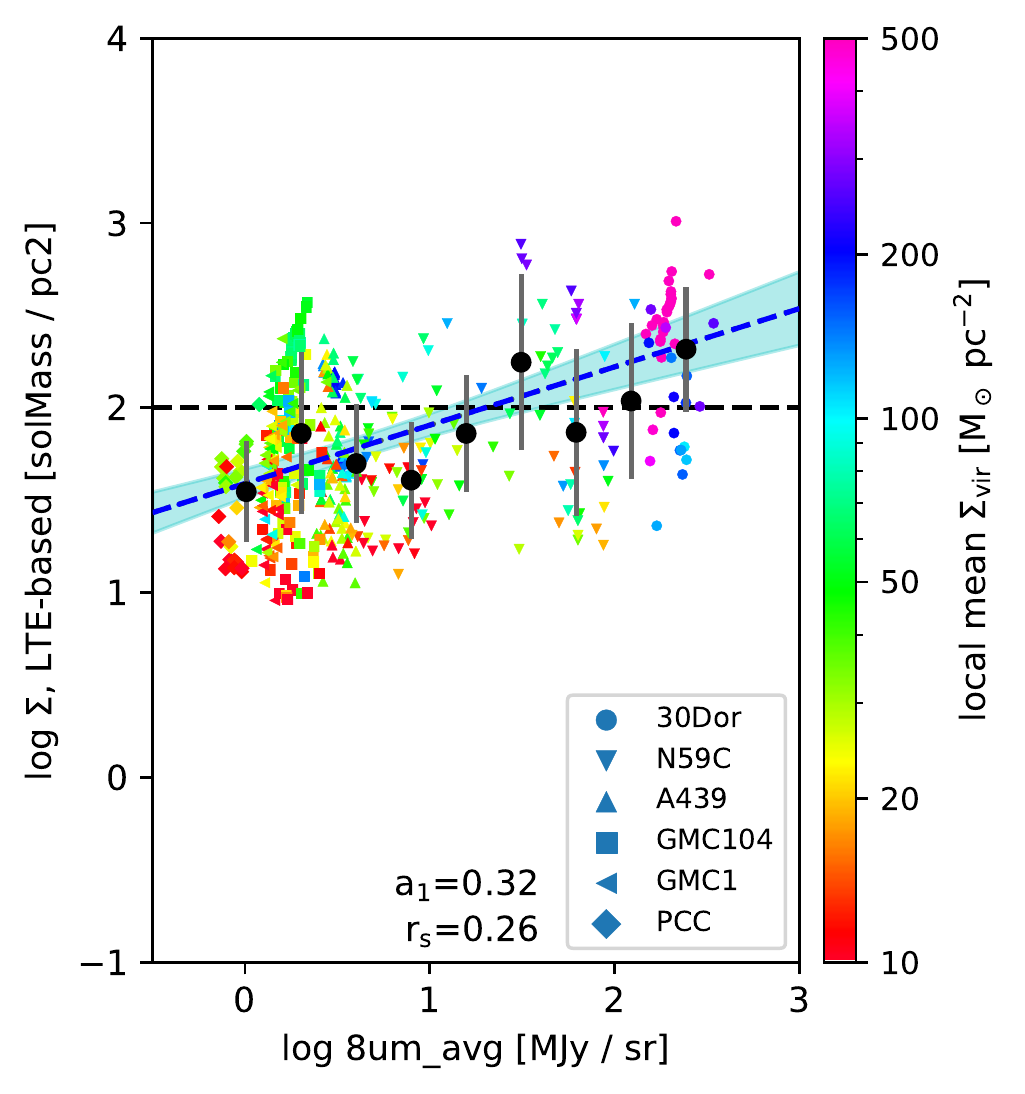}
    \caption{Similar to the last panel of Figure~\ref{fig:comp:8um}, but with points color coded by the local virial surface density $\Sigma_{\rm vir} \propto \sigma_v^2/R$, which scales monotonically with the normalization of the $R$--$\sigma_v$ relation, $v_0$.  Increasing trends in $\Sigma_{\rm vir}$ are seen both horizontally and vertically in this plot, indicating that both IR emission and mass surface density appear to correlate with $v_0$ when the other variable is held constant.}
    \label{fig:comp:svir}
\end{figure}

If the IR brightness is largely responding to $\Sigma$, as the virial interpretation would imply, we should find a good local correlation between 8$\mu$m intensity and $\Sigma$.  To investigate this we compare in Figure~\ref{fig:comp:8um} the 8$\mu$m intensity averaged within each \ttco\ dendrogram structure with the surface density of the structure measured in three different ways: from the virial theorem ({\it left panel}), CO brightness ({\it middle panel}), or LTE analysis ({\it right panel}).  While a correlation is apparent in all three panels, it becomes {progessively weaker for mass estimators that are less} sensitive to $\sigma_v$ ($\Sigma_{\rm vir} \propto \sigma_v^2$, and $\Sigma_{\rm CO} \propto \sigma_v$ for an optically thick CO line with a constant saturation brightness temperature.)  This suggests that the correlation between 8$\mu$m intensity and $\Sigma$, {while significant,} is driven in part by sensitivity of {common mass estimators} to the CO line width.

On the other hand, 8$\mu$m intensity alone does not appear to fully account for changes in $v_0$.  This is apparent from Figure~\ref{fig:comp:svir}, which shows that at fixed 8$\mu$m intensity, $\Sigma_{\rm vir} \propto v_0^2$ {(indicated by the marker colors)} increases with {independent} measures of $\Sigma$ (in this case, the LTE-based value, but the CO-based surface densities show the same pattern).  Conversely, at fixed $\Sigma$, $v_0$ increases with 8$\mu$m intensity.  Although {our crude LTE treatment of CO excitation may introduce intrinsic correlations between the two axes (e.g., by underestimating the column density in cold regions)}, we interpret this result as pointing to significant roles for both feedback {(represented by the abscissa)} and self-gravity {(represented by the ordinate)} in setting the normalization of the $R$--$\sigma_v$ relation {(represented by the the marker color)}.  {With a slope significantly below unity, the} data are also consistent with a steep dependence of the SFR surface density on $\Sigma$, a point we return to in \S\ref{sec:disc}.

\subsection{Cloud-averaged correlations}\label{sec:cldavg}

\begin{figure*}
    \centering
    \includegraphics[width=0.48\textwidth]{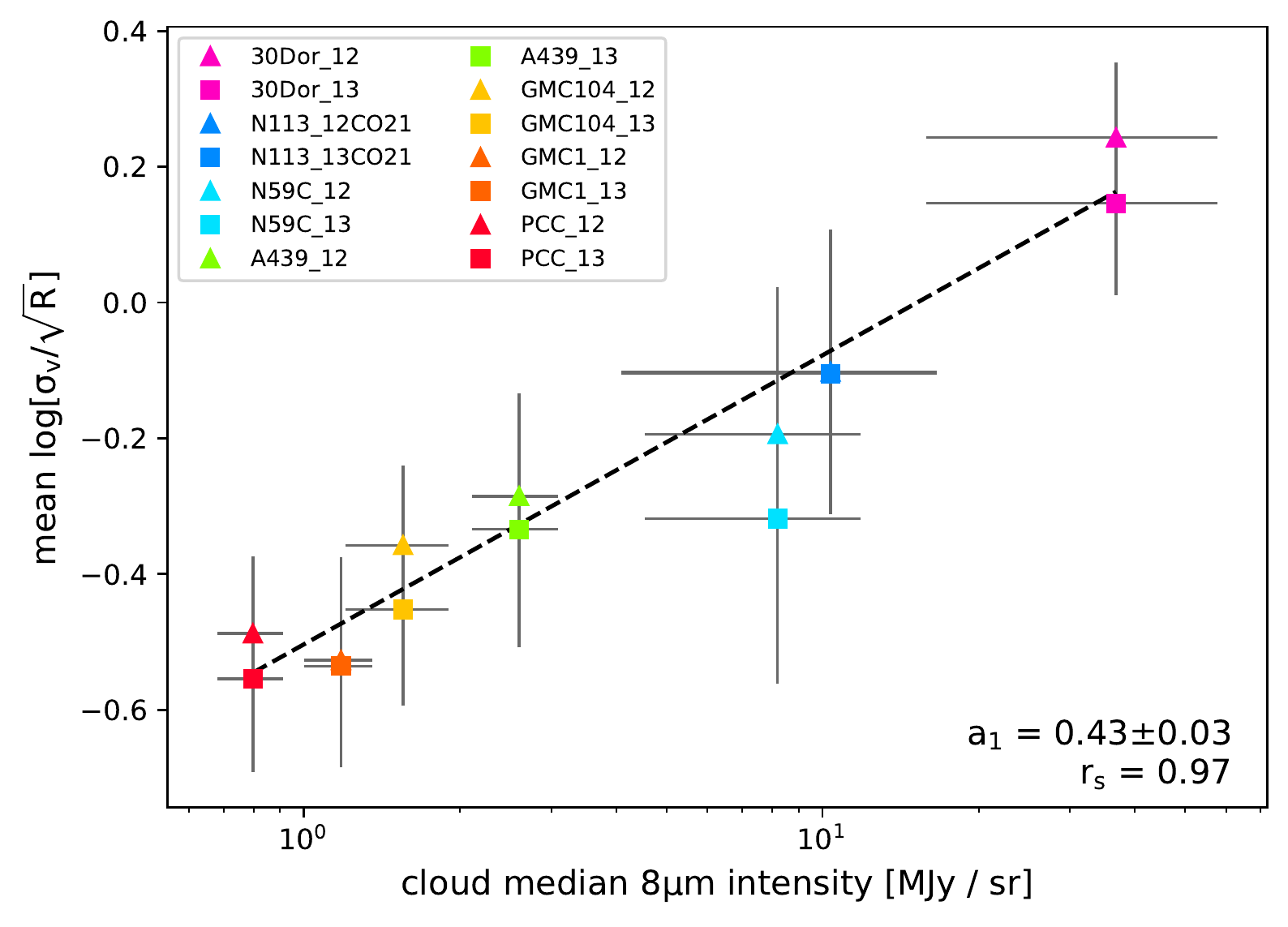}
    \includegraphics[width=0.48\textwidth]{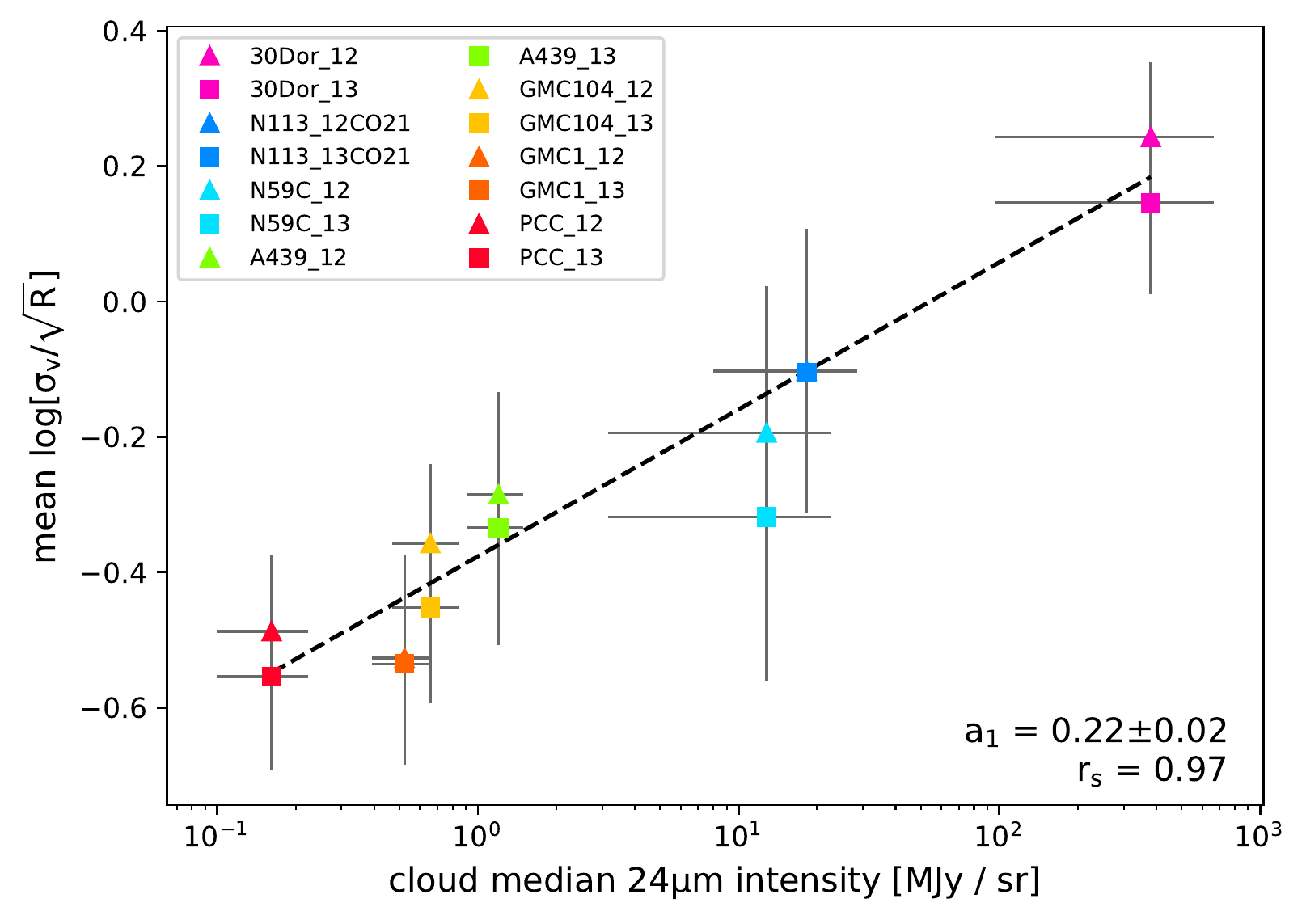}\\
    \includegraphics[width=0.48\textwidth]{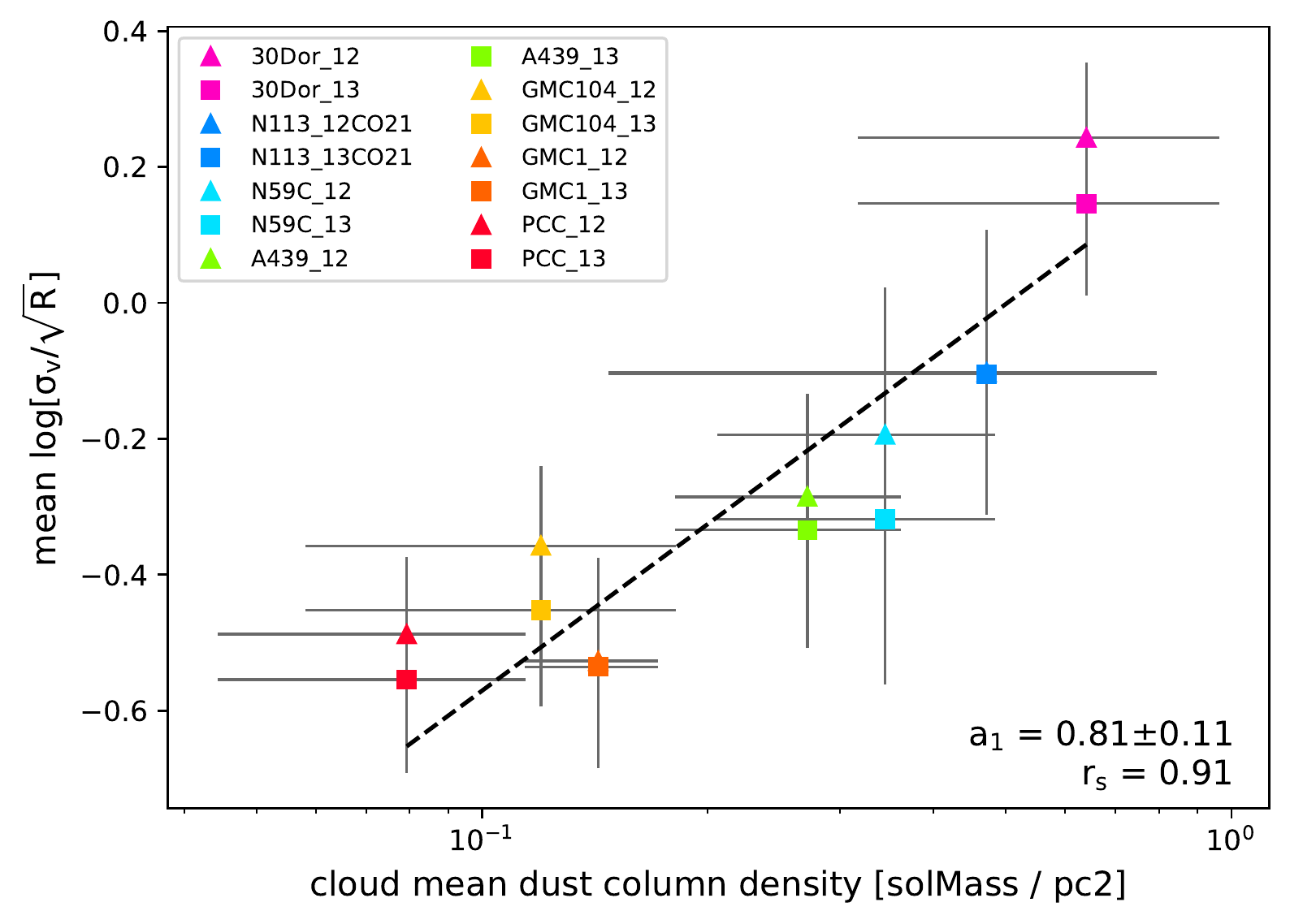}
    \includegraphics[width=0.48\textwidth]{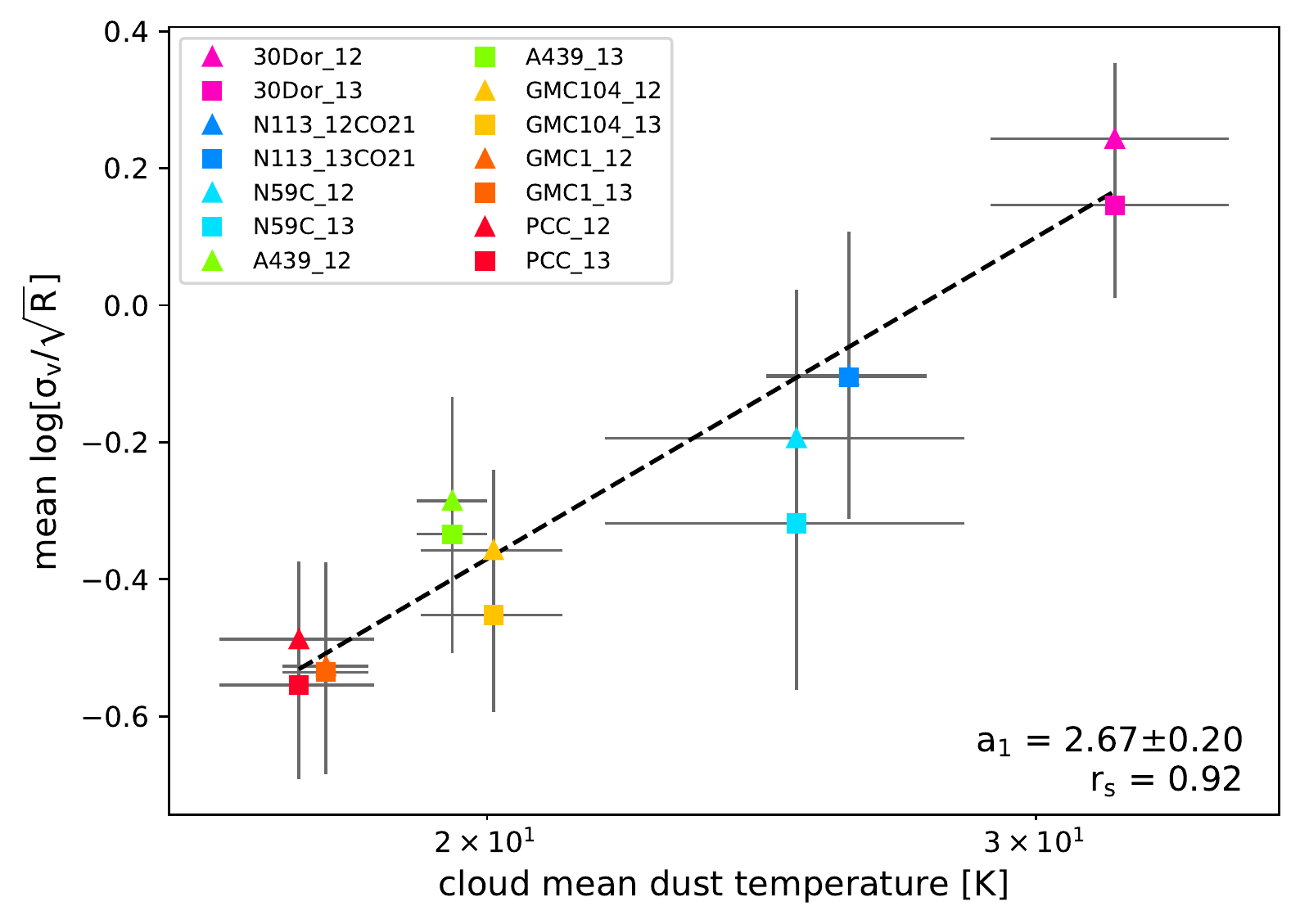}\\
    \caption{Correlation of cloud mean $v_0$ against ({\it upper left}) cloud median 8\,$\mu$m intensity; ({\it upper right}) cloud median 24\,$\mu$m intensity; ({\it bottom left}) cloud mean $N_{\rm dust}$; ({\it bottom right}) cloud mean $T_{\rm dust}$.  Triangles show values for \twco, while squares show values for \ttco.  {Preliminary results for N113 are also shown; the \twco\ and \ttco\ points nearly overlap because the abscissa values are identical and the ordinate values differ by $<$3\%.}  All relations show similarly good Spearman correlation coefficients, with differences in slope ($a_1$) reflecting differences in the dynamic range of the abscissa.}
    \label{fig:corrplts}
\end{figure*}

We can also examine the relationship between $v_0$, IR emission, and mass surface density on cloud-averaged scales, as seen in Figure~\ref{fig:corrplts}.  {Here the error bars are conservatively drawn to indicate the standard deviation of the individual values of $v_0$ across the set of dendrogram structures, or across the cloud in the case of the abscissa.}
We see again that the cloud-averaged $v_0$ (shown as the vertical axis on each panel) correlates well with the median 8$\mu$m intensity; the upper left panel of Figure~\ref{fig:corrplts} shows this is consistent across all clouds and both line tracers (\twco\ and \ttco).  
{The observed trend is driven in part by the much higher $v_0$ values in 30 Dor, but including values for the N113 cloud, based on a preliminary analysis of ALMA \twco\ (2--1) and \ttco\ (2--1) maps (Sewi{\l}o et al., in preparation), further strengthens the case for a general trend across clouds.}

Considering that 8$\mu$m is a complex tracer (attributed to FUV-irradiated PAH molecules) that could be intrinsically related to molecular gas column density, we can check whether this correlation persists with other tracers of star formation activity.  As the two panels on the right {side of Figure~\ref{fig:corrplts}} show, a comparably good correlation is seen with 24$\mu$m intensity and dust temperature, both of which should be responsive to the FUV radiation field in ways that are independent of the PAH emission.  At the same time, $v_0$ also shows a good correlation with dust column density inferred from FIR emission ({\it bottom left panel}).  This provides further evidence that both star formation activity and mass surface density contribute to setting the characteristic $v_0$ for a cloud.

\begin{figure*}
    \includegraphics[width=\textwidth]{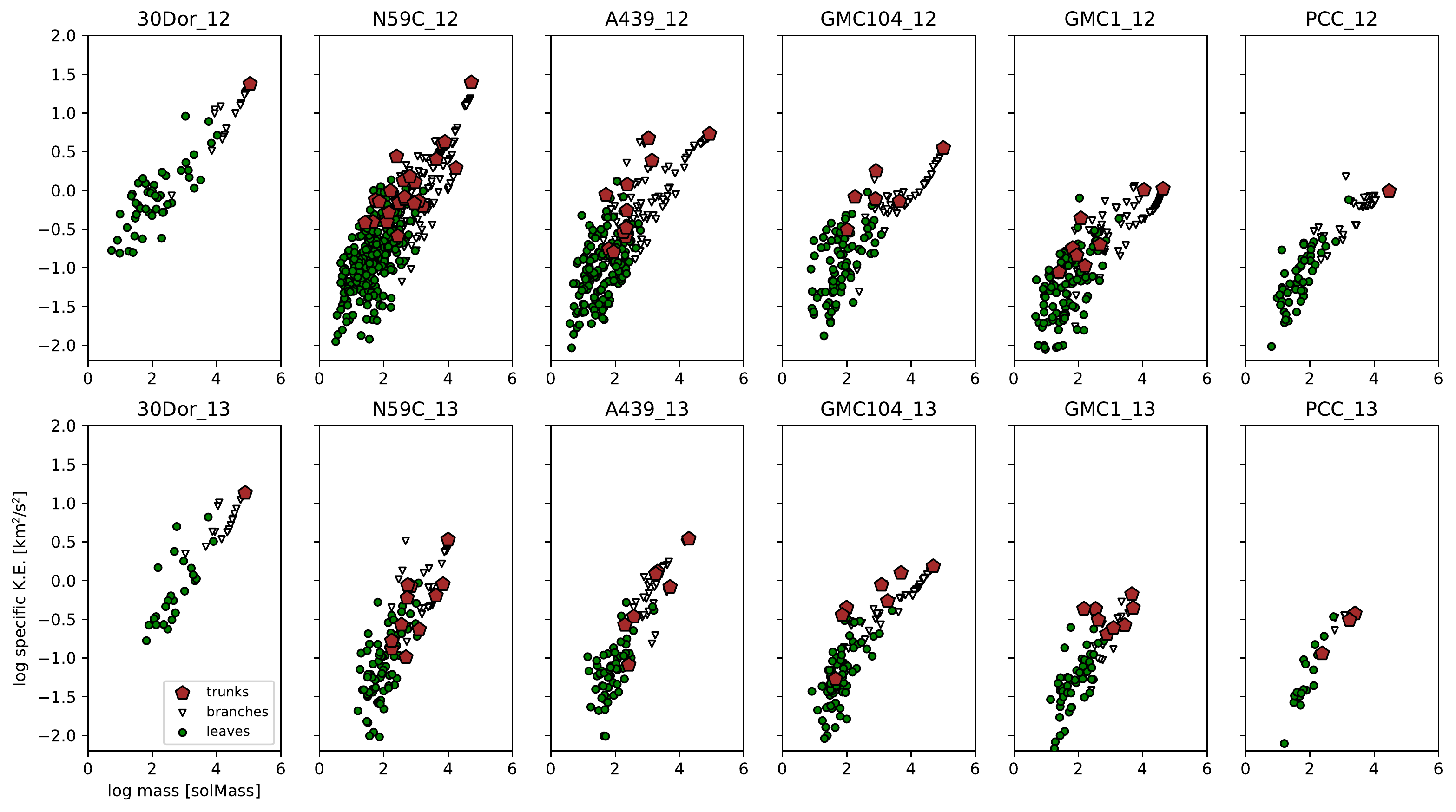}
    \caption{Specific kinetic energy of dendrogram structures (expressed as $\sigma_v^2$) as a function of structure equivalent radius.  Each panel shows structures identified in a single cloud with a single tracer (\twco\ or \ttco).  Trunks, branches, and leaves of the dendrogram are distinguished by different plot symbols.  The clouds are ordered by decreasing star formation activity, as traced by median 8$\mu$m surface brightness.  The close connection between star formation activity and specific kinetic energy is apparent in this figure.}
    \label{fig:energy}
\end{figure*}

\section{Discussion and Conclusions}\label{sec:disc}

The main results of our study can be summarized as follows.  Firstly, while there is substantial scatter in the line width at a given size within a cloud, the scatter among clouds is even larger, allowing one to meaningfully assign a mean value of the normalization $v_0$ to each cloud in our sample.  Secondly, the characteristic $v_0$ for a cloud correlates independently with both its mean IR brightness and its mean surface density---although we emphasize that the surface density is inferred from CO or \ttco\ line intensity, which may be sensitive to variations in radiative excitation and gas velocity dispersion.  Our results are consistent with at least two interpretations.  The first is that higher line widths are due to energetic feedback from recent star formation, which tends to occur in high density regions.  The second is that the higher line widths are due to gravity-driven motions in high column density structures, with some of these structures collapsing to form stars.

The ambiguity between these two interpretations reflects the ongoing debate over which of the possible sources of turbulence in the interstellar medium dominates: gravitational collapse, stellar feedback, or large-scale galactic dynamics \citep[e.g.,][]{Padoan:16,Ballesteros:11a,Krumholz:16}.  It is reasonable to expect a combination of these factors to be at work, with the balance between them shifting as a function of spatial scale.  Our ALMA observations probe scales of 1--50 pc, substantially smaller than the diameters of ionized superbubbles \citep{Chu:95} or the inferred disk thickness \citep{Elmegreen:01}.  Thus we lack sensitivity to energy injection on larger scales, although large-scale flows can still generate a turbulent cascade to the smaller scales that we probe \citep{Klessen:10}.  A model by \citet{Krumholz:16} offers a simple prediction for distinguishing feedback from gravitationally dominated turbulence on kiloparsec scales.  Relying on a state of vertical dynamical equilibrium and marginal Toomre stability, they predict ${\rm SFR} \propto \sigma_v^2$ for a feedback-driven model and ${\rm SFR} \propto f_g^2\sigma_v$ for a gravity driven model, where $f_g$ is the gas fraction.  

However, these scalings may not apply on the scales of individual molecular clouds, where equilibrium may not hold and the gas density can deviate significantly from the critical Toomre value.  
{On the scale of individual molecular clouds, feedback simulations predict that the integrated star formation efficiency increases with $\Sigma$, since more feedback is required to halt collapse, and the ratio of the self-gravity force to the rate at which feedback injects momentum depends on surface density \citep[e.g.,][]{Fall:10,Raskutti:16,Grudic:18}.  This is qualitatively consistent with our finding of a steep dependence of star formation rate on $\Sigma$ (Figure~\ref{fig:comp:svir}), although a quantitative comparison will require taking a detailed census of recent star formation in these clouds.}

In the absence of a simple analytic prediction to distinguish {feedback-driven} from gravitationally-driven turbulence, we can examine whether the highest dispersion molecular gas is associated with small or with large structures.  We noted previously in \citetalias{Wong:17} that in the PCC, and to some extent in the 30 Dor cloud, some of the highest dispersions are found in the smallest resolved structures (the dendrogram leaves, green symbols in Figure~\ref{fig:rdv12}).  This is most easily apparent in the $R$--$\sigma_v$ diagram, or equivalently a plot of specific kinetic energy ($\sigma_v^2$) as a function of size scale.  We present such a plot in Figure~\ref{fig:energy}, now including all six clouds.  We note that the highest specific K.E. is generally associated with the largest structures, especially when using the \ttco\ line, which is less sensitive to opacity broadening (see below).  The most quiescent clouds, PCC and GMC1, show a somewhat flatter upper envelope in the distribution of points, indicating significant energy injection on small scales.  The other clouds, especially GMC104 and A439, seem more consistent with a simple energy cascade from large to small scales.  It therefore appears difficult to ascribe a single dominant mechanism for turbulent energy injection across the entire sample.

We note that interpretations of $v_0$ in terms of feedback or gravity rely on our ability to infer physical quantities from measured observables.  In doing so we should be aware of the following caveats.
\begin{enumerate}
    \item {\it Opacity broadening:} \citet{Hacar:16} have analyzed the overestimate of the intrinsic velocity dispersion that results from high line opacity.  The overestimate amounts to up to a factor of 2--3 in $\sigma_v$ (and thus up to an order of magnitude in $\Sigma_{\rm vir}$) in the case of the \twco\ line, and may be responsible for many of the points which scatter above the virial equilibrium line in Figure~\ref{fig:comp:bnd} ({\it left}).  We note, however, that we obtain consistent results for $v_0$ from {both the \twco\ and \ttco\ analyses, while the latter} should be less susceptible to (though not immune from) opacity effects.
    \item {\it Mass uncertainties:} We have estimated structure masses using either a constant $X$-factor ($\Sigma_{\rm CO}$) or a simple LTE analysis ($\Sigma_{\rm LTE}$), both of which involve simplifying assumptions and require adopting highly uncertain abundance and line ratios.  In particular, $\Sigma_{\rm LTE}$ is based on an assumed $T_{\rm ex}$ that is likely underestimated due to beam dilution (\S\ref{sec:lte}).  Although a systematic error in mass would not affect our characterization of the size-line width relation, it would affect our ability to disentangle the influences of line width and surface density on 8$\mu$m emission.
\end{enumerate}

Although we are unable to identify the dominant driving mechanism for turbulence in our clouds, there are some obvious next steps that will bring us closer to doing so.  Direct comparison with numerical simulations, translated into the observational domain using radiative transfer modeling, will help interpret the scatter in line width, particularly on small scales.  Ongoing wide-field mapping of the LMC with the Morita 7-m array of ALMA will extend our analysis to larger scales and thus better constrain the form of the $R$--$\sigma_v$ correlation.  In addition, improved characterization of the dust mass, ionized gas properties, and the young stellar population, made possible with near-infrared and optical surveys and (in the near future) with the James Webb Space Telescope (JWST), will constrain more tightly the available energy from stellar feedback and provide mass estimates that are independent of CO emission.  These advances should enable substantial progress is characterizing the energy flow within turbulent molecular clouds.

{Images and analyses presented in this paper are available for download at the URL \url{https://mmwave.astro.illinois.edu/almalmc/}, or an updated URL can be found at the \href{https://arxiv.org}{arxiv.org} listing for this paper.}

\acknowledgments
{We thank Mike Grudi\'c, Michael Fall, and the anonymous referee for helpful comments on an earlier draft of this paper.}
We thank Dyas Utomo for assistance with the dust maps.
The Mopra radio telescope is part of the Australia Telescope National Facility which is funded by the Australian Government for operation as a National Facility managed by CSIRO.
This paper makes use of the following ALMA data: ADS/JAO.ALMA \#2011.0.00471.S, ADS/JAO.ALMA \#2013.1.00832.S, and ADS/JAO.ALMA \#2016.1.00193.S. ALMA is a partnership of ESO (representing its member states), NSF (USA) and NINS (Japan), together with NRC (Canada), NSC and ASIAA (Taiwan), and KASI (Republic of Korea), in cooperation with the Republic of Chile. 
The Joint ALMA Observatory is operated by ESO, AUI/NRAO and NAOJ.
The National Radio Astronomy Observatory is a facility of the National Science Foundation operated under cooperative agreement by Associated Universities, Inc.
R.I. acknowledges support from NSF AAG award 1313276.
A.H. acknowledges support from the Centre National d'Etudes Spatiales (CNES).
K.T. was supported by NAOJ ALMA Scientific Research Grant Number 2016-03B.
The work of M. S. was supported by NASA under award number 80GSFC17M0002.
Part of this research was conducted at the Jet Propulsion Laboratory, California Institute of Technology, under contract with the National Aeronautics and Space Administration.
R.S.K. and	S.C.O.G. acknowledge financial support from the German Science Foundation (DFG) in the Collaborative Research Center SFB 881 {\em The Milky Way System} (subprojects B1, B2, and B8).
This research made use of {\tt astrodendro}, a Python package to compute dendrograms of astronomical data, SCIMES, a Python package to find relevant structures in dendrograms of molecular gas emission using the spectral clustering approach, and Astropy, a community-developed core Python package for astronomy.

\facility{Mopra, ALMA, Herschel, Spitzer}

\software{CASA \citep{McMullin:07}, {\tt astrodendro} (\url{http://www.dendrograms.org}), Kapteyn \citep{KapteynPackage}, Scipy \citep{Jones:01}, SCIMES \citep{Colombo:15}, Astropy \citep{Astropy:13}.}

\bibliographystyle{aasjournal}
\bibliography{almacyc4}

\end{document}